%% file: hsc_dr3_lowres.tex
\documentclass{pasj01}
\usepackage{url}

\def\uncatcodespecials{\def\do##1{\catcode`##1=12}\dospecials}%
{\catcode`\`=\active\gdef`{\relax\lq}}
\def\setupcode 
    {\tt %
     \spaceskip=0pt \xspaceskip=0pt 
     \catcode`\`=\active
     \uncatcodespecials \obeyspaces
     \catcode`\{=1\catcode`\}=2
    }%
\def\SETUPCODE 
    {\tt %
     \spaceskip=0pt \xspaceskip=0pt 
     \catcode`\`=\active
     \uncatcodespecials \obeyspaces
    }%
\def\docode#1{#1\endgroup}%
\def\code{\begingroup\setupcode\docode}%

\begin{document}
\SetRunningHead{HSC et al.}{HSC-SSP PDR3}

\title{Third Data Release of the Hyper Suprime-Cam\\Subaru Strategic Program} 

\input{author}

\altaffiltext{\ }{\ }
\altaffiltext{*}{\small The corresponding author is Masayuki Tanaka.}
\email{masayuki.tanaka@nao.ac.jp}

\KeyWords{Surveys, Astronomical databases, Galaxies: general, Cosmology: observations}

\maketitle
\definecolor{gray}{rgb}{0.6, 0.6, 0.6}
\newcommand{\commentblue}[1]{\textcolor{blue} {\textbf{#1}}}
\newcommand{\commentred}[1]{\textcolor{red} {\textbf{#1}}}
\newcommand{\commentgray}[1]{\textcolor{gray} {\textbf{#1}}}


\begin{abstract}
  The paper presents the third data release of Hyper Suprime-Cam Subaru Strategic Program (HSC-SSP),
  a wide-field multi-band imaging survey with the Subaru 8.2m telescope.
  HSC-SSP has three survey layers (Wide, Deep, and UltraDeep) with different area coverages and depths, 
  designed to address a wide array of astrophysical questions.
  This third release from HSC-SSP includes data from 278 nights of observing time and covers about 670 square degrees in all five broad-band filters
  at the full depth ($\sim26$~mag at $5\sigma$) in the Wide layer.
  If we include partially observed area, the release covers 1,470 square degrees.
  The Deep and UltraDeep layers have $\sim80\%$ of the originally planned integration times, and are considered done, as
  we have slightly changed the observing strategy in order to compensate for various time losses.
  There are a number of updates in the image processing pipeline.  Of particular importance is the change
  in the sky subtraction algorithm; we subtract the sky on small scales before the detection and measurement stages, which
  has significantly reduced false detections.  Thanks to this and other updates, 
  the overall quality of the processed data has improved since the previous release.
  However, there are limitations in the data (for example, the pipeline is not optimized for crowded
  fields), and we encourage the user to check the quality assurance plots as well as a list
  of known issues before exploiting the data.
  The data release website is \url{https://hsc-release.mtk.nao.ac.jp}.
\end{abstract}

\section{Introduction}

The Hyper Suprime-Cam Subaru Strategic Program (HSC-SSP; \cite{aihara18a}) is a three-tiered imaging survey
aimed to address a wide range of astrophysical questions ranging from cosmology to solar system bodies.
The survey uses Hyper Suprime-Cam (HSC; \cite{miyazaki18}), a wide-field imaging camera installed at
the prime focus of the Subaru 8.2m telescope on the summit of Maunakea, Hawaii.
In its widest component (the Wide layer), we cover about 1200 deg$^2$ of the sky mostly around
the celestial equator in five broad-band filters ($grizy$; \cite{kawanomoto18}) with integration times of 10-20~min.
A particular emphasis is put on the $i$-band, with 
which we measure precise shapes of galaxies for weak-lensing cosmology.  The second component, the Deep layer,
covers four separate fields, each of which is $\sim7$ deg$^2$, both in the broad-band filters and three narrow-band filters.
The last component, UltraDeep, is two fields centered at the COSMOS and Subaru XMM-Newton Deep Field (SXDS).
We take very long integrations ($\sim5-10$ hours in each band) to peer deep into the distant Universe.
These components are designed to be used in conjunction with each other in order to enable a wide array of
scientific explorations.

The survey was originally awarded 300 nights of observing time.
This is the largest observing program ever approved at Subaru.  While we have made good progress with
the observations, we have suffered from various issues such as bad weather, telescope trouble, and
seismic activities in Hawaii during the course of the survey.
In order to compensate for the time lost due to these issues, an additional
30 nights were recently allocated to the survey, giving a total of 330 nights.  We have also made changes to the observing strategy
as we discuss in Section \ref{sec:survey_progress} to catch up with the original survey plan.

HSC-SSP has made two public releases so far \citep{aihara18b,aihara19}.  In addition to these major releases,
we have made several incremental releases to add further value.  The public data release 2 (PDR2) has accompanied
five incremental releases and the following products were made available in each release: (1) revised bright star masks and shape
catalog \citep{mandelbaum18}, (2) photometric redshifts \citep{tanaka18,nishizawa20},  (3) emission-line
object catalog \citep{hayashi20} and deblended images ({\tt heavyFootprint}), (4) narrow-band data from
CHORUS (Cosmic HydrOgen Reionization Unveilled with Subaru; \cite{inoue20}),
and (5) galaxy density map from \citet{shimakawa21}.  This paper presents a new major release, PDR3.
PDR3 was originally anticipated to be the final data release, but due to the additional time awarded to the survey,
PDR4 will be the final release as we discuss later in Section \ref{sec:summary}.

The structure of the paper is as follows.  We first give a brief overview of the release in Section \ref{sec:the_release},
followed by updates in the processing pipeline in Section  \ref{sec:pipeline_updates}.  We then summarize data products
included in the release in Section \ref{sec:data}.  Section \ref{sec:data_quality_and_known_issues} evaluates
the quality of the data and discusses issues that the user should be aware of.  Section \ref{sec:status_of_collaborating_surveys} gives
updates on our collaborating surveys.  Finally, we discuss prospects for the final data release, and conclude the paper
in Section \ref{sec:summary}.  All magnitudes quoted in the paper are AB magnitudes \citep{oke83}.

\section{Overview of the Release}
\label{sec:the_release}

\subsection{Updates from PDR2}

This release includes data taken between March 2014 and January 2020 from 278 nights in total (including nights
lost to weather and other reasons).  The release includes data from an additional $\sim100$ nights since PDR2 and hence represents
a major increase in terms of the data volume and area coverage.  There are also changes in the data products.
Major updates from PDR2 can be summarized as follows.

\begin{figure*}
  \begin{center}
    \includegraphics[width=18cm]{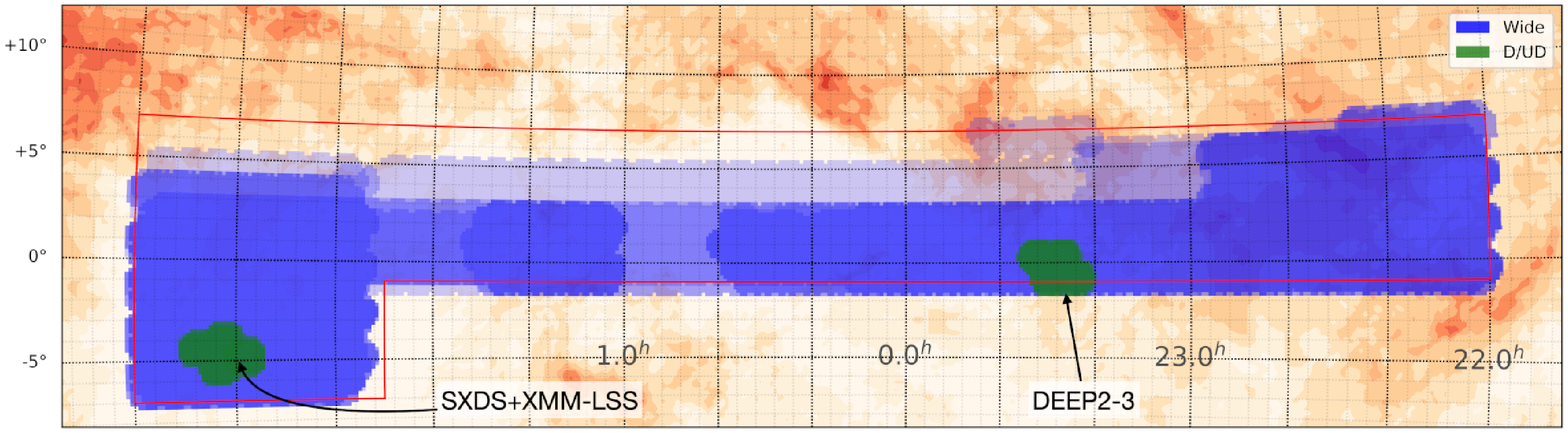}\vspace{1cm}
    \includegraphics[width=18cm]{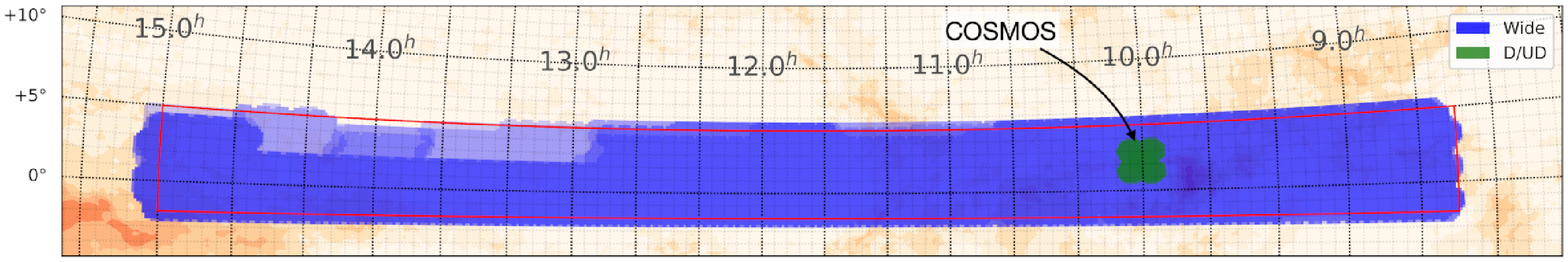}\vspace{1cm}
    \includegraphics[width=18cm]{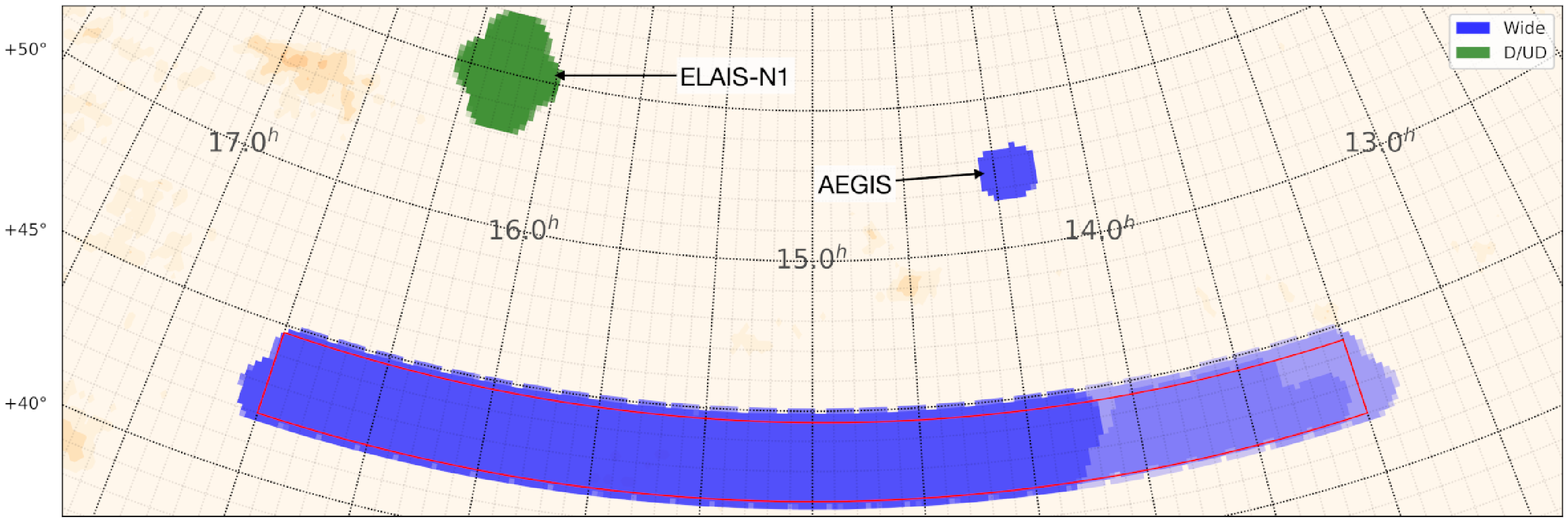} 
  \end{center}
  \caption{
    The area covered in this release.
    The blue and green areas show the Wide and Deep/UltraDeep layers, respectively.
    For the Wide layer, the shading indicates the number of filters in which data are available;
    the darkest blue corresponds to all five filters.
    The red boxes indicate the approximate boundaries of the three disjoint regions that will make up the final Wide survey.
    Note that AEGIS is a calibration field observed at the Wide depth.
    The Galactic extinction map from \citet{schlegel98} is shown in the background.
  }
  \label{fig:sky_coverage}
\end{figure*}

\begin{itemize}
\item
  The Deep/UltraDeep (D/UD for short) layers are now $\sim80\%$ complete and we choose to stop there
  (i.e., we do not take any further data in D/UD).  Most of the Deep fields have $\sim1.5$ times longer
  integration times compared to PDR2.  UD-SXDS is much deeper ($\times 3$ times longer integration times)
  than PDR2 due to a transient survey carried out in that field.  The overall increase in D/UD-COSMOS
  is relatively minor because the transient survey in that field was performed earlier in the survey\citep{yasuda19}.
  NB1010 data are newly available in UD.
\item
  There is also major progress in the Wide layer. The full-color full-depth (FCFD, i.e., observed to
  the required depth in all five filters) area increased from 300 deg$^2$
  to 670 deg$^2$.  The area coverage is shown in Fig.~\ref{fig:sky_coverage}.
  Statistical properties of the data are summarized in Table~\ref{tab:exptime}.
\item
  We have changed the observing strategy as we detail in Section \ref{sec:survey_progress}.  In short,
  we have relaxed the seeing constraint in the $i$-band, reduced the exposure times in the $izy$ bands in the Wide layer,
  and declared the D/UD layers done at 80\% of the planned exposure times.
\item
  In order to astrometrically calibrate many visits available in the UD fields, we switched from the {\tt meas\_mosaic}
  algorithm \citep{bosch18}, which was used until PDR2, to {\tt jointcal}.
  {\tt jointcal} uses sparse linear algebra for efficient memory usage.
  The astrometric accuracy remains similar to the previous releases.
\item
  We use a new photometric calibration algorithm, the Forward Global Calibration Method (FGCM; \cite{burke18}),
  which forward-models the atmosphere and system response as a function of time and position (Section \ref{sec:fgcm}).
  There is, however, a small error in the implementation, which resulted in spatially varying offsets
  in the photometric zero-point at a few percent level across the survey field.  These offsets are corrected for using the location
  of the stellar sequence in color-color diagrams (Section \ref{sec:stellar_sequence_regression}).
\item
  The pipeline performs global sky subtraction for extended object science as in the previous release.
  We have improved the global sky subtraction algorithm (Section \ref{sec:improved_global_sky_frame}), but
  extended wings of stars and bright galaxies still introduce false detections and measurement failures.
  Thus, we introduce a second, local sky subtraction
  just before object detection and detailed measurements (Section \ref{sec:local_sky_subtraction}).
  While there is a single object catalog, there are two types of coadd images (local sky vs. global sky) and
  the user should choose which one to use depending on their science goals.
\item
  We use the fifth order Lanczos kernel, as opposed to the third order we have used before, to warp images for coaddition.
  This improves the PSF model on the coadds and reduces the fractional PSF size residual defined in
  Section \ref{sec:shape_measurements} to about 0.1\% \citep{li21}.
\item
  Utilizing the effective filter transmission introduced in \citet{aihara19}, we have estimated
  corrections to translate $r/i$-band magnitudes into $r2/i2$ band magnitudes (Section \ref{sec:effective_filter_response}).
\item
  As in PDR2, we do not include detailed shape measurements as well as deblended images ({\tt heavyFootprint}) in the release.
  The shape measurements in PDR2 are still withheld, but as we discuss in Section  \ref{sec:shape_measurements},
  the shapes are not useful, and we instead plan to release a shape catalog based on a newer internal data release
  to the community.  The deblended images from PDR3 will be made available in August 2022.
\end{itemize}

\begin{table*}[htbp]
  \begin{center}
    \begin{tabular}{l|ccccccccc}
      \hline\hline
      {\bf Wide}            &  $g$                  &  $r$                 &  $i$                 &  $z$                 &  $y$                 & & & & \\
      exposure (min)        &  $10^{+2}_{-2}$        & $10^{+2}_{-2}$        & $20^{+3}_{-6}$        & $20^{+3}_{-10}$        & $20^{+3}_{-10}$        & & & & \\
      seeing (arcsec)       &  $0.79^{+0.09}_{-0.08}$ & $0.75^{+0.13}_{-0.09}$ & $0.61^{+0.05}_{-0.05}$ & $0.68^{+0.08}_{-0.06}$ & $0.68^{+0.10}_{-0.08}$ & & & & \\
      depth (mag)           &  $26.5^{+0.2}_{-0.2}$   & $26.5^{+0.2}_{-0.2}$  & $26.2^{+0.2}_{-0.3}$   & $25.2^{+0.2}_{-0.3}$   & $24.4^{+0.2}_{-0.3}$   & & & & \\
      saturation (mag)      &  $17.4^{+0.6}_{-0.4}$   & $18.1^{+0.5}_{-0.5}$  & $18.3^{+0.5}_{-0.3}$   & $17.5^{+0.5}_{-0.4}$   & $17.0^{+0.5}_{-0.7}$   & & & & \\
      area (deg$^2$)        &  1332                  & 1298                 & 1264                  & 1299                  & 1209                  & & & & \\
      \hline
      {\bf Deep/UltraDeep}  &       $g$            &         $r$          &       $i$            &        $z$           &        $y$           &       $NB387$        &      $NB816$         &      $NB921$         & $NB1010$ \\
      exposure (min)        & $70^{+21}_{-21}$      & $66^{+17}_{-17}$       & $98^{+46}_{-32}$      & $177^{+130}_{-46}$      & $93^{+23}_{-23}$      & $68^{+13}_{-13}$       & $120^{+30}_{-15}$     & $168^{+14}_{-28}$     & $705^{+45}_{-345}$\\
      seeing (arcsec)       & $0.83^{+0.05}_{-0.12}$ & $0.77^{+0.04}_{-0.04}$ & $0.66^{+0.07}_{-0.06}$ & $0.78^{+0.02}_{-0.03}$ & $0.70^{+0.04}_{-0.05}$ & $0.82^{+0.07}_{-0.08}$ & $0.70^{+0.07}_{-0.08}$ & $0.67^{+0.04}_{-0.04}$  & $0.77^{+0.02}_{-0.02}$\\
      depth (mag)           & $27.4^{+0.2}_{-0.2}$  & $27.1^{+0.1}_{-0.2}$   & $26.9^{+0.2}_{-0.3}$   & $26.3^{+0.1}_{-0.3}$   & $25.3^{+0.2}_{-0.2}$   & $25.0^{+0.2}_{-0.2}$  & $26.0^{+0.2}_{-0.2}$   & $25.9^{+0.2}_{-0.2}$   & $24.2^{+0.2}_{-0.5}$\\
      saturation (mag)      & $18.0^{+0.4}_{-0.5}$  & $18.2^{+0.4}_{-0.4}$   & $18.6^{+0.3}_{-0.4}$   & $17.7^{+0.3}_{-0.3}$   & $17.4^{+0.3}_{-0.3}$   & $14.8^{+0.4}_{-0.3}$  & $16.8^{+0.4}_{-0.4}$   & $16.9^{+0.4}_{-0.3}$   & $14.8^{+0.2}_{-0.2}$\\
      area (deg$^2$)        & 36                   & 36                   & 36                   & 37                   & 36                   & 30                   & 33                   & 33 & 5\\
      \hline \hline
    \end{tabular}
  \end{center}
  \caption{
    Approximate exposure time, seeing, $5\sigma$ depth,
    and saturation magnitudes for each filter and survey layer,
    averaged over the area included in this release.
    The depth and saturation magnitudes are for point sources.
    The numbers are the median and the quartiles of the distribution,
    except for the area, which shows the total area covered in at least one exposure.
    The NB1010 coverage is small but that is because it is used only in the UD fields.
    The numbers for the Wide layer are close to the full-depth values, while
    those for the D/UD are closer to the Deep depth due to the spatial averaging
    (Deep is wider than UD).  UD is roughly 0.8 magnitude deeper than Deep.
    The user is referred to quality assurance (QA) plots available at the data release site
    for the spatial variation of some of these numbers.
  }
  \label{tab:exptime}
\end{table*}

\subsection{Survey Progress and Changes in the Survey Strategy}
\label{sec:survey_progress}

\begin{figure}
  \begin{center}
  \includegraphics[width=9cm]{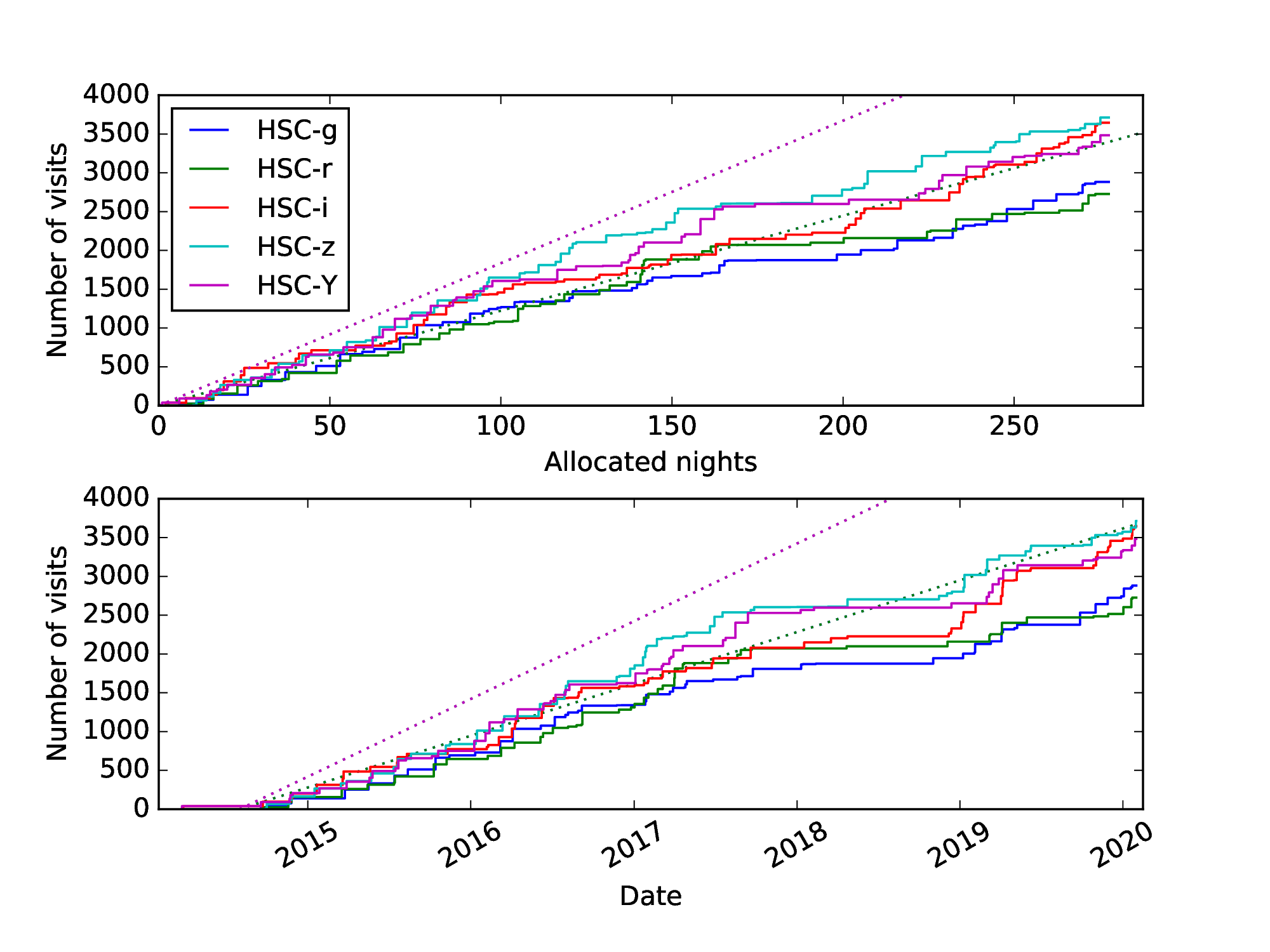} 
  \end{center}
  \caption{
    {\bf Top}: Cumulative number of visits for the Wide layer as a function of the number of allocated nights.
    The upper dotted line is for the $g$ and $r$-bands and it indicates the mean survey speed required to complete
    the survey in 300 nights.
    The lower dotted line is for the other bands, but note that 6 dithers per pointing is assumed.
    The solid lines are the achieved survey speed for each filter as shown in the legend.
    {\bf Bottom}: Same plot as above but as a function of date.
  }
  \label{fig:survey_progress}
\end{figure}

Here, we briefly summarize where we stand in terms of the survey progress.  Fig.~\ref{fig:survey_progress}
compares the expected survey speed and actual survey progress.  This is for the Wide layer, but
as we spend 2/3 of the observing time for Wide, it is a good proxy for the overall survey progress.
There is a big plateau in 2018 when survey progress was very slow.  This is for a combination of
reasons; bad weather, telescope troubles, and earthquakes triggered by volcanic activity in Hawaii.
The progress in 2019 was approximately at the expected survey speed, but overall we are significantly behind
the original survey plan.
We have been awarded an additional 30 nights to compensate for the loss, but in order to further catch up
with the plan, (1) we decided to stop at $\sim80\%$ of the original integration time in the D/UD fields,
(2) we relaxed the seeing constraint in the $i$-band in the Wide layer from $\sim0.7$ to $\sim0.9$ arcsec,
although we still try to observe in the $i$-band under good seeing conditions whenever possible,
and (3) we reduce the exposure time from 20~min to 16.7~min (from 6 to 5 dithers) in the $izy$ bands in Wide.
We applied these changes in observation strategy after November 2018.
We made major progress in the $i$-band in 2019 thanks to the relaxed seeing constraint, but the median
seeing is still as good as $\sim0.7$ arcsec for the data taken after November 2018.
With these changes, we expect to cover $\sim1,200$ deg$^2$ in all the bands at the full depth by the end of the survey.

\begin{figure}
  \begin{center}
  \includegraphics[width=8cm]{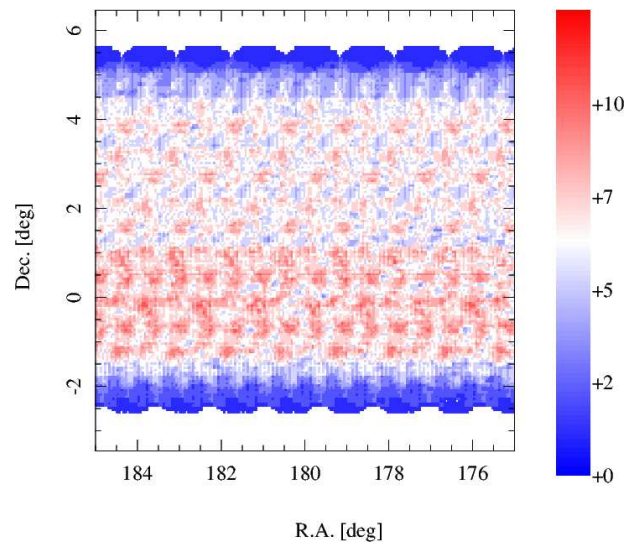} 
  \end{center}
  \caption{
    A portion of the Wide area in this release.  The color coding shows the number of visits in the $i$-band.
    The bottom half of this area was observed with 6 dithers, while the top half was with 5 dithers.
  }
  \label{fig:input_counts}
\end{figure}

This change in the dithering strategy is illustrated in Fig.~\ref{fig:input_counts}.
Due to the reduced exposure time, the 5-dither region is shallower than the 6-dither region by $\sim0.1$~mag.
This is not a major effect and the seeing may be more important for the depth, especially for compact sources.
The reduced dithers also mean that there is less spatial averaging of the PSF, which may 
have a non-negligible effect on the PSF model, especially from weak-lensing perspectives.
\citet{li21} carry out PSF model tests using the $i$-band data in the Wide area around the region in Fig.~\ref{fig:input_counts},
which has large enough areas of both 5-dither and 6-dither regions to provide sufficient statistical
power to examine the differences. The fractional PSF size residual defined as
$f_{\delta\sigma}=(\sigma_{\rm PSF}-\sigma_*)/\sigma_*$, where $\sigma_*$ is the measured size of
a star and $\sigma_{\rm PSF}$ is the size of the PSF model evaluated at the position of the star,
is computed as a function of the $i$-band magnitude.
Fig.~20 of \citet{li21} shows that small differences between the two dithering
strategies as a function of magnitude.
However, these differences  seem to be driven by a few regions with extreme size residuals.
Once these areas are excised out, 
the fractional size residuals are similar
irrespective of the dithering strategy. These tests were carried out
using the internal S19A release, but the same conclusion should hold in this release, as there is no
change in the way we model the PSF.

\subsection{Shape Measurements}
\label{sec:shape_measurements}

As we did in PDR2, we withhold detailed shape measurements as well as deblended images from this release.
We plan to release the shapes once their characteristics 
are fully understood and ready to be used for weak-lensing science.  We will make the deblended images
available to the community in August 2022.

The galaxy shapes from PDR2 are still withheld.  Detailed investigations have shown that
the PDR2 shapes are problematic; this is likely due to the global sky subtraction algorithm introduced in PDR2.
Although the modifications to the sky subtraction vastly improved the over-subtraction of background around bright nearby galaxies,
it resulted in a considerable population of faint objects with anomalously large size measurements.
We discuss this issue further in Section~\ref{sec:overestimated_cmodel_fluxes}.
We did not see this effect in image simulations performed to calibrate the weak lensing shape measurements in PDR2,
meaning that the PDR2 shapes were unusable for precision weak lensing studies.
As we discuss in Section~\ref{sec:local_sky_subtraction}, we have gone back to the local sky subtraction
scheme for detections and measurements after PDR2, and the weak lensing working group is utilizing
the shapes from the S19A internal data release \citep{li21}, which will eventually be released
to the community.

\subsection{Previous Internal Releases}
\label{sec:previuos_internal_relesaes}

A public data release from HSC-SSP is based on an internal data release made $\sim1$ year ago.
We briefly summarize our internal releases since PDR2 in Table~\ref{tab:data_releases}.
We changed the observing strategy between S18A and S19A (Section \ref{sec:survey_progress}).  That means that the definition
of FCFD changed and the increase in the FCFD area in S19A is partially due to the change.
We also note that the decrease in the number of objects in S19A, particularly in D/UD, is because we changed the sky subtraction
scheme and the number of spurious sources dropped significantly
(Section~\ref{sec:local_sky_subtraction}).

\begin{table*}[htbp]
  \begin{center}
    \begin{tabular}{l|ccccrrc}
      \hline
      Release               & Date       & Layer          & N      & \multicolumn{1}{c}{Area}        & \multicolumn{1}{c}{N}      & hscPipe \\
                            &            &                & filter & \multicolumn{1}{c}{(deg$^2$)}   & \multicolumn{1}{c}{object} & version \\
      \hline \hline
      Public Data Release 3 & 2020-08-03 & Deep/UltraDeep & 9      &                             37  &                 19,051,243 & 8\\
      (=S20A)               &            & Wide           & 5      &                    1470   (670) &                507,215,729 & 8\\
      \hline
      S19A                  & 2019-09-25 & Deep/UltraDeep & 9      &                             37  &                 18,090,313 & 7\\
                            &            & Wide           & 5      &                    1289   (560) &                433,472,409 & 7\\
      \hline
      Public Data Release 2 & 2019-05-31 & Deep/UltraDeep & 8      &                              37 &                 20,451,226 & 6\\
      (=S18A)               &            & Wide           & 5      &                    1114   (305) &                436,333,410 & 6\\
      \hline
      \hline   
    \end{tabular}
  \end{center}
  \caption{
    Summary of this public release and previous internal data releases. 
    The fifth column gives the survey area in deg$^2$ covered in at least in one filter and one exposure.
    The full-color full-depth area in the Wide survey is shown in parentheses. 
    The sixth column shows the number of primary objects.
  }
  \label{tab:data_releases} 
\end{table*}

\section{Pipeline Updates}
\label{sec:pipeline_updates}

The PDR3 data have been processed with \code{hscPipe v8}, a customization of
the LSST Science Pipelines \citep{juric17,bosch18,bosch19,ivezic19}.  This is an updated version of the pipeline 
used in PDR2. This section summarizes the improvements in the order of the processing flow.

It is useful to remind the reader of important notions in our image processing at this point.
{\tt visit} is an single exposure with HSC and has an integer number uniquely assigned.
A visit thus has 112 CCD images.  In the joint calibration stage onwards, we use {\tt tracts},
which are equi-area rectangular regions on the sky, each of which is $\sim1.7$ degree on a side
with $\sim1$ arcmin overlap with adjacent tracts.  A tract is split into $9\times9$ {\tt patches},
each of which is 4200 pixels on a side ($\sim12$ arcmin) with an overlap of 100 pixels ($\sim17$ arcsec)
on the edges.  The tracts and patches are introduced to parallize the processing and are the most
fundamental areal units.

\subsection{Jointcal}
\label{sec:jointcal}

The first is an update in the astrometric calibration.  We used to use
an algorithm termed 
\code{meas_mosaic} \citep{bosch18} to solve for astrometry in the joint calibration step, but
its memory usage becomes prohibitively expensive when we try to fit all visits
in the UD fields.  In order to improve the memory efficiency, we switch to using the
LSST \code{jointcal}\footnote{\url{https://github.com/lsst/jointcal/}} package,
which uses sparse linear algebra for efficient memory usage.
This is an older version of the code used in \citet{leget21}, with considerable
customization to integrate it with the rest of the LSST/HSC pipelines, and
a less sophisticated model chosen (conservatively) to mimic what we have used in the past
\citep{bosch18}: a single per-visit, full-focal-plane, high-order
polynomial to capture both optical distortions and atmospheric effects,
composed with an affine transform for each CCD, fit to each band separately.
The order of the per-visit polynomial has been reduced from 9th order
to 7th order, and the objective function of the fit is updated via an
efficient Cholesky rank-1 update each time an outlier is rejected, instead of
in batches.
The quality of the astrometry is unfortunately not competitive with \citet{leget21}
(our astrometry has a 2-3 times larger scatter), but it is still similar to the previous releases.

\subsection{FGCM}
\label{sec:fgcm}


Photometric calibration is performed with the Forward Global Calibration
Method (FGCM; \cite{burke18}).  FGCM calibrates the full PDR3 dataset with
a forward model approach that uses atmospheric model parameters in
conjunction with scans of the instrument throughput as a function
of wavelength.  FGCM was originally developed for use with the Dark Energy
Survey data~\citep{burke18,sevilla21}, and has been incorporated into
the LSST/HSC pipeline for use with HSC and LSST data.

The FGCM model begins with measurements of the instrument throughput, including
the mirrors, filters, and detectors.  Filter scans are taken from \citet{kawanomoto18},
and we average in the azimuthal direction prior to use
in the model fit.  We use the fiducial mirror and detector
throughput measurements as reported at the Subaru website\footnote{
\url{https://subarutelescope.org/Observing/Instruments/HSC/sensitivity.html}}.
The chromatic response of the mirror changes with time as the surface
oxidizes.  The FGCM model approximately accounts for this variation by
allowing for a different rate of throughput decline in each band, such that the
bluer bands have an overall throughput that degrades more rapidly than the
redder bands.  In the PDR3 calibration run, no adjustments are made for the
intra-band throughput variation.  Finally, we note that the fiducial detector
response does not account for the differing anti-reflection (AR) coatings
across the CCD surface \citep{kamata14}
and therefore different chromatic response in the g-band.
The effect of the different AR coatings is visible in the chromatic residuals from the fit
and it is about $\pm2$\% between the first and third quartiles of the color distribution
of the stars.

The atmospheric model of FGCM is provided by MODTRAN \citep{berk99}, which has been run to
create a look-up table with the atmospheric throughput as a function of zenith
distance at the elevation of the Subaru telescope.  Light is attenuated as it
travels through the atmosphere due to absorption and Rayleigh scattering by
molecular constituents (particularly $O_2$ and $O_3$), absorption by
precipitable water vapor (PWV), Mie scattering by airborne particular aerosols,
and finally gray (achromatic) scattering by larger ice crystals and water
droplets in clouds.  We allow the aerosols and PWV to vary linearly over each
night, with a constant aerosol optical index and $O_3$ contribution per night.
See Section 3 of \citet{burke18} for details of the FGCM
atmospheric model.

The FGCM fit minimizes the variance in repeated observations of stars, and we use
stars with a signal-to-noise ratio larger than 10 within the 2 arcsec aperture, which roughly corresponds to $i<23$.
As configured, we calibrate the 2 arcsecond aperture fluxes.
Due to significant residuals in the background of the processed CCD images,
we additionally subtract off a per-star local background estimate.
Tests with a subset of the
data show that this reduces residuals as a function of magnitude, as well as
tightening the repeatability in the photometry of a given star.  The local background subtraction introduces
a few percent flux change at faint magnitudes of $i=22-23$
and is smaller at brighter magnitudes of $i\sim20$
(but still up to 0.5\% depending on the local source density) due to the over-subtraction
of the sky.  Unfortunately, we did not notice this effect until after the calibrations were
produced, and therefore the final output zero-points used in the coaddition do not account
for this effect, resulting in spatially varying photometric zero-points at a 0.5\% level,
which we correct for at a later stage (Section~\ref{sec:stellar_sequence_regression}).

As an additional constraint on the fit, we use the PanSTARRS1 (PS1; \cite{schlafly12,tonry12,magnier13,chambers16})
network of calibrated
stars to ensure uniformity across the disconnected regions of the PDR3 footprint.
We use the updated color terms from Section 4.8 of the PDR2 paper.  Only stars detected with
a signal-to-noise ratio greater than 50 in the PS1 $i$-band (roughly $i<20$) are used to constrain the fit.
The details of how a network of standard stars is incorporated into FGCM will be
described in detail in Rykoff et al. (in prep).

\begin{figure*}
    \begin{center}
        \includegraphics[width=8cm]{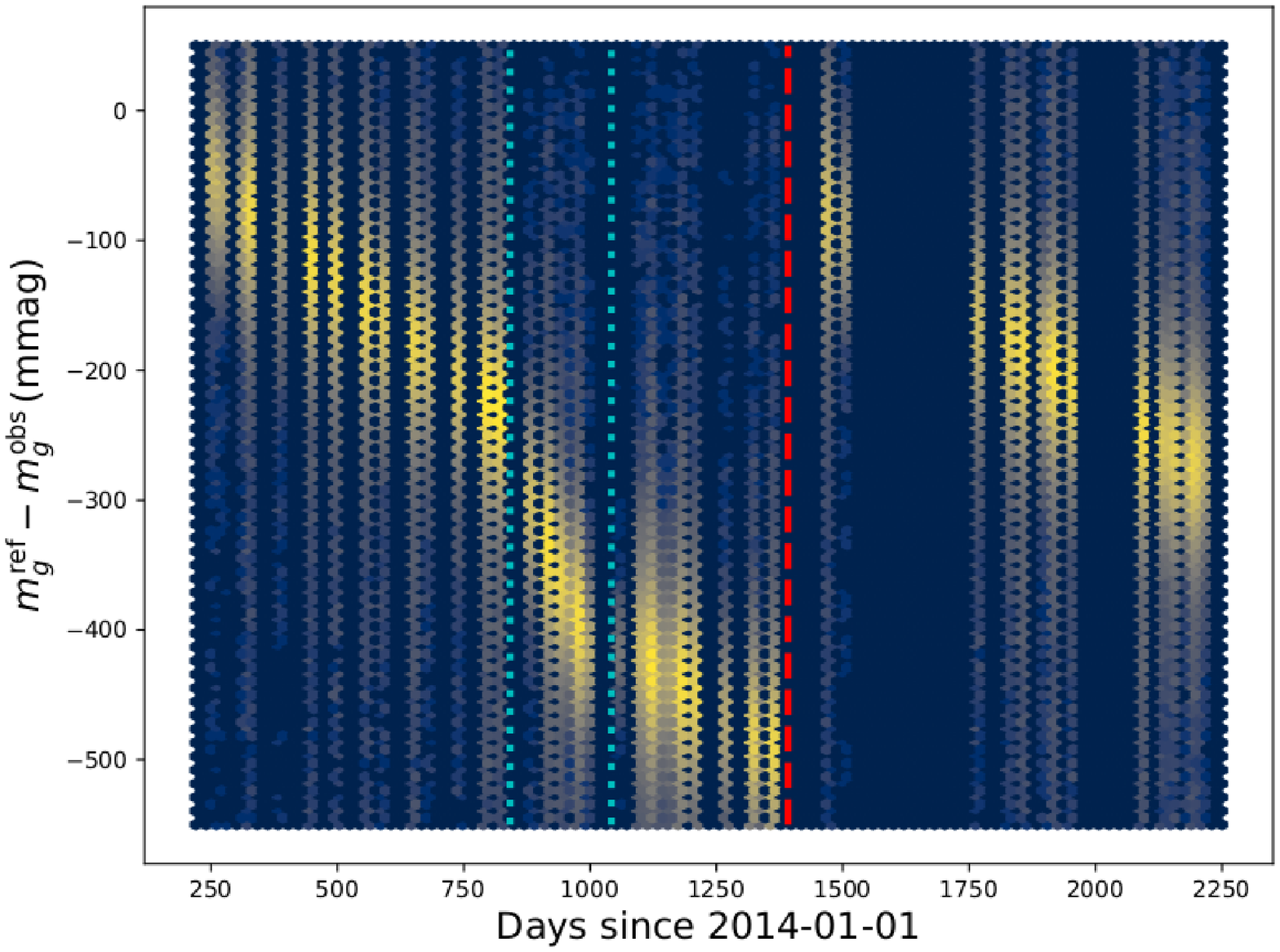}\hspace{0.5cm}
        \includegraphics[width=8cm]{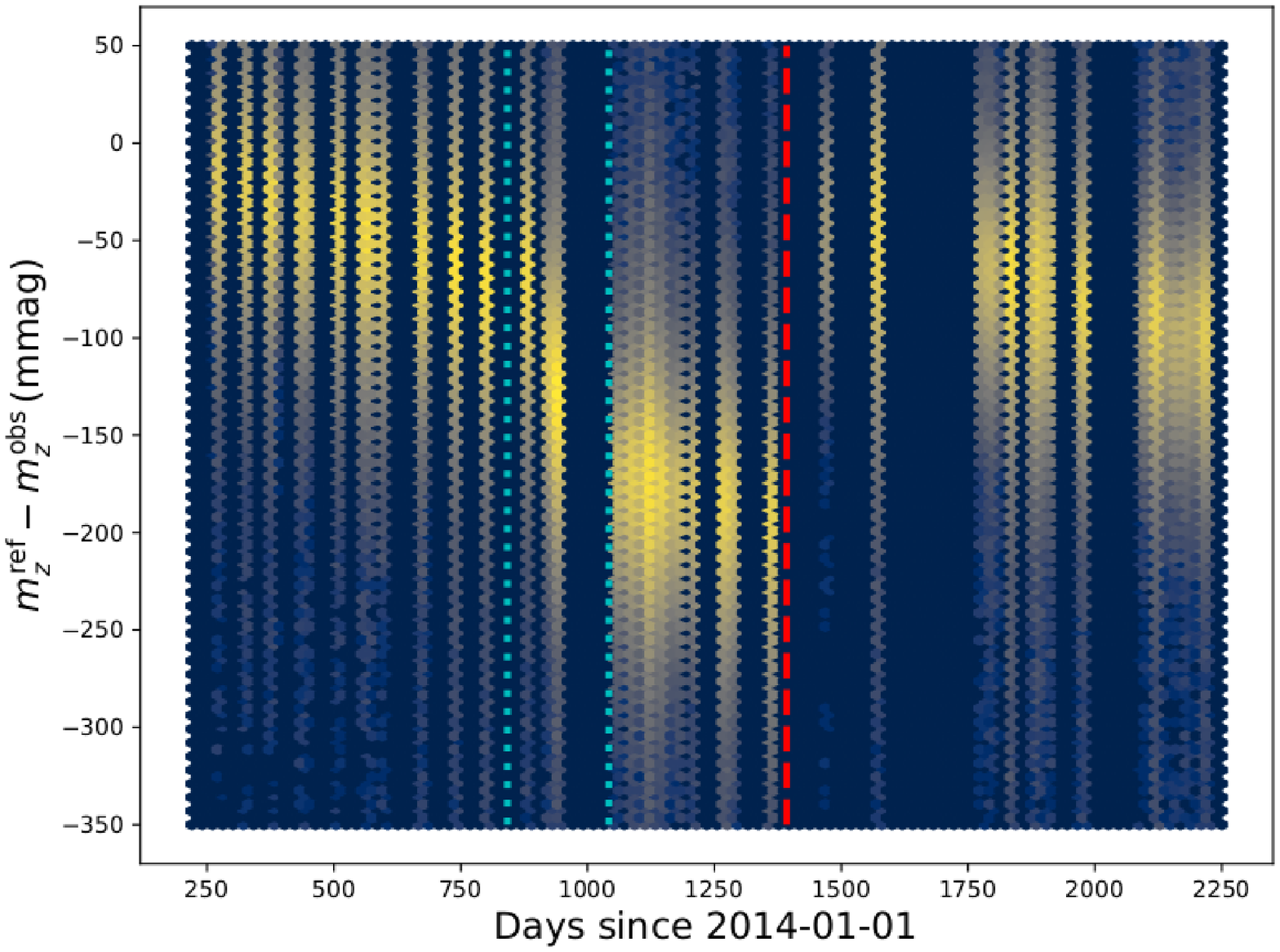}
    \end{center}
    \caption{
      Comparison of individual star observations in the $g$ (left) and $z$-bands  (right)
      between the observed HSC magnitude and the PS1 reference magnitude.  We show these two filters
      as an example, and the other filters show the same behavior.  Each point contributing
      to the heat map corresponds to a single observation of a single star at a moment in time
      (on the $x$ axis).  The value $m^{\mathrm{obs}}_{g/z}$ is the observed magnitude, corrected
      for atmospheric transmission and absolute throughput, but excluding any fit terms from
      relative system throughput.  The value $m^{\mathrm{ref}}_{g/z}$ is the PS1 reference star magnitude,
      with the default color terms applied.
      The red dashed line shows the date that the mirror was recoated.  The cyan dotted lines show
      the approximate time period that the rapid decay is observed in all bands, due perhaps to
      volcanic fog.
    }
  \label{fig:chromatic_term}
\end{figure*}

For good photometric results, we need to accurately constrain the degradation of
the mirror surface over time.  The default model in FGCM is for a piecewise-linear decay function,
with discontinuous changes when the instrument/telescope underwent changes.  We set October 2017,
at the date of the Subaru primary mirror recoating as such a discontinuous point.
Fig.~\ref{fig:chromatic_term} illustrates the throughput degradation from the fit.
After the mirror recoating, the system throughput improves by $\sim~5\%$ ($g$) and
$\sim~2\%$ ($z$).  As expected, bluer bands show a larger change.
Interestingly, we notice that the rate of decay increased between 22 April 2016 and 08 November 2016
in all bands. These dates coincide with increased activity of the Maunaloa volcano, and such
increased decay may be due to volcanic fog.

One particular challenge of performing a global calibration of the PDR3 dataset
is that we replaced the $r$ and $i$ filters with the $r2$ and $i2$ filters with improved
uniformity, respectively.
Although these filters cover similar wavelengths, they
have different chromatic response (with much smaller focal plane
variations in the replacement filters), as well as different throughputs.  The
FGCM model includes additional fit terms to constrain the relative offset of
the $r$ and $r2$ (and $i$ and $i2$) filters, using reference stars and
the limited number of stars with observations in multiple filters.
Unfortunately, these terms are somewhat degenerate with the temporal variation
of the system throughput.  Therefore, after the processing is complete we
noticed that there are significant differences in the stellar locus in color-color space,
corresponding to regions with $i$ and $i2$ observations.  These observations
and the mitigations we applied to correct for them are described in Sections \ref{sec:effective_filter_response}
and \ref{sec:stellar_sequence_regression}.

\begin{figure}
    \begin{center}
        \includegraphics[width=8cm]{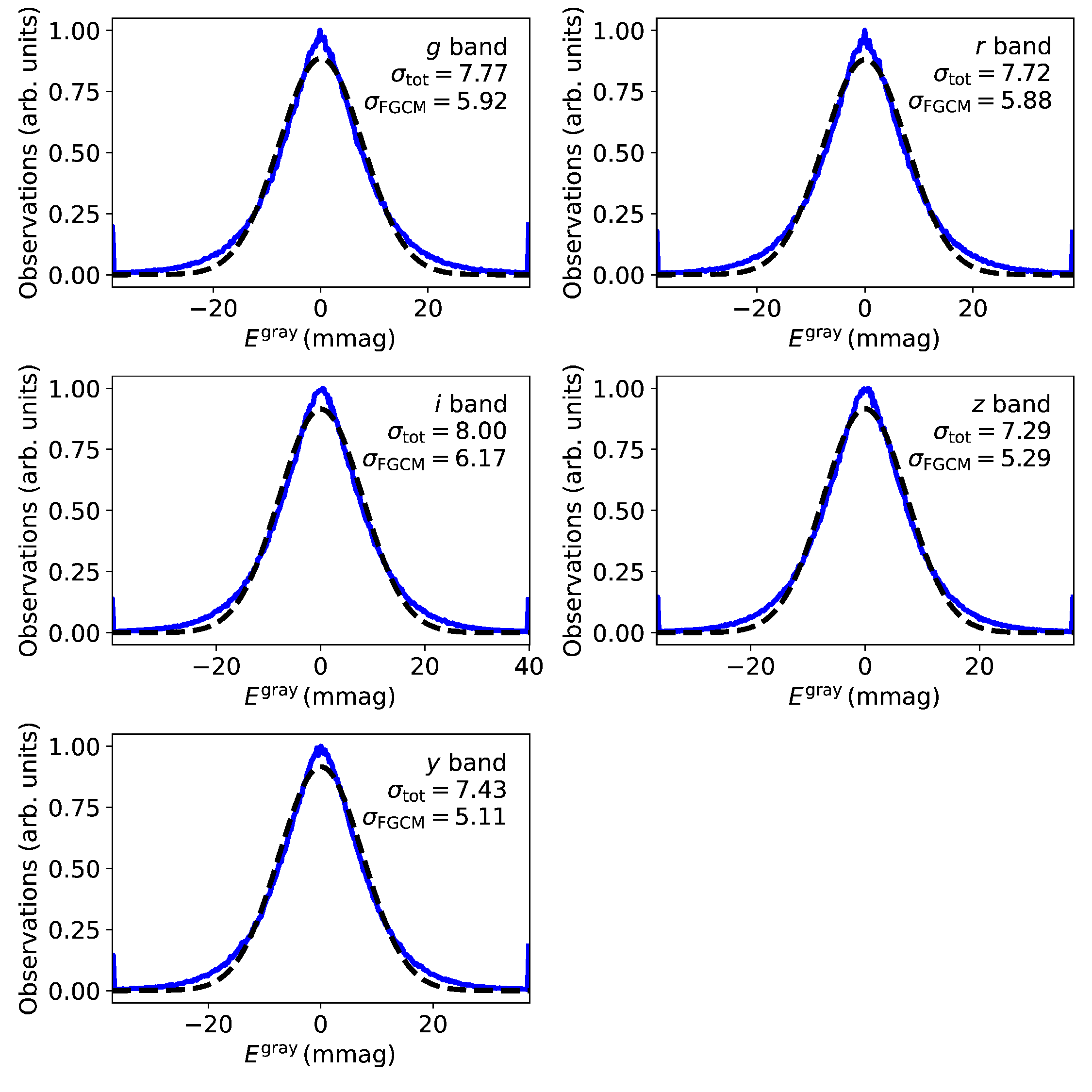}\\
        \includegraphics[width=8cm]{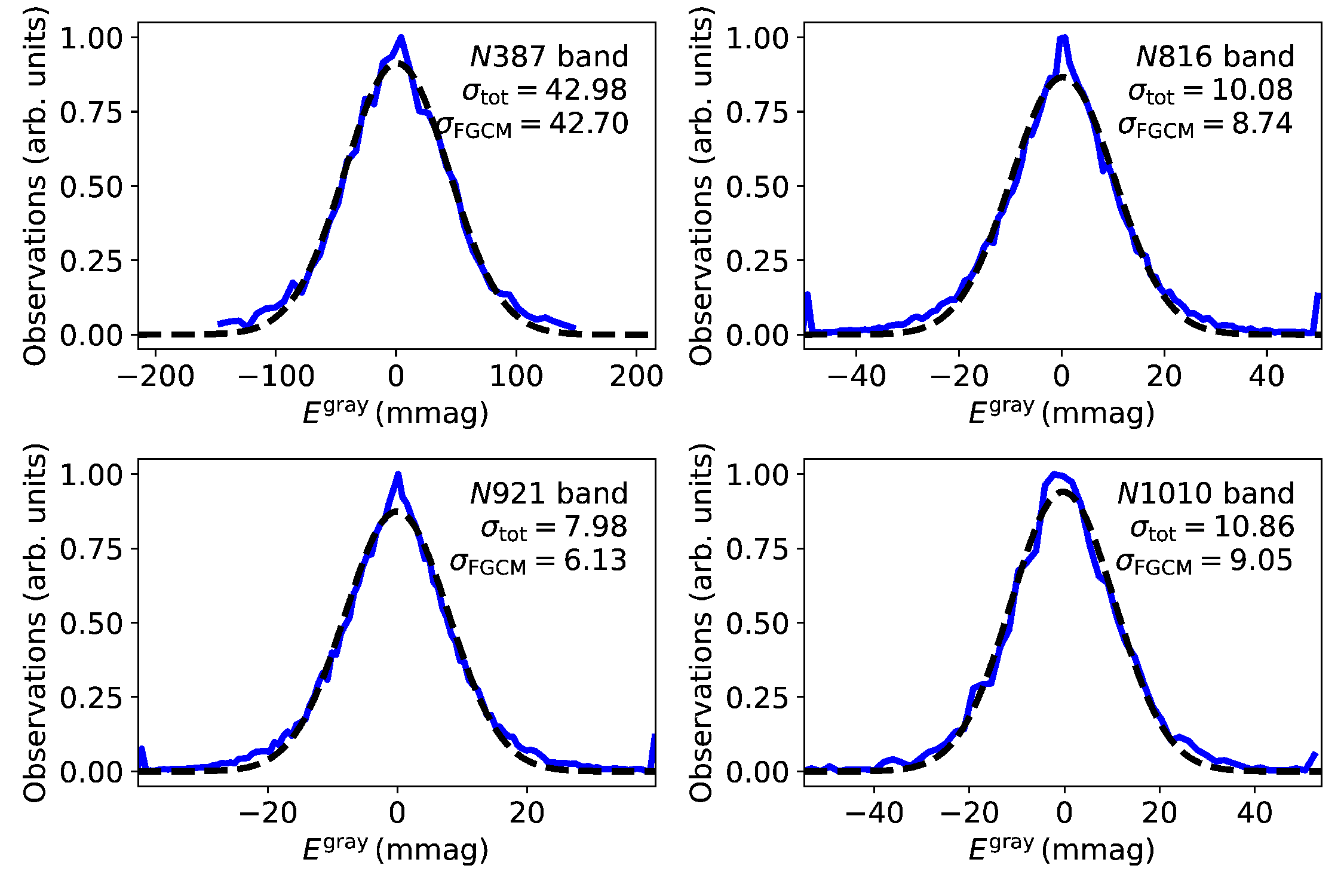}
    \end{center}
    \caption{
      The distribution of magnitude differences of repeated measurements of reserved stars not used in the FGCM fit
      for each band.
      Only stars with photometric errors less than 0.01~mag are used.  The horizontal axis is $E^{\mathrm{gray}}$,
      the difference between fully calibrated individual visits and the mean magnitude of
      all the visits of the same star.  See, e.g., Equation 30 and Figure 11 from \citet{burke18}.
      The blue solid line is a histogram of the data (arbitrarily normalized), and
      the black dashed line is a Gaussian fit to this histogram.  The calibration dispersion,
      $\sigma_{\mathrm{FGCM}}$ (in mmag), is computed by subtracting the median photometric error of
      the sample in quadrature.  Overflow counts are accumulated in the extreme ends of the horizontal bins.
      The figure includes both Wide and D/UD data.
    }
  \label{fig:repeatability}
\end{figure}

After five cycles of fitting the atmospheric parameters and rejecting outliers and
non-photometric observations, we achieve fairly good repeatability in the photometric calibration.
This is demonstrated in Fig.~\ref{fig:repeatability}.  We find rms values of
$5.92,\ 5.88,\ 6.17,\ 5.29,\ \rm and\ 5.11\,\mathrm{mmag}$ for the $g$, $r$, $i$, $z$, $y$ bands respectively
for a sample of 10\% of stars that were reserved from the fit.  For the narrow-bands,
the repeatability rms values are $42.7,\ 8.74,\ 6.13,\ \rm and\ 9.05\,\mathrm{mmag}$ for NB387, NB816,
NB921, and NB1010 respectively.  NB387 is not calibrated very well as we discuss in Section~\ref{sec:persisting_issues};
it is a difficult filter to calibrate.

\subsection{Increased Lanczos Kernel Order for Warping}
\label{sec:increased_lanczos_order_for_warping}


After the astrometric and photometric joint calibrations, we warp individual CCD images
for coaddition.  We used to use the third order Lanczos kernel for warping, but
as we discussed in the PDR2 paper, there is an about $0.3-0.4\%$ bias in our PSF model on the coadds,
in the sense that the model PSF is larger than the observed PSF.  This bias level is close to
the required accuracy for the first year HSC shape catalog \citep{mandelbaum18}, but exceeds
the accuracy needed for successive weak-lensing analyses over a wider area.
We found that using a fifth order Lanczos kernel improves the size residual
to $\sim0.1\%$, sufficient for year-3 weak-lensing analyses \citep{li21}.
For this reason, we adopt the fifth order kernel in PDR3.

\subsection{Artifact Rejection}
\label{sec:artifact_rejection}

CCD images have various artifacts such as cosmic rays, satellite trails, and optical ghosts around bright stars.
Some of them are identified and interpolated over in the CCD processing, but the remainder are left
in the processed image.  The processing pipeline makes an attempt to identify those remaining artifacts
and remove them in the coaddition stage.

The current artifact rejection algorithm, \texttt{CompareWarp}, detects artifacts in the image differences,
produced by subtracting a PSF-matched 2-sigma-clipped coadded image as a template of the static sky from each warped and
PSF-matched visit. As described in \citet{Alsayyad18} and the PDR2 paper, artifact candidates are
then detected on these image differences and labelled as follows.  
If an artifact candidate is seen in a small percentage of visits, it is labeled transient,
and is masked during coaddition, or ``clipped''.  If seen in many visits, it is considered
part of the static sky, labeled persistent and is not clipped.
The algorithm works better when more visits are available, but several visits available in the Wide layer
are still effective in identifying artifacts.  There are, however, a few typical failure modes, which
we will discuss in Section \ref{sec:persisting_issues}.

If footprint of an artifact
candidate falls entirely within the footprint of a source in the template, it is not clipped.
This feature is designed to avoid the introduction of a visit discontinuity within a source footprint
(e.g., part of a footprint falls in a CCD gap in a visit).
In particular, point sources with visit discontinuities within their footprints are not well modeled by
the PSF models, which do not take into account clipped pixels.
Two configuration parameters control the size of the artifact footprints on the difference
image and the size of the source footprints on the template.  
The parameters adopted in PDR2 introduce increased scatter in the PSF photometry with respect to PS1
in regions where all but one visit had good seeing (Section 6.6.6 in the PDR2 paper).
This scatter is particularly apparent ($> 50$ mmag) in several tracts
in the VVDS field, however we estimate that this misconfiguration in PDR2
contributed an extra 1-4 mmag scatter in the stellar locus to other fields.

\begin{figure}
    \begin{center}
        \includegraphics[width=8cm]{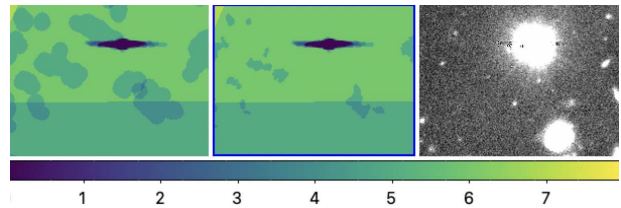}
    \end{center}
    \caption{
      Count of visits contributing to each pixel in PDR2 (left),
      PDR3 (center), and the $i$ coadded image (right) of a region
      centered at R.A.=$\rm22^h31^m57^s.3$, Dec.=$\rm+00^\circ30'34''.8$ and $\rm35"$ across.
      Artifact detections are not grown as large in this data release, leading to less total area
      clipped and more conservative preservation of the wings of stars.
      The bright star in the figure is $i\sim15$.
    }
  \label{fig:nImage}
\end{figure}

In this data release, we change these configuration parameters to reduce the effect on photometry.
With respect to the previous data release, the new configuration shrinks the artifact candidate
footprints relative to the footprints of sources detected on the static template coadd.
These parameters are empirically chosen to minimize false positives---as measured the width of
the stellar locus in color space, an independent metric of the fidelity of the PSF photometry---without
increasing the false negatives.  False negatives, the number of compact artifacts leaking
into the coadd, are estimated by the source density of detected sources in the \texttt{CompareWarp}
coadd that do not have a corresponding match in the 2-sigma-clipped template. Shrinking
the size of the artifacts also provides the added benefit of reducing the total number of
clipped pixels (Fig.~\ref{fig:nImage}).

\subsection{Improved Global Sky Frame}
\label{sec:improved_global_sky_frame}

A global sky subtraction scheme was introduced in PDR2 to improve the deficiencies of the background
subtraction we performed on individual CCDs in PDR1.  The algorithm is described in detail in the PDR2 paper,
but in short, (1) it first puts the subtracted sky back in to processed CCD images, (2) grids a visit image into
superpixels of 1k x 1k pixels (168 arcsec on a side) taking a clipped mean with object footprints excluded,
(3) fits the superpixels with 2d polynomials to construct a sky background model, (4) subtracts it from
a visit image, and then (5) subtracts a scaled sky frame, which is prepared separately by stacking
many science visits with large dithers.  The sky frame contains spatial structure on a scale smaller than 1k pixels.
This algorithm was run just before the coaddition stage.

While the algorithm works well to preserve extended wings of bright objects, it has a few side effects
such as longer compute time, increased false detection, and poor galaxy photometry for a fraction of
faint sources as we discuss later (Section~\ref{sec:overestimated_cmodel_fluxes}).
Also, only the spatial structure smaller than the superpixel (168 arcsec) is included in the sky frame,
and we suffer from missing CCDs when constructing a sky frame as the superpixel is smaller than the size of a CCD
(11.5 arcmin $\times$5.7 arcmin).

For PDR3, we modify the algorithm to include large-scale static features in the sky frame.
We first subtract the background using 8k x 8k superpixel ($\sim23'$ on a side).  This is meant to subtract exposure
specific large-scale gradients such as scattered light and is done on a per-visit basis.  It is also important for
reducing effects of missing CCDs because the superpixel size is larger than the size of a single CCD.
We then combine many visits, typically over 50, to generate a normalized sky frame.

\begin{figure}
    \begin{center}
        \includegraphics[width=8cm]{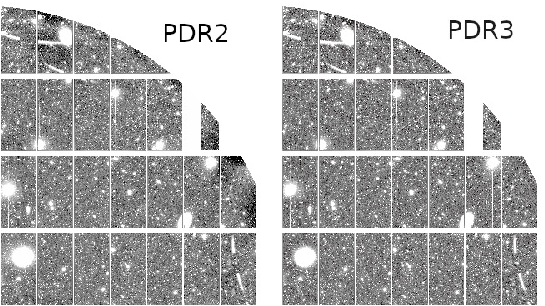}
    \end{center}
    \caption{
      Blow-up of a visit image after the sky subtraction using the superpixels (i.e., image to be used to
      construct a sky frame) processed with the PDR2 (left) and PDR3 (right) algorithms. There are regions
      with over-subtracted sky near the field edge in PDR2, while those regions are much flatter in PDR3,
      allowing us to construct a less biased sky frame.
      This is the $i$-band, but a similar feature can be seen in the other bands.
    }
  \label{fig:skyframe}
\end{figure}

Fig.~\ref{fig:skyframe} demonstrates the improvement in this release.  In PDR2, a sky frame was often biased
around dead CCDs or dead amplifiers located at the field edge because the superpixel size is smaller than the size of
a CCD or amplifier and we had to extrapolate.
This propagated to the sky frame and resulted in over/under-subtracted sky near the field edge, which then propagated to the coadded
image.  On the other hand, the sky background is flat to the edge of the field in PDR3, demonstrating the improved sky frame.

For science exposures, we first subtract the 8k x 8k background in the same manner as above,
followed by the subtraction of the scaled sky frame.  There is a small-scale background fluctuation left
at a very low flux level at this stage.  To eliminate it, we perform additional small-scale sky subtraction using
256 x 256 superpixels ($\sim43$ arcsec).
Care is taken to mask objects well; we subtract the sky, detect sources, define their footprints
(which are pixels above $2.5\sigma$ grown by 2.4 times the size of PSF) and mask them,
put the sky back in, and repeat.  We iterate 3 times to ensure that we have sufficiently large masks around
objects.  This way, we keep the wings of bright sources, while subtracting the sky background
on a relatively small scale.

\subsection{Local Sky Subtraction for Detection and Measurements}
\label{sec:local_sky_subtraction}

In PDR2, the deblending in UD-COSMOS became prohibitively long and memory-intensive because too many source
footprints are connected with each other and the deblender had to run many times.
This is due to the well-preserved wings of objects (i.e., a consequence of the global sky subtraction).
This also causes reduced detection efficiencies at faint magnitudes.
While we have improved the global sky subtraction scheme as described in the previous subsection,
the extended wings of bright objects still remain as a major issue.
To mitigate it, we choose to subtract the sky on small scales and intentionally subtract wings of bright
objects before the detection and measurement steps, so that object footprints are connected less frequently
and the deblender does not have to run too many times.
The choice of the local sky subtraction scale is primarily driven by object detection efficiency and footprint size.
After some experiments, we found that a superpixel size of 128 ($\sim21.5$ arcsec) is a reasonable trade-off
between the detection efficiency and over-subtraction of moderately extended sources, and we adopt it here.

\begin{figure*}
    \begin{center}
        \includegraphics[width=8cm]{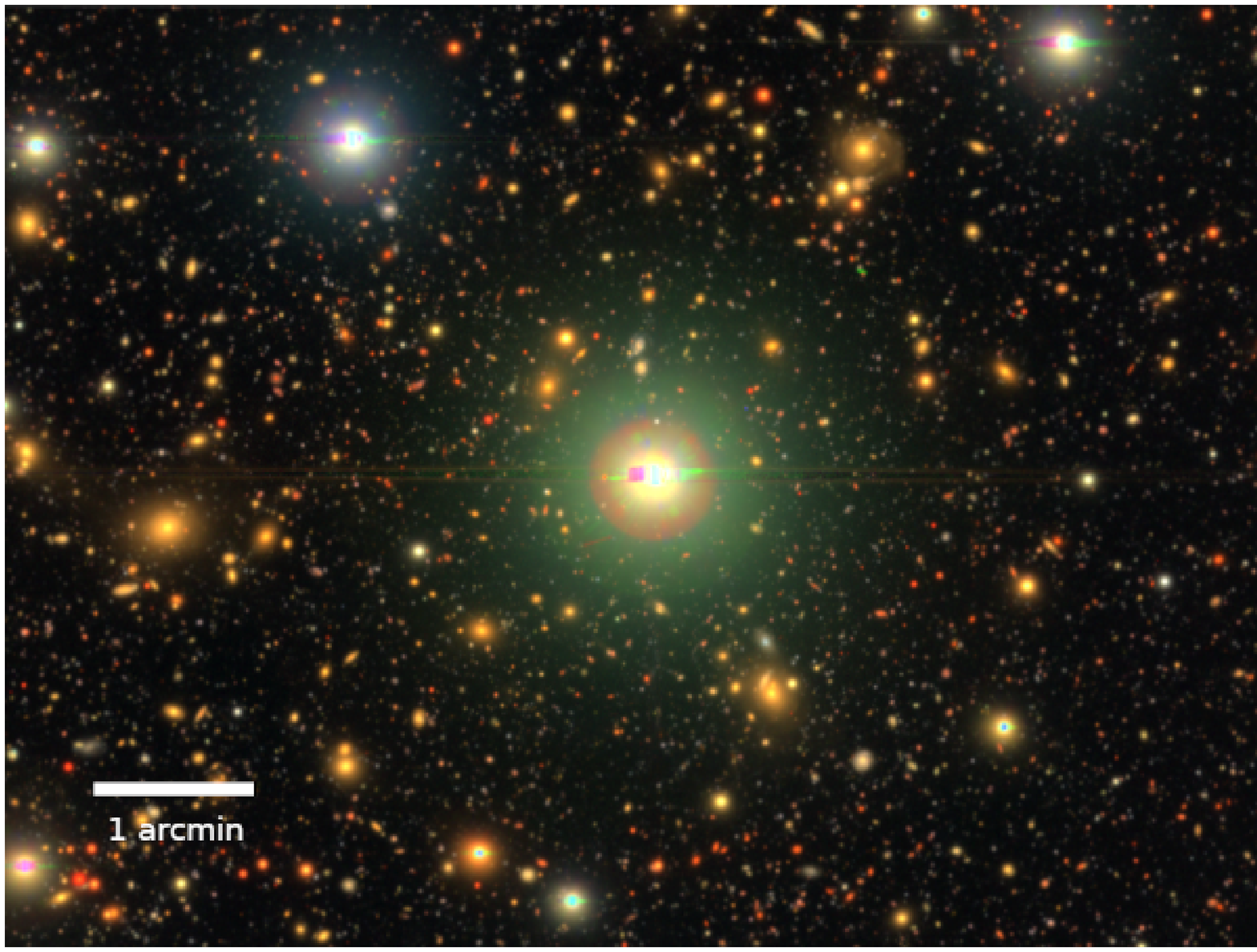}
        \includegraphics[width=8cm]{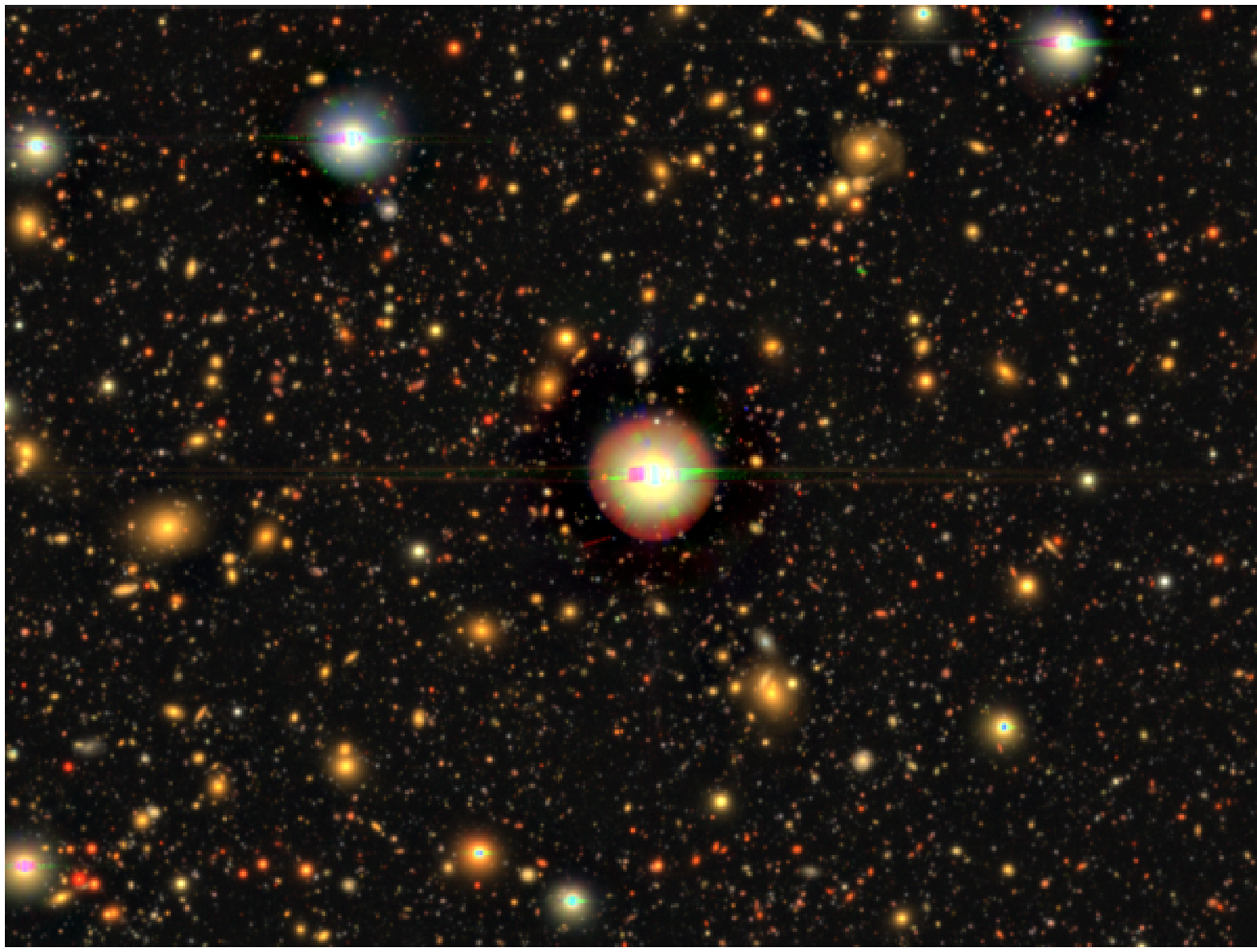}
    \end{center}
    \caption{
      $gri$ color composite of a portion of the UD-COSMOS field ($8'\times6'$).
      The left is the coadd with the global sky subtraction applied and the right is
      with the local sky subtraction.  The main differences are seen around bright stars,
      and galaxies are largely unaffected.
    }
  \label{fig:skysb}
\end{figure*}

We recall that we perform the global sky subtraction in the processing prior to the local sky subtraction
described here.  The coadd images with the global sky subtraction are stored as patch images (under {\tt deepCoadd/}),
so that the user interested in large extended objects can exploit the coadds for image analyses.
Then, the object detection and measurements are performed on coadds with the local sky subtraction applied;
these coadd images are stored as {\tt calexp} under {\tt deepCoadd-results/}.  Fig.~\ref{fig:skysb} illustrates
the difference between the two sky subtraction schemes.  As can be seen, only bright sources lose their wings
with local sky subtracted and sources with $\lesssim30$ arcsec extent remain unaffected.

\subsection{Algorithm fix in detection and peak merge}
\label{sec:algorithm_fix}

After the object detection in each band, we merge footprints and peak positions within
them across the bands in order for us to perform consistent photometry.
As described in Section 3.4 in \citet{2018PASJ...70S...5B}, peaks detected in different bands are merged
sequentially in so-called ``priority order,'' which we define as $irzyg$ for the broadband filters, followed by
the narrow-band filters.  We merge a peak from the current band into the set of peaks detected in
the higher-priority bands (found in source footprints that overlap) per the following rules:

\begin{itemize}
\item If a peak in the new band is at least 1'' from all peaks from the previous step, we add a new peak.
\item If a peak in the new band is less than 0.3'' from the nearest peak from the previous step, we mark
  the peak as having been detected in the new band while maintaining the position from the previous step.
\item If a peak in the new band is between 0.3'' and 1'' from the nearest peak from the previous step,
  it is ignored. We cannot conclusively identify the peak as either the same source or a distinct source.
\end{itemize}

In rare cases, a software bug prevented this prescription from being followed.
If the new peak is in a footprint that connected two previously disconnected footprints in the higher-priority
bands, the new lower-priority peak could \textit{replace} the set of higher-priority peaks if it were
between 0.3'' and 1'' away.  For PDR3, the routine has been corrected to always chose
the higher-priority peak set for the merged catalog.

\subsection{Effective Filter Response}
\label{sec:effective_filter_response}

The $i$-band filter has strong radial dependence in its transmission curve \citep{kawanomoto18}.
This spatially varying transmission function is the primary cause of the donut-like background structure
seen in a visit image in that filter.  A new filter, $i2$, was manufactured and has been in use since
the middle of the survey in 2016.  We thus have images taken with both $i$ and $i2$ filters and
the processing pipeline combines them together in the coaddition process.  These filters have similar
but not identical transmission curves, and they introduce two issues; (1) spatially varying response
function in regions observed in the $i$-band, and (2) a fraction of the survey area is observed
in $i$, some in $i2$ and the rest in both, further introducing spatial variation of the response function.
As we discussed in Section 6.6.4 of the PDR2 paper, colors of objects differ between the $i$ and $i2$ regions,
which illustrates the latter problem here.

To mitigate these problems, we use the effective response function introduced in PDR2 (Section 4.5 of the PDR2 paper)
to correct for the spatially varying filter transmissions.  We use an effective response function of an object
given where it fell in the focal plane in the individual visits and estimate the expected difference between that
response and the fiducial $i2$ response as a function of object color.

We need two ingredients here; the underlying spectral energy distribution (SED) of an object and its observed broad-band color.
We want to use a wide enough range of SEDs to compute the magnitude offset for various objects.  Here, we
use a set of SED templates from the photometric redshift code of \citet{tanaka15}, which has a large number of objects
covering a wide range of SED types over a wide redshift range.
We have confirmed that the corrections
are very similar if we use the \citet{pickles98} stellar library instead.  We then convolve these SEDs with a set of 'target' filters
and the fiducial $i2$ filter.  A target filter is either $i$ or $i+i2$ combined.
The latter just a weighted average of $i$ and $i2$ filters.
We pre-compute the target filters in the $i$-only and $i/i2$-mixed regions and estimate
the magnitude corrections to the fiducial filter ($i2$).
We also convolve these SEDs with the fiducial $grizy$ filter set so that we can compute broad-band colors for each SED.
Each SED gives a slightly different amount of correction, and we average the correction as a function of
the $g-i$, $r-i$, $i-z$ and $i-y$ colors.  At this point, we have a look-up table for the correction as a function
of the target filter and observed color.

For each object, we have its observed color from the HSC data and also can compute the effective response function
(=target filter) because we know where on the focal plane the object is located in each visit.
We can then estimate the magnitude offset from the look-up table 
and apply it to each object to translate the $i$-band (or mixture of $i$ and $i2$-bands) photometry into the fiducial $i2$ photometry.
We perform exactly the same analysis for the $r$ vs. $r2$ bands as well.

\begin{figure}
    \begin{center}
        \includegraphics[width=8cm]{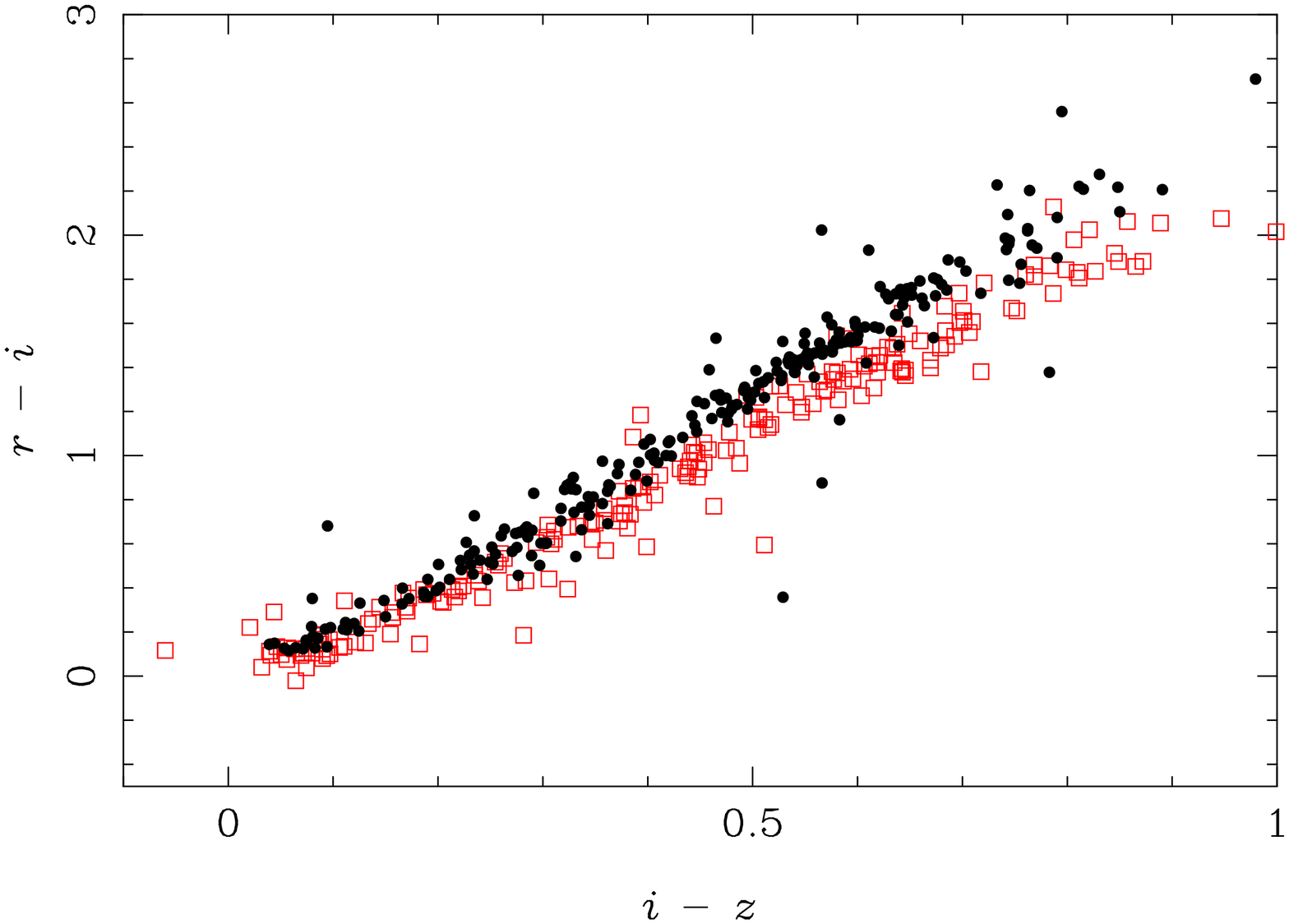}\\
        \includegraphics[width=8cm]{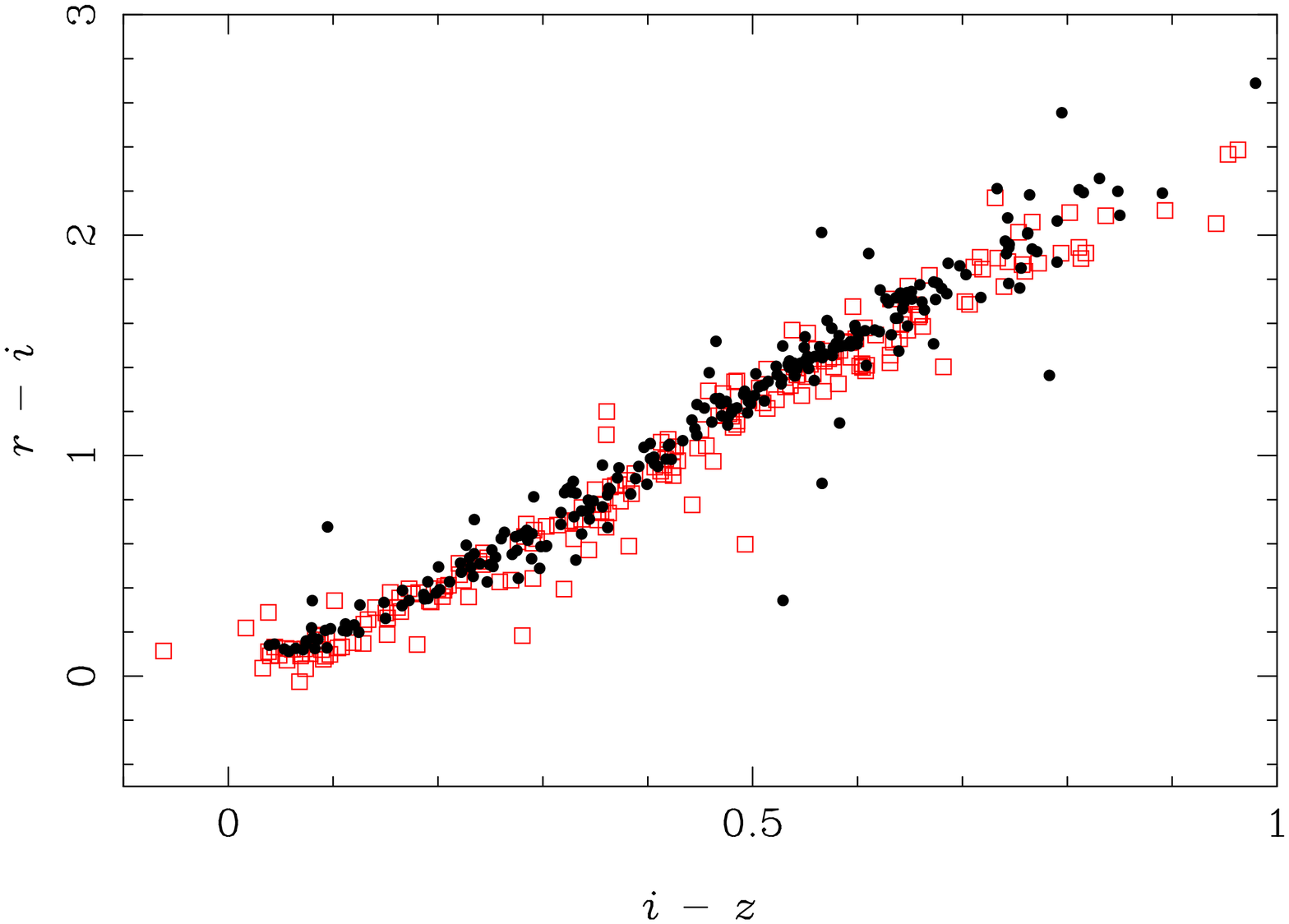}\\
    \end{center}
    \caption{
      $r-i$ plotted against $i-z$ for stars with $i<23$.  The filled points are stars observed with the $i2$ filter,
      while the open squares are observed with the $i$ filter.  The top panel is the original data and the bottom panel is
      after the correction for the filter response difference.
    }
  \label{fig:riz_corr}
\end{figure}

Fig.\ref{fig:riz_corr} illustrates the correction, showing the $riz$ color-color diagram of stars.
In the top panel, there is a clear difference between the
stellar sequences of stars measured with the $i$ and $i2$ filters.  If we apply the correction and translate
the $i$ band photometry into $i2$, the two stellar sequences agree well (bottom panel).
We provide the correction table for each object in this data release.
Note that the correction is not available
for objects observed in a single filter because we cannot compute their color and hence cannot infer their intrinsic SEDs.
Note as well that the correction
is for objects with typical SEDs and should be used with care for objects with exotic colors.

\subsection{Stellar Sequence Regression}
\label{sec:stellar_sequence_regression}

\begin{figure*}
    \begin{center}
        \includegraphics[width=12cm]{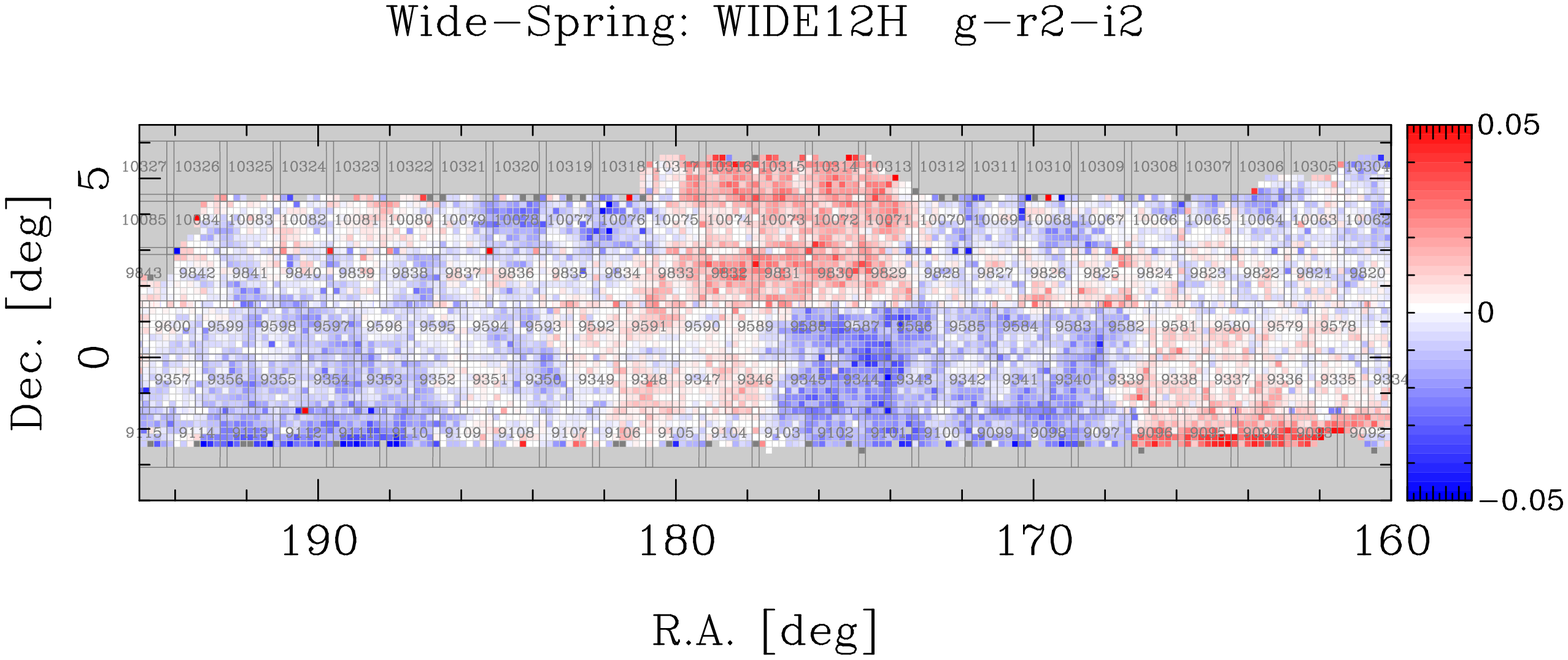}\\\vspace{0.5cm}
        \includegraphics[width=12cm]{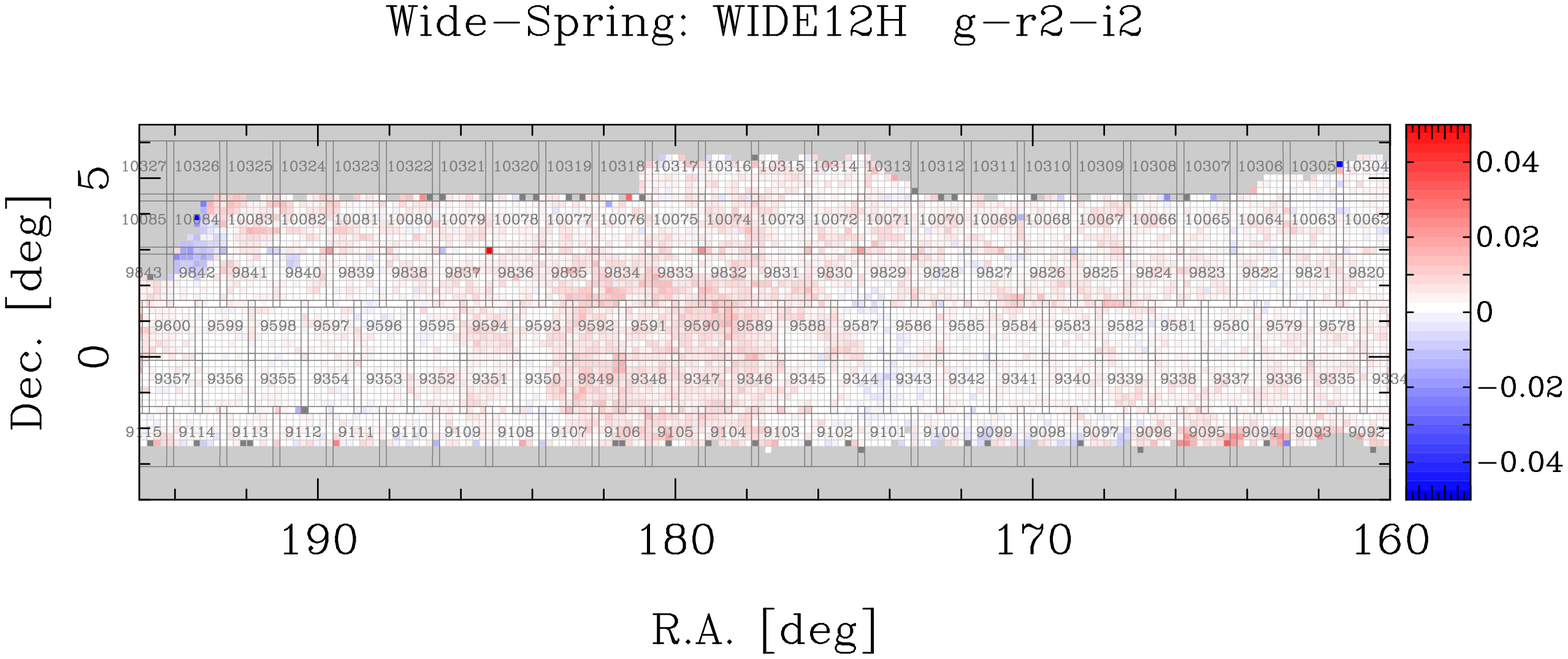}\\
    \end{center}
    \caption{
      Stellar sequence offset in $gri$ around the R.A.=12$^h$ region before the zero-point correction (top) and
      after the correction (bottom).  The color-coding indicates the amount of offset as shown in the color bar,
      and the offset is computed for each patch.  The gray rectangles are the tract boarders, and
      the tract IDs are also shown.
    }
  \label{fig:stellar_sequence}
\end{figure*}

We mentioned in Section \ref{sec:fgcm} that FGCM resulted in a small inhomogeneity of photometric zero-points
across the survey field.
While it should be possible to correct for it in a principled way, we adopt
an empirical approach and correct for it using the location of the stellar sequence in color-color diagrams.
We use stars with $i<22.5$~mag and estimate the zero-point offsets in each filter in each patch.
We correct for the Galactic extinction in all bands using \citet{schlegel98} and assuming that the stars we use are all behind
the dust curtain. This is a reasonable assumption because the stars we use are typically $i\sim22$ and
thus likely are halo stars, but this could potentially be a source of systematic uncertainty.
We estimate the expected location of the stellar sequence in the $grizy$ multi-color space using the \citet{pickles98} stellar library.
The top panel of Fig.~\ref{fig:stellar_sequence} shows the observed offsets of the stellar sequence measured in $gri$.
There is a clear spatial structure with a peak-to-peak variation of $\sim0.05$ mag.
Note that some of the variation is due to the $r/i$ vs. $r2/i2$ difference.

In order to estimate the zero-point correction, we first apply the correction derived in the previous subsection
to the $r$ and $i$ band photometry.  We then draw color-color diagrams of the observed stars and estimate
average offsets from the expected location for each filter.  In this step, we assume that there is no offset in
the $i$-band.  After correcting for the offsets, there are still spatially varying stellar sequence offsets at
a level of $\sim2$\% remaining, suggesting that our correction was imperfect.
The next step is to attribute the remaining offsets to the $i$-band only.  We have confirmed that
the correction to the $i$-band is small ($\sim 1$\%) but it does reduce the spatial variation.

The bottom panel of Fig.\ref{fig:stellar_sequence} shows the improvement. The variation observed in the top
panel is reduced to $\sim0.015$ mag after the correction.
We encourage the user to apply these magnitude corrections to the cataloged magnitudes\footnote{
Note that the cataloged magnitudes in the pipeline outputs and database are not corrected for
the offsets.
There are database tables that serve the offsets and the offsets have to be
{\it subtracted} from the cataloged magnitudes.
}.

\section{Data}
\label{sec:data}

\subsection{Value-added Products}
\label{sec:value_added_products}

In addition to the data products from the pipeline, we serve value-added
products as in our previous releases, including the COSMOS Wide-depth stacks, a collection
of public spectroscopic redshifts, and random points.  The COSMOS Wide-depth stacks and
the random points are generated in the same way as in the previous release; the reader is
referred to the PDR2 paper.  In what follows, we briefly discuss the updated list of spectroscopic
redshifts, and photometric redshifts based on the spectroscopic catalog.
There is also a major update to the mask around bright stars, which we discuss in Section~\ref{sec:bright_star_mask}.

\begin{itemize}
  
\item {\bf Public spectroscopic redshifts:}
   We have updated the list of public spectroscopic redshifts from the literature.  The list includes
   redshifts from
   2dFGRS \citep{colless03},
   3D-HST \citep{skelton14,momcheva16},
   6dFGRS \citep{jones04,jones09},
   C3R2 DR2 \citep{masters17,masters19},
   DEEP2 DR4 \citep{davis03,newman13},
   DEEP3 \citep{cooper11,cooper12},
   DEIMOS 10k sample \citep{hasinger18},
   FMOS-COSMOS \citep{silverman15,kashino19},
   GAMA DR2 \citep{liske15},
   LEGA-C DR2 \citep{straatman18},
   PRIMUS DR1 \citep{coil11,cool13},
   SDSS DR16 \citep{ahumada20},
   SDSS IV QSO catalog \citep{paris18},
   UDSz \citep{bradshaw13,mclure13},
   VANDELS DR1 \citep{pentericci18},
   VIPERS PDR1 \citep{garilli14},
   VVDS \citep{lefevre13},
   WiggleZ DR1 \citep{drinkwater10}, and
   zCOSMOS DR3 \citep{lilly09}.
   As one-to-one correspondence between the spectroscopic objects and photometric objects is not always
   obvious, we match objects within 1 arcsec and all matched objects are stored in the database.
   In most cases, the  most likely match will be the object with the smallest matching distance.
   Each spectroscopic survey has its own flagging scheme to indicate the reliability of a redshift measurement.
   We have homogenized the flags to make it easy for the user to make a clean redshift catalog.
   See the online documentation for the definition of the homogenized flag.  We emphasize that
   the user should acknowledge the original data source(s) when using this product.

 \item {\bf Photometric Redshifts:}
   Photometric redshifts for objects in PDR3 have been computed using a few independent codes \citep{tanaka18,nishizawa20}.
   The performance of the codes are overall similar to the previous release.  There has been
   a significant delay in the delivery of the photometric redshift products to the internal team.
   We thus anticipate  a future incremental release to make these photometric redshifts available to
   the community in mid to late 2022.
   
\end{itemize}

\subsection{Bright Star Mask}
\label{sec:bright_star_mask}

The vicinity of a bright star suffers from optical ghosts and other artifacts, and object detection
and measurements there are unlikely reliable.  We generate a mask around a bright star to indicate
such a problematic region since PDR1.
Given that we changed the way we subtract the sky (Section \ref{sec:local_sky_subtraction}), we have
updated the bright star mask accordingly.  We have changed the mask size around stars and made
a few improvements in how we treat various artifacts around bright stars as we detail in this subsection.

\subsubsection{Reference catalog}

We use Gaia DR2 \citep{gaia18} as a reference catalog of bright stars.  To ensure good photometric quality, we require
that the S/N in the $G$-band is greater than 50, and those in the $B_p$ and $R_p$ bands are
both greater than 20.
In addition, we apply cuts on the $B_p+R_p$ flux excess as discussed in \citet{gaia18} to reduce contamination of nearby sources
in the prism photometry.  These cuts leave us with stars down to $G\sim19$.

In our previous releases, the size of the mask around bright stars was made independent of filter for simplicity.
However, this resulted in sub-optimal masks
around very blue/red stars.  We now account for the magnitudes of bright stars in each of the HSC bands.
As bright stars are often saturated in the HSC images, we use the Gaia prism photometry ($B_p$ and $R_p$)
to infer magnitudes of stars in the HSC bands.  We use the \citet{pickles98} stellar library to make
a mapping between the Gaia $G$-band and each of the HSC broad-bands as a function of $B_p-R_p$.
We apply a magnitude cut of $<18$ to the Gaia stars in each of the HSC filters.  For instance,
we define the $i$-band mask for stars with $i<18$.  The same applies in all HSC filters.

\subsubsection{Artifacts}

\begin{figure*}
  \begin{center}
    \includegraphics[width=16cm]{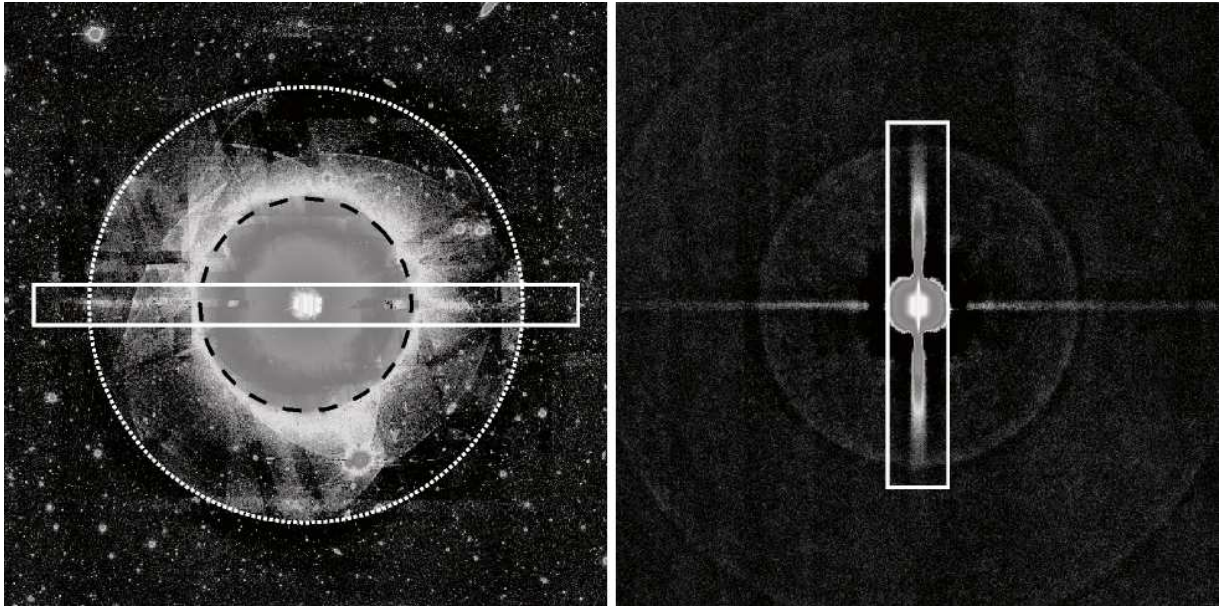}
  \end{center}
  \caption{
    {\bf Left:} Image of a 5th magnitude star in the $i$-band.  This is a coadd image with the local sky subtraction applied
    (i.e., image used for object detection and measurement).
    The inner and outer circles have radii of 160 and 320 arcsec, respectively.
    These circles correspond to the ghost edges as discussed in the text.  The horizontal rectangle indicates
    the blooming feature.
    {\bf Right:} Stacked $y$-band image of 5th magnitude stars.  The vertical box encloses the channel-stop feature.
    The horizontal stripe is blooming and the ghost edges at 160 and 320 arcsec are also clearly seen.
  }
  \label{fig:mask_features}
\end{figure*}

There are several different types of optical/detector artifacts, each of which has its own shape and size.
We consider the following features:

\begin{itemize}
\item {\it halo}:
  A star has an extended smooth halo around it, which can be considered as part of PSF.
  The size of this halo depends on the brightness of a star and also on the algorithm that we use to subtract the sky.
\item {\it ghost}:
  In addition to a halo, a very bright star is surrounded by larger optical ghosts due to reflection inside the camera.
  As there are multiple paths of reflection, each ghost feature is out of focus by a different amount, and thus
  has a different angular size on the focal plane.  A ghost feature is roughly constant in surface brightness and
  it is partly subtracted by the sky subtraction.  However, its edge can be sharp and thus remains clearly on the coadded
  images.  We wish to mask the outermost edge to avoid it affecting object detection.
\item {\it blooming (bleeding)}:
  A common CCD feature.  When a pixel exceeds its full-well capacity, electrons overflow to adjacent pixels along
  the column.  The maximum length of the blooming feature is set by the detector size, which is 4k in case of HSC.
\item {\it channel-stop}:
  At long wavelengths ($\sim1\mu m$), a CCD becomes optically thin and a fraction of incident photons go through
  the silicon layer and get reflected by elements in the top layer of the underlying circuitry, probably by the channel-stop.
  This causes   a diffraction pattern around a bright star perpendicular to the blooming direction.
  In HSC, this feature is seen only in the $y$-band and $NB1010$.
  Fig. \ref{fig:mask_features} illustrates the ghost, blooming, and channel-stop features on 5th magnitude stars.
\item {\it dip}:
  Although we account for the object footprint when we estimate the sky background, a portion of a bright star is still
  subtracted as part of the sky, resulting in slightly negative background around the vicinity of a star.
  Because of the negative sky, the object detection there is slightly inefficient and the density of detected objects
  drops.  We call this feature a dip.
\end{itemize}

\noindent
We briefly discuss how we determine the mask size for each component in what follows.
Our goal here is to define a mask around bright stars to minimize effects of these artifacts,
in particular to avoid spurious sources.  At the same time, we want to make the masks no larger
than they need to be to maximize the usable area of the survey.

\subsubsection{Halo and ghost}

\begin{figure}
  \begin{center}
    \includegraphics[width=8cm]{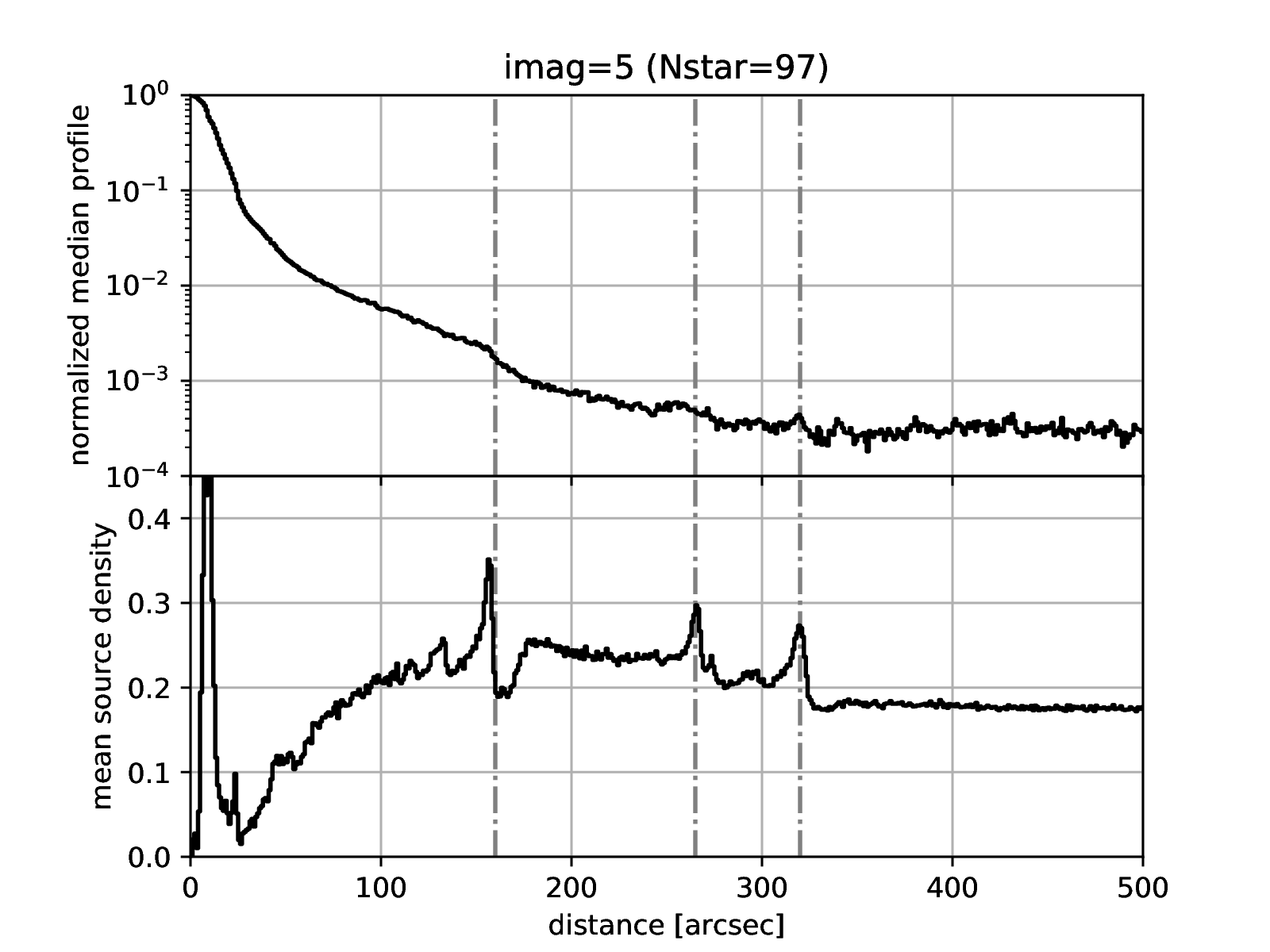}
  \end{center}
  \caption{
    {\bf Top:} Normalized intensity profile of 5th magnitude stars in the $i$-band.
    The intensity is normalized at the center and the vertical lines show radii of 160, 265, and 320 arcsec, respectively,
    where we observe sharp edge of the ghost feature.
    {\bf Bottom:} Number density of detected sources as a function of distance from 5th magnitude stars.  The increased
    number of sources at the ghost edges can be clearly seen.
  }
  \label{fig:mask_detection}
\end{figure}

We carry out two-way analyses; image-based and catalog-based.  For the image analysis, we normalize the radial profile
of each bright star using the coadd image with the local sky subtraction algorithm applied.
We then take the median of the profiles of many bright stars with
similar magnitudes.  This is done for various magnitude bins.
The top panel of Fig.\ref{fig:mask_detection} shows the averaged radial profile of 5th magnitude stars in the $i$-band.
The ghost extends roughly to 300 arcsec and settles down to the background level at a larger distance.
The background level does not reach zero; this is because we do not mask objects around the bright stars.
We perform the same analysis in the other filters, but we find that the dependence on filter at fixed magnitude
is small.

We move on to examine the effect of the ghosts more carefully using the coadd catalog.  We here look at
the object detection; the source density increases where there are optical artifacts and hence it is a sensitive
probe of artifacts.  The bottom panel of
Fig.\ref{fig:mask_detection} shows the mean source density as a function
of distance from 5th magnitude stars.  As can be seen, there are detection spikes at specific radii.  These spikes correspond
to the ghost edges, whose radii are independent of stellar brightness.  The outermost ghost is located
at 320 arcsec.  This is the most significant component for stars brighter than $\sim$7th~mag.  It gets fainter
for fainter stars and the inner ghost at 160 arcsec is dominant at $7-9$th~mag.
We note that the exact size and shape of the reflection ghost depends on position of the star relative to
the telescope boresight.  Thus, there are spurious sources outside the mask discussed here
in some cases because of that effect.
In order to deal with such cases, we make two additional masks that are 20 and 50 \% larger
than the ghost sizes of 160 and 320 arcsec, and they are assigned separate mask bits.

The reflection ghosts are not prominent for stars fainter than 9th~mag and the halo component
becomes the dominant component.  We measure the halo size using the same image and catalog based methods
over a range of magnitudes and fit an exponential function to represent the relationship between
the halo radius and stellar brightness:

\begin{equation}
  r_{halo}=1.105\times10^3 \times \exp(-0.347 {\rm mag}) + 4.950\ \rm arcsec.
\end{equation}

\noindent
We find that the halo size does not strongly depend on filter at fixed magnitude.
We thus adopt this relation in all filters.

\subsubsection{Blooming}

Blooming, or bleed trail, is a common phenomenon in CCDs and appears along the columns.
The blooming length depends on the source brightness and seeing as well as the source position.  It can be up to 11 arcmin long,
which corresponds to the detector size.  As the blooming feature appears at the same location (albeit with differences
in length) with respect to a star in all visits, the pipeline cannot completely remove it.
Recall that we always observe at the same position angle on the sky (i.e., we do not perform rotational dithers).
We carry out the same catalog analysis as for the halos and ghosts above and compare the numbers of objects detected along the horizontal
and vertical directions.  A detection excess along the horizontal direction is due to spurious detections along the blooming feature,
and for a bright star, this excess is seen over as large as 600 arcsec.  We parameterize the half-length of the excess
as a function of stellar brightness:

\begin{equation}
  s_{blooming} = -8.5714 \times {\rm mag}^2 + 40.857 \times {\rm mag} + 778.57
\end{equation}

The width of the blooming feature is primarily driven by the seeing, but it can be larger than the seeing FWHM
at very bright magnitudes because a large number of photons may saturate not just the core of a star but also
its outer parts.  Visual inspections of the stacked images of bright stars reveal that the width of 20 pixels
is sufficient and we adopt it here for all stars.

\subsubsection{Channel-stop}

As discussed above, a CCD can be optically thin at wavelengths close to the Silicon band-gap.  The front structure,
probably the channel-stop,
introduces a diffraction pattern at long wavelengths, which is aligned perpendicular to the blooming feature.  The length
of this feature depends on the stellar brightness.  We parameterize the half-length as a function of magnitude
in a similar fashion to the blooming feature.

\begin{equation}
  s_{channel-stop}=5.982\times10^2\times\exp(-1.640 {\rm mag}) -17.66
\end{equation}

\noindent
We fit a similar function to the width of this feature:

\begin{equation}
  w_{channel-stop}=28.3\times10^2\times\exp(-0.1640 {\rm mag}) -2.03
\end{equation}

\subsubsection{Dip}

A small region around a star suffers from over-subtracted background.  The source density is again a useful
probe of its effect.  We find that, interestingly, the reduced source density around the vicinity of a star
is limited to within 40 arcsec regardless of the stellar brightness.  This is probably due to our sky subtraction
scheme; we adopt a superpixel size of 128 pixels to estimate the background, which corresponds to 21.5 arcsec.
Depending on where a bright star is located inside a superpixel, up to 2 superpixels can be affected by
the extended stellar halo, which can explain our observation here.  The dip mask is fixed to 40 arcsec
regardless of brightness for all stars brighter than 18th magnitude.

\subsubsection{Summary of masks}

\begin{figure}
  \begin{center}
    \includegraphics[width=8cm]{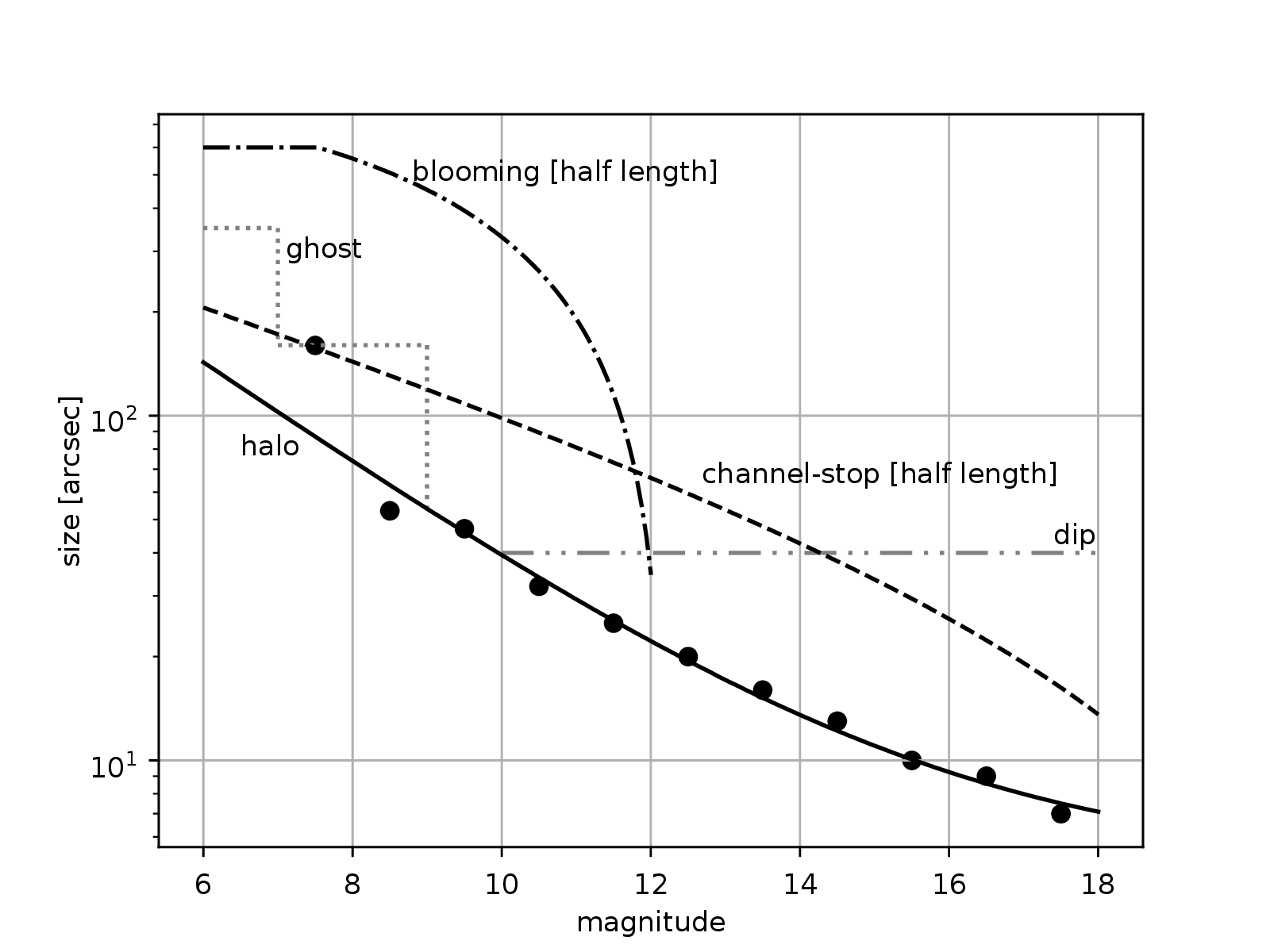}
  \end{center}
  \caption{
    Mask size as a function of stellar magnitude for each component.
    The points show the sizes of the halo feature in each magnitude bin and the solid curve is a fit to the points.
    The dotted, dashed, dot-dashed, dot-dot-dashed curves are for ghost, channel-stop, blooming and dip features, respectively.
    Using the union of these masks (perhaps except for the dip mask) is the most conservative approach.
  }
  \label{fig:mask_summary}
\end{figure}

Fig. \ref{fig:mask_summary} summarizes the sizes of masks discussed here as a function of stellar magnitude.
The reflection ghost is the largest component
at very bright magnitudes and the dip feature is the dominant feature for stars fainter than 10th magnitude.
However, the dip mask introduces
a significant loss in the survey area; about 80\% of the Wide area is left after applying the halo, ghost, and
blooming masks, but the fraction reduces to 50\% when the dip mask is also applied because faint stars are
numerous.  If one does not care about
a $\sim10\%$ detection loss around the detection limit around bright stars, the dip mask can be ignored.
All these mask flags are available for each object in each filter at the database.
At this point, the masks are defined only for broad-band filters.  For narrow-band filters, one can
use the broad-band mask that is closest in wavelength (e.g., $y$-band is a good proxy for NB1010).

\section{Data Quality and Known Issues}
\label{sec:data_quality_and_known_issues}

\subsection{Photometry and Astrometry}
\label{sec:photometry_and_astrometry}

\begin{figure*}[htb]
    \begin{center}
        \includegraphics[width=14cm]{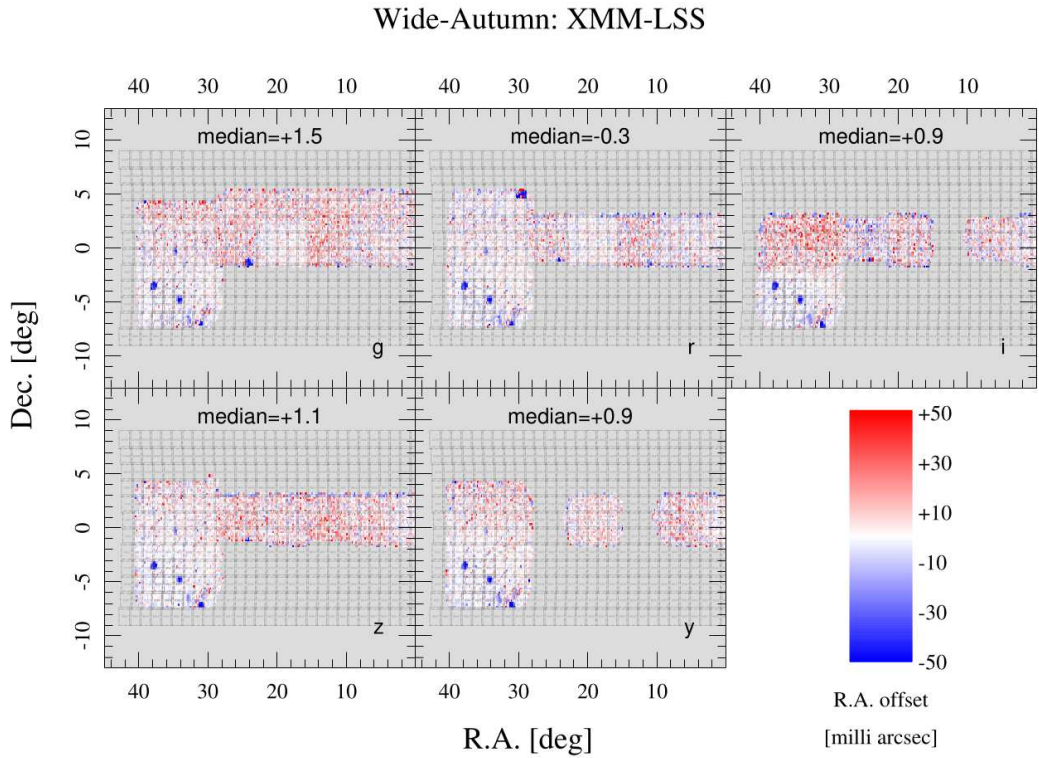}
    \end{center}
    \caption{
      Astrometric offset in the R.A. direction for a portion of the Wide survey.
      We first match objects between HSC and Gaia in a patch and then
      compute the mean offset.  We show the mean offset for each patch using the color-coding shown in the bottom-right panel.
      The panels are for the $grizy$ filters
      from top-left to bottom-right.  The number in each panel is the median R.A. offset in
      milli arcsec over the area shown for each filter (the numbers are small due to spatial
      averaging).  The gray squares are the tract borders.
    }
  \label{fig:astrometry}
\end{figure*}

The overall astrometric and photometric accuracy of this release is similar to that of PDR2.
There are changes in the way we calibrate the data (astrometric calibration by {\tt jointcal}
and photometric calibration by FGCM;
Sections~\ref{sec:jointcal} and \ref{sec:fgcm}).  But our primary calibration source has not changed; we still
use the PanSTARRS1 DR2 catalog \citep{schlafly12,tonry12,magnier13,chambers16}
astrometrically calibrated against Gaia DR1 \citep{gaia16a,gaia16b}.  We plan to 
change the astrometric reference catalog to a newer Gaia release in our future public release.
As in the two previous releases, we present a full set of quality assurance (QA) plots
on the data release website, and the user should check these QA plots before utilizing the data for science.
Here we present some of the key example diagnostics.

Fig.~\ref{fig:astrometry} shows the astrometric calibration accuracy in a portion of the Wide field.
We compute an offset in R.A. by comparing the HSC and Gaia DR1 coordinates of common stars brighter than $G=20$.
Overall, our astrometric calibrations are good, but there are multiple small regions with
rather large negative offsets.  We have visually inspected some of these regions, but there
are nothing obviously wrong in the images.  It turns out that these are regions in which the astrometry
in our own reference catalog (PS1 calibrated against Gaia) is problematic.
The bad astrometry in the reference
data unfortunately propagated to HSC.  We hope to fix it in our future release.
Fig.~\ref{fig:astrometry} shows only the offset, but we also estimate rms of the astrometric residual,
which is a good indicator of overall astrometric accuracy.  We find that our astrometry is rms$\sim13$ mas.

\begin{figure*}
    \begin{center}
        \includegraphics[width=14cm]{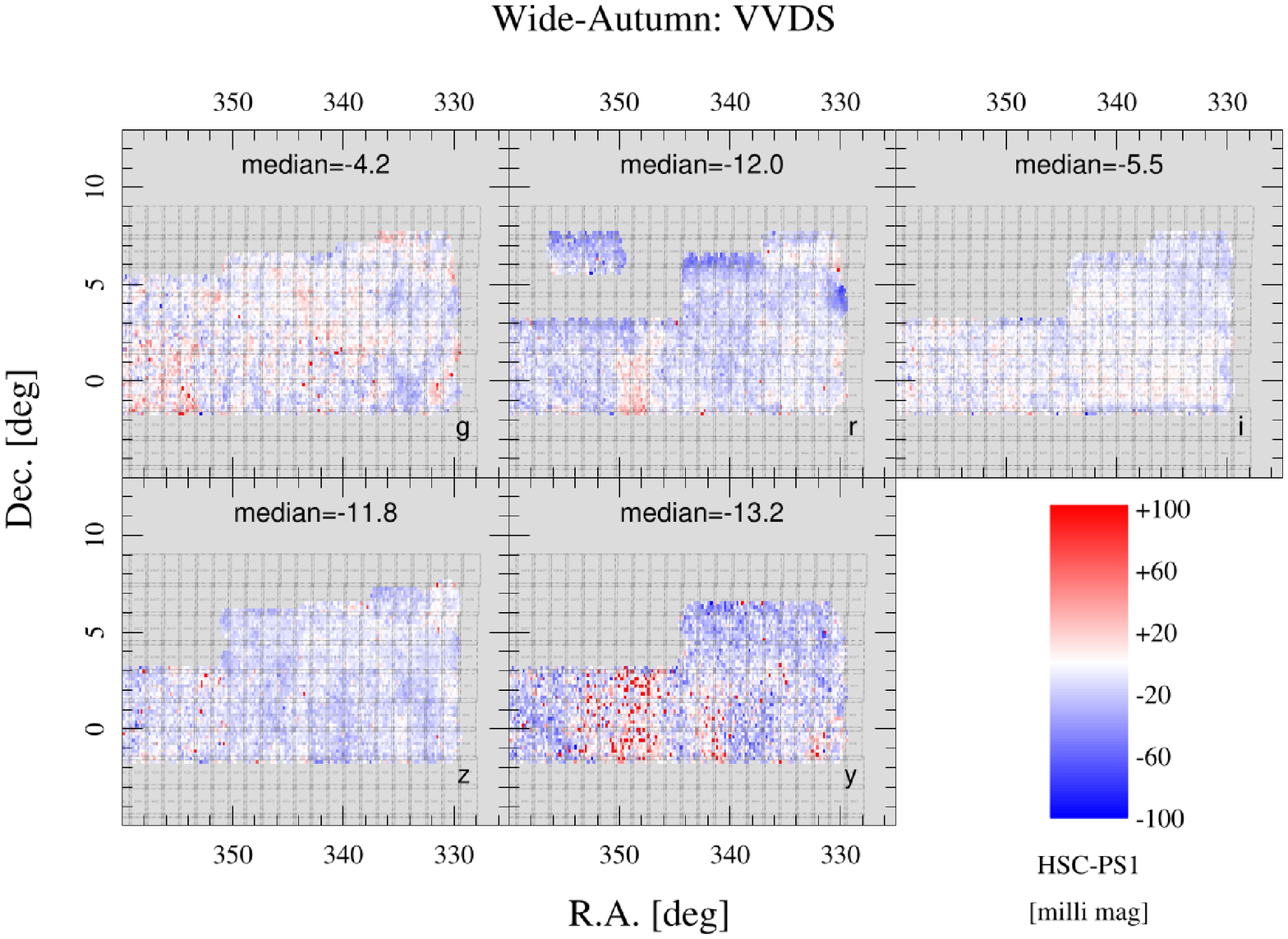}
    \end{center}
    \caption{
      The mean offset between
      the HSC PSF photometry and PS1 photometry for bright point sources in five broad-band filters as labeled in each panel.
      This is for a portion of the Wide survey.  As in Fig.~\ref{fig:astrometry}, the offset is computed for each patch and
      the color coding shows the amount of offset as shown in the bottom-right panel.  The numbers in the panels show
      the median offset over the area shown.
    }
  \label{fig:photometry}
\end{figure*}

Moving on to photometry,
Fig.~\ref{fig:photometry} shows a comparison between HSC and PS1 for bright point sources.
Here we focus on stars $1.5$~mag or more brighter than the typical $5\sigma$ magnitude limits of PS1 in each filter
(23.3, 23.2, 23.1, 22.3, 21.3 in $grizy$, respectively).
There is a systematic bias in our photometry with respect to PS1 at a $0.5-1.5$ \% level depending on filter.
A similar trend is seen in other regions of the Wide survey, although it is not seen in the D/UD fields.
The reason why we see the offset only in Wide is not understood yet.
There are patches that show a significant offset, especially in the $y$-band.
These patches seem to coincide with regions that show a somewhat large difference between
the size of model PSF and the size of observed PSF (plot not shown), which suggests that the root cause of the problem
is the PSF modeling in HSC.  Further investigations are in progress as of this writing.

\begin{figure*}
    \begin{center}
        \includegraphics[width=14cm]{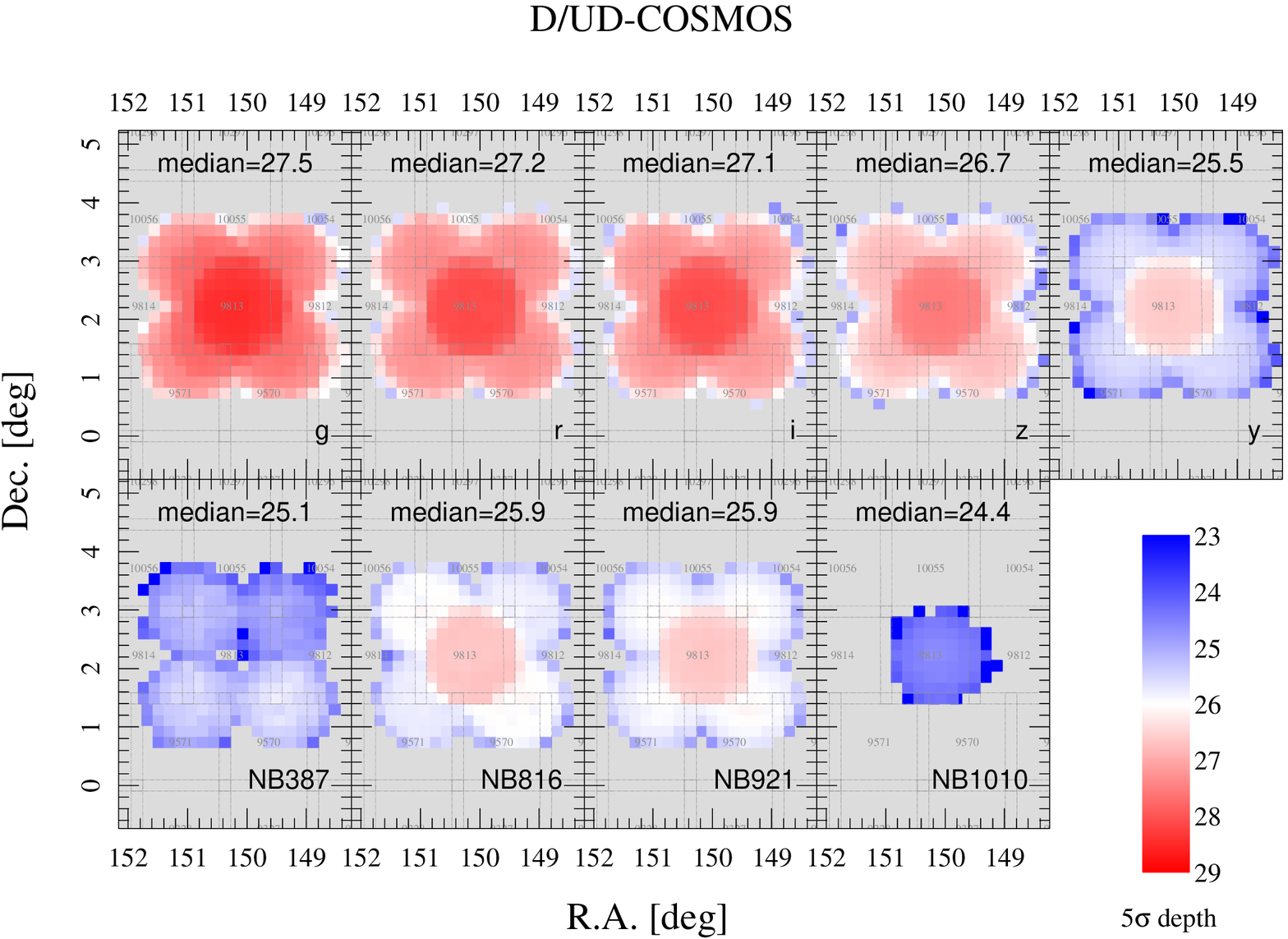}
    \end{center}
    \caption{
      As in Fig.~\ref{fig:astrometry} but the color represents the $5\sigma$ depth for point sources estimated
      using the PSF photometry.  This is the D/UD-COSMOS region with one UD pointing at the center surrounded by four D pointings.
      In addition to the broad-bands, 4 narrow-band filters are also shown.
    }
  \label{fig:depth}
\end{figure*}

Finally, Fig.\ref{fig:depth} shows the $5\sigma$ depths for point sources in the D/UD-COSMOS field.
This is based on the flux uncertainty of the PSF photometry as quoted by the pipeline and could be slightly optimistic.
The central UD region reaches down to 27 to 28th magnitude in the broad-bands; this is one of the deepest
imaging data sets of the COSMOS field.  It is even deeper than the final 10-year depth of Rubin LSST \citep{ivezic19}.
Because the field includes both Deep and UltraDeep imaging, there is a significant spatial variation of the depths,
which the user should be aware of when utilizing the D/UD data.

\subsection{Known Issues}
\label{sec:known_issues}

As we have discussed above, the overall quality of the data in this release is good.
In this section, we summarize a few issues
that the user should be aware of.

\subsubsection{Issues that persist from PDR2}
\label{sec:persisting_issues}

While we have put efforts in improving the data processing pipeline, there are issues persisting
from our previous releases, some of which are fundamentally difficult problems.  One such fundamental
problem is the over-shredding of nearby galaxies.  For instance, star-forming regions and arms of a spiral
galaxy can be deblended into separate pieces.  It is difficult to distinguish multiple galaxies blended
with each other from a single galaxy with structure.  Another fundamental difficulty with the deblending
is that it occasionally fails to cleanly deblend sources in crowded regions such as the dense core of galaxy clusters.
There is no easy solution to this problem.  There is a multi-color deblender \citep{melchior18}, which we do not
use in this release.  We hope it will help, although it remains to be seen how subtle color differences
the algorithm can exploit (cluster galaxies have similar colors).
We also note that the pipeline is not optimized for
very crowded fields such as star clusters or dwarf galaxies around the Milky Way Galaxy.  The user should
carefully check the images and catalogs in crowded regions before using them.

The NB387 data should also be used with caution because
this is a rather difficult filter to calibrate.  We infer NB387 magnitudes by applying color-terms to
the PS1 $gr$ photometry when we calibrate its zero-point.  We use the stellar library from \citet{pickles98}
to derive the color-terms.  Most of the stars in this library are close to solar metallicity, 
but the stars that we use for calibration are faint halo stars, which are likely to have subsolar metallicities.
The NB387 filter is quite sensitive to the metallicity of stars
and the PS1 $gr$ photometry does not fully capture the metallicity variation.  As we have reported on
the known issue page for PDR2 at the data release website, the photometric zero-point can be off as
much as 0.45 mag.  This is from a comparison with SDSS spectroscopic stars and is still a tentative number.
Further investigations are needed.

There are artifacts left in the coadd images.  Most obvious artifacts are optical ghosts around bright stars.
We do make an attempt to remove them (Section \ref{sec:artifact_rejection}), but we are not always successful,
especially in regions where only a few visits have been taken.  Also, our algorithm cannot remove artifacts if they
stay at similar positions on the sky.  This is often the case in the D/UD fields, where the dithers are relatively
small (several arcmin) and optical ghosts stay at similar positions.  An additional algorithm would be needed
to fully eliminate the artifacts.

\subsubsection{Overestimated CModel fluxes}
\label{sec:overestimated_cmodel_fluxes}

We perform CModel photometry in our measurements (Section 4.9.9 of\cite{bosch18}).  This is for galaxy photometry and
is expected to deliver good colors of extended sources.
In PDR2, there was a population of apparently faint objects with large CModel fluxes.
Visual inspections of those objects on the coadds indicate that their CModel fluxes are likely over-estimated.
We suspect that the over-estimated fluxes are at least partly due to the global sky subtraction applied in PDR2;
it preserves wings of bright objects and it could contaminate the outer parts of nearby objects' footprints.
It could also be due to the deblender; HSC images are very deep and the deblender often faces a difficult
problem of deblending multiple sources simultaneously (especially when the extended wings are preserved),
and the deblender may leave residual fluxes in crowded regions, which then contaminate the outskirts of nearby objects.
As CModel is relatively sensitive to fluxes in the outer parts, the flux can be over-estimated.
The fact that the Kron photometry, which is also sensitive to fluxes in the outskirts, shows similar
over-estimated fluxes supports this hypothesis.

\begin{figure}
    \begin{center}
      \includegraphics[width=8cm]{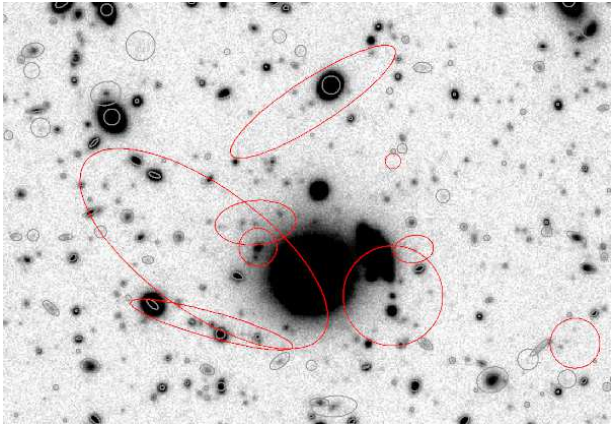}
    \end{center}
    \caption{
      Blow-up of UD-COSMOS in the $i$-band.  The ellipses show the CModel ellipses with sizes scaled to
      twice the half-light radii for illustrative purposes.  The red ellipses are objects
      with a magnitude difference larger than 1~mag between CModel and 2 arcsec aperture.
    }
  \label{fig:cmodel_ellipse}
\end{figure}

To further explore this, we show in Fig.~\ref{fig:cmodel_ellipse} a small piece of the UD-COSMOS field.
The ellipses illustrate inferred object sizes from CModel, and as can be seen, most objects have ellipses with reasonable sizes.
However, there are ellipses that appear much too large given the visual sizes of objects.  They are indicated as
the red ellipses and they are the problematic objects with over-estimated fluxes.
A close inspection of these objects shows that some of them are even false detections.
In this release (PDR3), we switch back to the local sky subtraction scheme and the fraction of these objects decreases
compared to PDR2, but they are still there.

\begin{figure*}[tbh]
    \begin{center}
      \includegraphics[width=5cm]{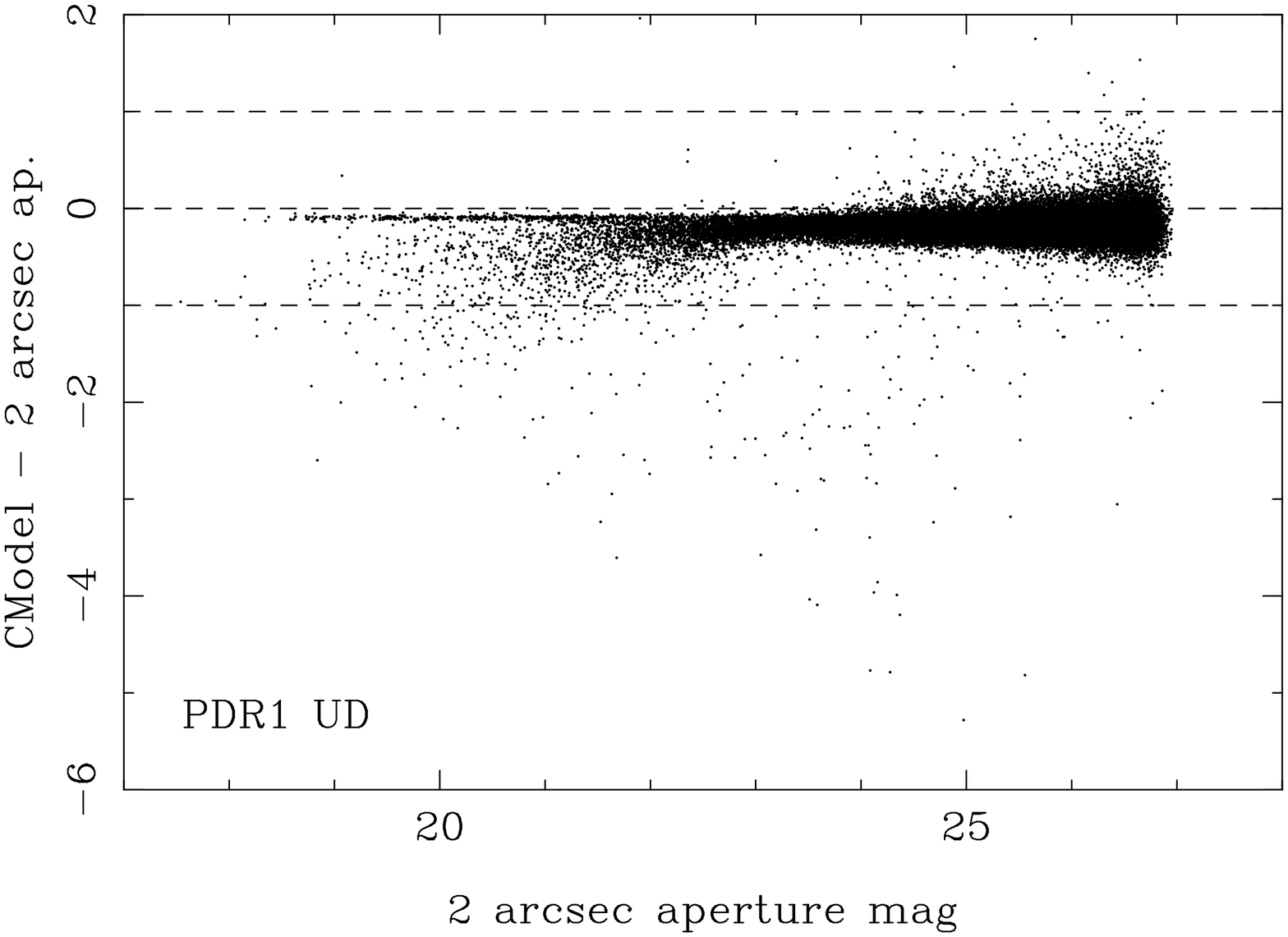}\hspace{0.5cm}
      \includegraphics[width=5cm]{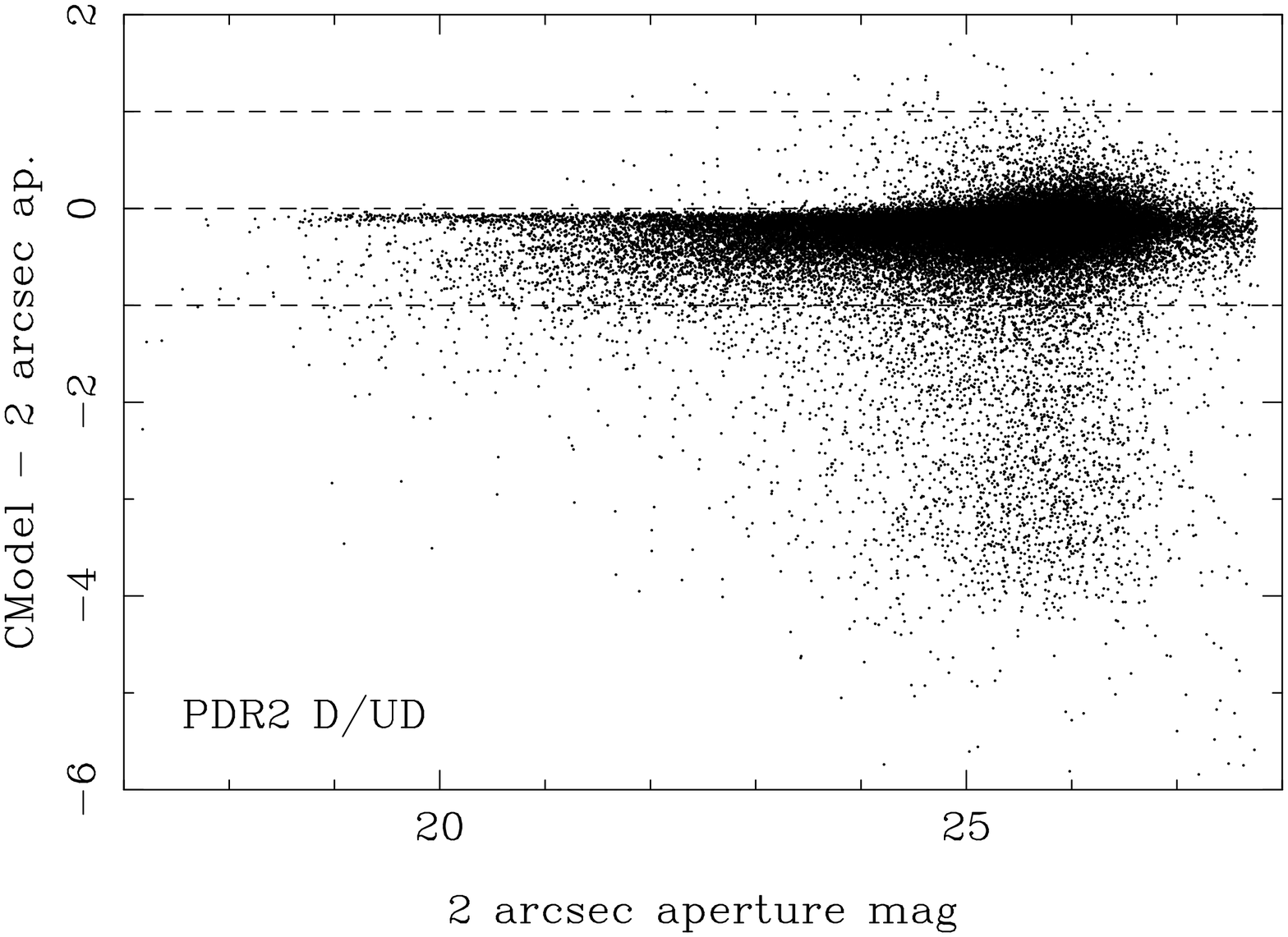}\hspace{0.5cm}
      \includegraphics[width=5cm]{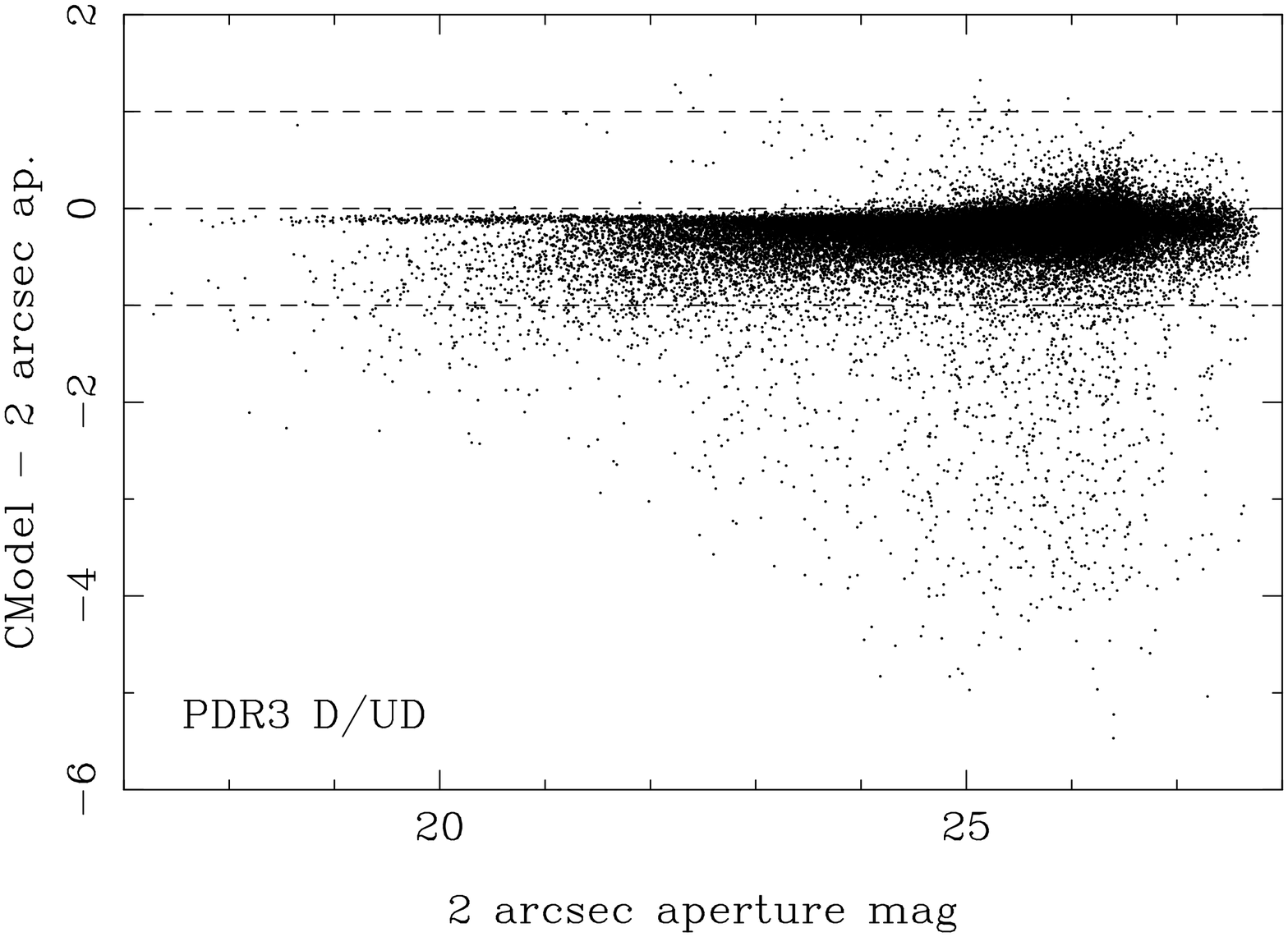}
    \end{center}
    \caption{
      Magnitude differences in $i$-band between 2 arcsec aperture and CModel as a function of the 2 arcsec
      aperture magnitude in three data releases, PDR1, PDR2, and PDR3 from left to right.  Objects with
      aperture photometry with $S/N>5$ in UD (PDR1) and D/UD (PDR2 and 3) are plotted.  For
      comparison, 50,000 objects are randomly chosen and plotted in each panel.
      The dotted horizontal lines indicate $\pm1$~mag.  The figures for Wide look very similar
      and the outlier fraction is only slightly reduced (see text).
    }
  \label{fig:cmodel_issue}
\end{figure*}

Let us be quantitative about these objects.  A way to identify them is to compare the fixed-aperture
(e.g., 2 arcsec) photometry and CModel photometry because the fixed-aperture photometry is insensitive to fluxes in
the outer parts of footprints and is thus relatively robust against flux contamination from nearby sources.
Fig.~\ref{fig:cmodel_issue} makes this comparison in D/UD.
The figures are for the $i$-band, but the same trends can be seen in all the filters.
In PDR1, there is a reasonable agreement between
the two photometric measurements and the difference between them is a smooth function of magnitude.  This is expected
because the size of objects is a function of magnitude and the amount of flux missed from the fixed-aperture
changes with magnitude.
In contrast, in PDR2, there are many objects around 25-26th magnitudes with large magnitude differences.
At this faint level, most sources are compact and the aperture photometry should be a reasonable proxy for
total flux.  Thus, CModel is likely too bright by $\sim3$ magnitudes or more for many of these sources.
The fraction of objects with a magnitude difference of more than 1~mag and detected
at $S/N>5$ is about 6\% in D/UD and 5\% in Wide\footnote{
These numbers should be regarded as rough numbers because we apply only a minimal set of flag cuts here:
only primary objects with no CModel/aperture measurement failure.
}.  In this release (PDR3), the number of such objects is significantly reduced
as shown in the right panel and
the fraction decreases to 3\% and 2\% in D/UD and Wide, respectively.
Care must be taken when comparing these numbers, but it is clear that PDR3 is better behaved than PDR2.
The fraction is only 0.5\% in PDR1 UD, although the data were rather shallow at that time.

We encourage the user to check the consistency between CModel and aperture photometry
for faint sources.
If the user is interested in relatively faint sources (e.g., $\gtrsim22$~mag),
the PSF-matched aperture photometry is more reliable; it is fixed-aperture photometry performed on
images smoothed to a common seeing size across bands and is a robust measure of colors.
It does not capture the total flux, but it is fairly
robust against deblending effects and residual flux in the outskirts.

\subsubsection{FGCM and $r/i$ vs. $r2/i2$}
\label{sec:fgcm_issues}

As already discussed in Section~\ref{sec:fgcm}, there is an issue with the photometric zero-point
introduced by the local sky subtraction performed internally in FGCM.  This has been mitigated by
the stellar sequence regression (Section~\ref{sec:stellar_sequence_regression}) and we encourage
the user to apply the zero-point offsets as well as the $r/i$ into $r2/i2$ offsets computed in
Section~\ref{sec:effective_filter_response}.  Note that these offsets have not been applied to
the photometry in the database (it is as quoted by the pipeline).  The user can apply these correction
by joining the magnitude offset tables.  See the data release website for details.

\subsubsection{Missing data in DEEP2-F3}
\label{sec:missing_data}

Due to a processing error, seven $i$-band visits (31.5 min in total) are excluded
from tract=9463 (DEEP2-F3) from the coaddition stage and the tract has $i2$-band data only.
The coadd is thus shallower than it should be by $\sim0.2$~mag.
This will be fixed in our next release.

\subsubsection{Catastrophic flux calibration failures}
\label{sec:flux_calibration_failures}

The joint flux calibration is performed using FGCM, but a tiny fraction of the CCD images
have incorrect calibrations applied due to bad outlier rejection (25 CCD images out of
2,822,678 CCD images used for Wide).  We identify these CCDs as an image with a strong spatial
gradient in the flux calibration, with at least one pixel having an inverted flux (i.e., the calibration
applied to that pixel is negative).
There are probably more CCDs that are less problematic but still bad.
This problem is visible particularly in warps because these are the images that are flux-calibrated by FGCM.
The warps are then added together to generate coadds.  We have visually inspected the potentially affected
coadds, but we see no obvious artifact.  It may be that the artifact
rejection algorithm (Section~\ref{sec:artifact_rejection}) rejected those problematic warps
during the coaddition.  It is unlikely that the coadd images as well as catalogs are significantly affected, but
warps are affected.  We provide a list of problematic warp images at the data release site.

\subsubsection{Pixel flags in the UltraDeep fields}
\label{sec:pixel_flags_in_the_ultradeep_fields}

A flag, {\tt pixelflags\_crcenter} is set for an object if a cosmic ray is detected within
3 pixels from the center of that object in any of the individual visits going into the coadd.
In the CCD processing, cosmic rays are identified and interpolated, but because interpolation
does not normally work well at object centers, we recommend excluding objects with this flag
set to select objects with clean photometry in the Wide area. 
 
\begin{figure}[tbh]
    \begin{center}
      \includegraphics[width=7.5cm,angle=270]{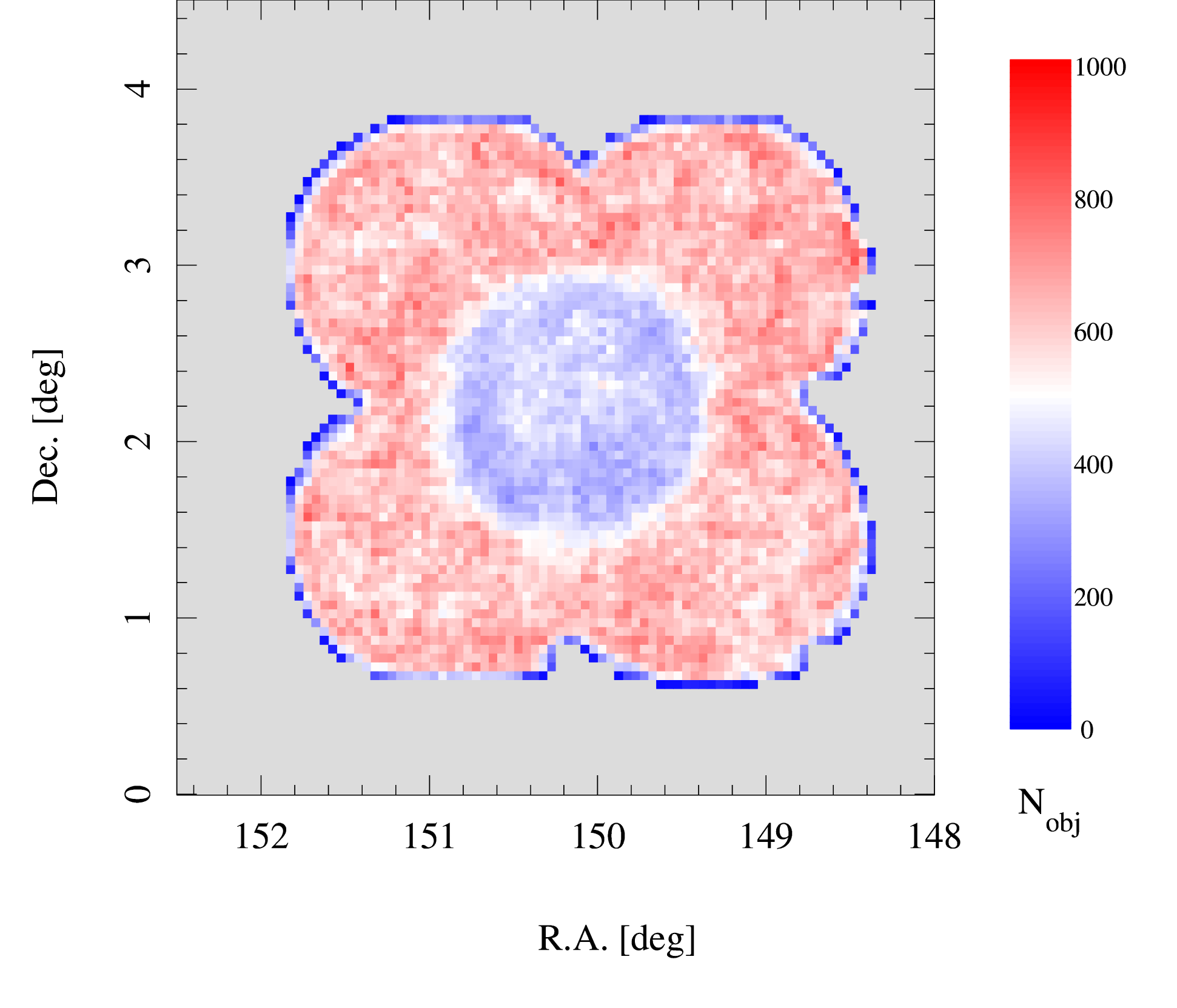}
    \end{center}
    \caption{
      Number of detected objects on a 0.05 deg grid after rejecting objects with the cosmic ray pixel flag set
      in the D/UD-COSMOS field.  The color-coding shows the number of objects.  Recall the central pointing
      is the UD pointing and the surrounding 4 pointings are D pointings.
    }
  \label{fig:cosmos_nobj}
\end{figure}

However, doing so can be problematic in the UD fields, which include 100 or more visits,
any one of which could be affected by a cosmic ray, giving a substantial chance of the object
being excluded.  Fig.\ref{fig:cosmos_nobj} illustrates the effect. The figure shows the number of
objects in the COSMOS field, on a grid of 0.05 deg, after excluding objects with
{\tt pixelflags\_crcenter} applied.  The number density in the UD region is only 2/3 that
in the surrounding D pointings, even though the UD data are considerably deeper.
The figure is for the $i$-band, but the other filters show exactly the same trend.
If we do not apply the flag cut, the UD region shows a considerably larger source density than
the Deep pointings, as expected.  We therefore suggest not to apply the cosmic ray and interpolated
pixel flags in D/UD, where there is a large number of visits and effects of cosmic rays are minor on coadds.

\section{Status of Collaborating Surveys}
\label{sec:status_of_collaborating_surveys}

The HSC-SSP survey has a number of collaborating surveys in other wavebands.
Here we give a brief update on each of them.
They all target the D/UD fields, where multi-wavelength data enable new galaxy evolution science.

\subsection{$U$-band}

The CFHT Large Area U-band Deep Survey (CLAUDS\footnote{\url{https://www.ap.smu.ca/~sawicki/sawicki/CLAUDS.html}};
\cite{sawicki19}) used the MegaCam imager on the Canada-France-Hawaii 3.6m telescope to obtain very deep $U$-band images
that overlap the D/UD fields. These new data, together with archival MegaCam images in some of the fields,
have been processed, resampled, and stacked to match the tract/patch grid, astrometric solution (updated to match
the Gaia-based astrometry of HSC-SSP PDR2/3), and pixel scale.  The CLAUDS data cover 18.60 deg$^2$ with median seeing of
FWHM=0.92 arcsec and to a median depth of $U = 27.1$ AB (5$\sigma$ in 2” apertures); selected areas in the COSMOS and SXDS
fields that total 1.36 deg$^2$ reach a median depth of $U = 27.7$ AB (5$\sigma$ in 2” apertures).  Altogether, the CLAUDS
images represent the equivalent of 113 classical-mode CFHT nights and are the deepest $U$-band data ever taken over this
combination of depth and area. 

Multiband ($U + grizy$ as well as NIR, where publicly available) photometry is carried out by the CLAUDS team
using SExtractor \citep{bertin96} and an adaptation of {\tt hscPipe} that has been adapted to handle  CFHT $U$-band images.
The combination of CLAUDS and HSC-SSP data enables research
in many broad science areas by significantly enhancing photometric redshift performance (particularly at $z\lesssim 0.8$ and
$z\gtrsim2$) and permitting the selection of $z\sim3$ Lyman break galaxies and quasars. Several science projects have already
been carried out with the joint CLAUDS+HSC-SSP datasets \citep{halevi19,moutard20,huang20,golob21,thibert21,harikane21},
while others are underway.   The CLAUDS team anticipates
releasing their $U+grizy$ catalogs and processed images to the public in late 2021 or early 2022 (G. Desprez et al. in prep.).

\subsection{near-IR}

Turning to near-infrared, the D/UD fields overlap with some of the major near-infrared imaging surveys
including the Deep eXtragalactic Survey of the UKIRT Infrared Deep Sky Survey (UKIDSS/DXS; \cite{kim11}),
the UKIDSS Ultra-Deep Survey (UDS; \cite{lawrence07}),
VISTA Deep Extragalactic Observations Survey (VIDEO; \cite{jarvis13}), and
Ultra Deep Survey with the VISTA Telescope (UltraVISTA; \cite{mccracken12}).
We designed the D/UD fields to maximize the overlap with these surveys, but a portion of the D/UD fields,
especially the flanking fields of the E-COSMOS region and DEEP2-F3, are missing the near-IR coverage.
Deep UKIRT Near-infrared Steward Survey (DUNES$^2$; Egami et al., in preparation) is filling the missing part.
With a total time allocation of $\sim$270 hours, DUNES$^2$ observed the E-COSMOS to
$J$\,$\sim$\,23.6, $H$\,$\sim$\,23.2, and $K$\,$\sim$\,23.2 mag at $5\sigma$ within 2 arcsec aperture,
DEEP2-F3 to $J$\,$\sim$\,23.3, and $K$\,$\sim$\,23.1 mag, and ELAIS-N1 to $H\sim23.2$ mag under seeing
conditions of FWHM\,$=$\,0.8--1.0 arcsec.  The actual field
coverage can be found at the DUNES$^2$ project website\footnote{
\url{http://gxn.as.arizona.edu/DUNES}}.

The DUNES$^2$ team has internally produced a $U$-to-$K$ band-merged catalog by combining the DUNES$^2$
and other existing near-infrared source catalogs with the CLAUDS $U$-band and HSC optical ($grizy$)
catalogs.  This internal catalog is primarily being used to improve photometric redshifts, derive galaxy
stellar mass functions (with a special focus on constraining the massive end), and identify massive
high-redshift ($z$\,$\sim$\,3) galaxy candidates, which are being actively followed up spectroscopically
using the Large Binocular Telescope (LBT) and Multiple Mirror Telescope (MMT) in Arizona (Y.-H.~Huang et al. in preparation).
This catalog has also been used to cross-check the quality/accuracy of the {\tt hscPipe}-based multi-band
catalog, which will be the official catalog from HSC-SSP (see below).  The DUNES$^2$ team plans to
publicly release the processed UKIRT/WFCAM images and associated source catalogs in early 2022.

A complementary survey, DeepCos (PI: Y.-T.~Lin), is also being carried out.  The survey originally aimed
to bring the DUNES$^2$ footprint to the depth of VIDEO and used WFCAM on UKIRT to further observe 
E-COSMOS and DEEP2-F3 in $J$ and $K$.
Due to operation issues with UKIRT, we have switched to using WIRCam on CFHT in 2020 and continued
the survey.  The survey now aims to provide a uniform $J_{\rm AB}=23.7$ coverage over $\sim11$ deg$^2$ area
in three of the Deep fields (XMM-LSS, E-COSMOS, and DEEP2-F3).  Such a data set will not only improve
the photo-$z$ accuracy of the D/UD fields, but will also play a critical role in the target selection for
an upcoming survey with Prime Focus Spectrograph (PFS; \cite{tamura18}).
The DeepCos survey is approximately 70\% done at this point and is expected to deliver coadded
WFCAM and WIRCam images by 2022.

\subsection{IR}

{\sl Spitzer} IRAC 3.6 and 4.5\,$\mu$m observations exist for a fraction of the D/UD fields.  These include
the $\sim$2 deg$^2$, {\sl Spitzer} COSMOS survey (S-COSMOS) \citep{sanders07} as well as multiple subsequent
deeper surveys within the same footprint.  There is also the warm mission SERVS survey \citep{mauduit12}, which
covered $\sim$2 deg$^2$ in ELAIS-N1 and $\sim$4.5 deg$^2$ in XMM-LSS. The SERVS coverage is fairly uniform with
exposure times of $\approx$1200s, translating to $\sim$23.1~mag at $5\sigma$ for point sources.  The entire
XMM-LSS field was later covered to the same depth with the {\sl Spitzer} DeepDrill survey \citep{lacy21}.
However, some of the D/UD fields are missing the IR coverage and the depths are shallow in some areas.

We have two collaborating surveys that increase the area and depth in the IR.
One is Spitzer Large Area Survey with Hyper Suprime-Cam (SPLASH; \cite{mehta18}).
SPLASH focuses on the two UD fields and it goes down to $\sim25.2$ at $5\sigma$ in 3.6$\mu m$ and $4.5\mu m$.
The survey is complete and a multi-band catalog in SXDS has been released to the community \citep{mehta18}.
The other collaborating survey is the Spitzer Coverage of the HSC-Deep with IRAC for Z-studies (SHIRAZ;
Annunziatella et al. in prep).  Nearly 500 hours have been awarded to SHIRAZ and it covers
the missing portions of the E-COSMOS and ELAIS-N1 fields as well as the whole DEEP-F3 field, where only very
shallow data existed previously \citep{timlin16}, increasing the {\it Spitzer}  coverage of the Deep fields
by $\sim7$ deg$^2$.  The depth now matches that of the SERVS and {\sl Spitzer} DeepDrill surveys
($\sim23$ mag in 3.6$\mu m$ and $4.5\mu m$).
By combining with the existing data, we have newly produced IRAC mosaics over $\sim$17 deg$^2$ for
all the Deep fields, except for XMM-LSS (which is published in \cite{lacy21}).  Details of the survey and
the data processing can be found in Annunziatella et al. (in prep.).

In order to fully exploit the data from the collaborating surveys, we have been putting efforts to combine
the HSC data with data from other wavelengths.
Instead of processing the raw data from the collaborating/external surveys, we start from coadds from each survey and run object
detection and measurements using {\tt hscPipe} to perform consistent object detection and photometry.  We have constructed
a combined catalog with deep multi-band photometry from $u$ through $K$ and it is being exploited by the collaboration, mostly
for distant galaxy studies.  The next step is to incorporate the Spitzer photometry into the catalog,
extending the wavelength coverage out to $8\mu m$, which will further enhance our science.
We plan to release the combined data set to the community in the future.

\section{Towards the Final Data Release}
\label{sec:summary}

We have presented the third data release from HSC-SSP in this paper.  PDR3 covers about 670 square degrees of
FCFD Wide area as well as four separate D/UD fields from 278 nights of observing time.  The dynamic range of
the area and depth spanned in this survey is a unique aspect of the survey.
The data are served at the same data release site as the previous releases,
\url{https://hsc-release.mtk.nao.ac.jp/}, where the user can find both the data and extensive metadata,
such as a comprehensive set of QA plots.  The list of known issues given in this paper and listed on the website
should be referred to before exploiting the data for science.

We have executed 95\% of the allocated observing time as of this writing and we expect to complete the survey
by the end of 2021.  The remaining observing time will be spent on the Wide layer and final data set will
include $\sim1,200$ square degrees of FCFD area.  We will not take additional D/UD data.
We anticipate that the next data release will be the final data release from HSC-SSP.  We do not have
a specific release date yet, but we will keep the community informed on our website.
Following the imaging survey, we plan to carry out a massive spectroscopic survey using PFS
on the Subaru Telescope.
The PFS survey will be based largely on the imaging data from HSC-SSP, and the combination
of imaging and spectroscopy will allow us to explore deeper into the Universe.

\section*{Acknowledgments}

The Hyper Suprime-Cam (HSC) collaboration includes the astronomical communities of Japan and Taiwan,
and Princeton University.  The HSC instrumentation and software were developed by the National
Astronomical Observatory of Japan (NAOJ), the Kavli Institute for the Physics and Mathematics of
the Universe (Kavli IPMU), the University of Tokyo, the High Energy Accelerator Research Organization (KEK),
the Academia Sinica Institute for Astronomy and Astrophysics in Taiwan (ASIAA), and Princeton University.
Funding was contributed by the FIRST program from Japanese Cabinet Office, the Ministry of Education,
Culture, Sports, Science and Technology (MEXT), the Japan Society for the Promotion of Science (JSPS),
Japan Science and Technology Agency  (JST),  the Toray Science  Foundation, NAOJ, Kavli IPMU, KEK, ASIAA,
and Princeton University.

This paper is based on data collected at the Subaru Telescope and retrieved from the HSC data archive system, which is operated by Subaru Telescope and Astronomy Data Center at NAOJ.  Data analysis was in part carried out with the cooperation of Center for Computational Astrophysics at NAOJ.  We are honored and grateful for the opportunity of observing the Universe from Maunakea, which has the cultural, historical and natural significance in Hawaii.

This paper makes use of software developed for Vera C. Rubin Observatory. We thank the Rubin Observatory for
making their code available as free software at http://pipelines.lsst.io/.

The Pan-STARRS1 Surveys (PS1) and the PS1 public science archive have been made possible through contributions by the Institute for Astronomy, the University of Hawaii, the Pan-STARRS Project Office, the Max-Planck Society and its participating institutes, the Max Planck Institute for Astronomy, Heidelberg and the Max Planck Institute for Extraterrestrial Physics, Garching, The Johns Hopkins University, Durham University, the University of Edinburgh, the Queen's University Belfast, the Harvard-Smithsonian Center for Astrophysics, the Las Cumbres Observatory Global Telescope Network Incorporated, the National Central University of Taiwan, the Space Telescope Science Institute, the National Aeronautics and Space Administration under Grant No. NNX08AR22G issued through the Planetary Science Division of the NASA Science Mission Directorate, the National Science Foundation Grant No. AST-1238877, the University of Maryland, Eotvos Lorand University (ELTE), the Los Alamos National Laboratory, and the Gordon and Betty Moore Foundation.

This work is also based on the following data and observations:
zCOSMOS observations carried out using the Very Large Telescope at the ESO Paranal Observatory under Programme ID: LP175.A-0839,
observations taken by the 3D-HST Treasury Program (GO 12177 and 12328) with the NASA/ESA HST, which is operated by the Association of Universities for Research in Astronomy, Inc., under NASA contract NAS5-26555,
data from the VIMOS VLT Deep Survey, obtained from the VVDS database operated by Cesam, Laboratoire d'Astrophysique de Marseille, France,
data from the VIMOS Public Extragalactic Redshift Survey (VIPERS), which has been performed using the ESO Very Large Telescope, under the "Large Programme" 182.A-0886, and the participating institutions and funding agencies are listed at http://vipers.inaf.it.
Funding for SDSS-III has been provided by the Alfred P. Sloan Foundation, the Participating Institutions, the National Science Foundation, and the U.S. Department of Energy Office of Science. The SDSS-III web site is http://www.sdss3.org/.  SDSS-III is managed by the Astrophysical Research Consortium for the Participating Institutions of the SDSS-III Collaboration including the University of Arizona, the Brazilian Participation Group, Brookhaven National Laboratory, Carnegie Mellon University, University of Florida, the French Participation Group, the German Participation Group, Harvard University, the Instituto de Astrofisica de Canarias, the Michigan State/Notre Dame/JINA Participation Group, Johns Hopkins University, Lawrence Berkeley National Laboratory, Max Planck Institute for Astrophysics, Max Planck Institute for Extraterrestrial Physics, New Mexico State University, New York University, Ohio State University, Pennsylvania State University, University of Portsmouth, Princeton University, the Spanish Participation Group, University of Tokyo, University of Utah, Vanderbilt University, University of Virginia, University of Washington, and Yale University.
GAMA is a joint European-Australasian project based around a spectroscopic campaign using the Anglo-Australian Telescope.
The GAMA input catalogue is based on data taken from the Sloan Digital Sky Survey and the UKIRT Infrared Deep Sky Survey. Complementary imaging of the GAMA regions is being obtained by a number of independent survey programmes including GALEX MIS, VST KiDS, VISTA VIKING, WISE, Herschel-ATLAS, GMRT and ASKAP providing UV to radio coverage. GAMA is funded by the STFC (UK), the ARC (Australia), the AAO, and the participating institutions. The GAMA website is http://www.gama-survey.org/.
Funding for the DEEP2 Galaxy Redshift Survey has been provided by NSF grants AST-95-09298, AST-0071048, AST-0507428, and AST-0507483 as well as NASA LTSA grant NNG04GC89G.  Funding for PRIMUS is provided by NSF (AST-0607701, AST-0908246, AST-0908442, AST-0908354) and NASA (Spitzer-1356708, 08-ADP08-0019, NNX09AC95G).
Funding for the DEEP3 Galaxy Redshift Survey has been provided by NSF grants AST-0808133, AST-0807630, and AST-0806732.
The LEGA-C survey is based on based on data products from observations made with ESO Telescopes at the La Silla  Paranal Observatory under programme ID 194.A-­2005(A-‐F).

\bibliographystyle{apj}
\bibliography{references}

\end{document}

%% file: author.tex
\author{
Hiroaki Aihara\altaffilmark{1,2},
Yusra AlSayyad\altaffilmark{3},
Makoto Ando\altaffilmark{4},
Robert Armstrong\altaffilmark{5},
James Bosch\altaffilmark{3},
Eiichi Egami\altaffilmark{6},
Hisanori Furusawa\altaffilmark{7},
Junko Furusawa\altaffilmark{7},
Sumiko Harasawa\altaffilmark{7},
Yuichi Harikane\altaffilmark{8,9},
Bau-Ching Hsieh\altaffilmark{10},
Hiroyuki Ikeda\altaffilmark{11},
Kei Ito\altaffilmark{12,7},
Ikuru Iwata\altaffilmark{7,12},
Tadayuki Kodama\altaffilmark{13},
Michitaro Koike\altaffilmark{7},
Mitsuru Kokubo\altaffilmark{3},
Yutaka Komiyama\altaffilmark{7,12},
Xiangchong Li\altaffilmark{2,1},
Yongming Liang\altaffilmark{12,7},
Yen-Ting Lin\altaffilmark{10},
Robert H. Lupton\altaffilmark{3},
Nate B Lust\altaffilmark{3},
Lauren A. MacArthur\altaffilmark{3},
Ken Mawatari\altaffilmark{7,8},
Sogo Mineo\altaffilmark{7},
Hironao Miyatake\altaffilmark{14,2},
Satoshi Miyazaki\altaffilmark{7},
Surhud More\altaffilmark{15,2},
Takahiro Morishima\altaffilmark{7},
Hitoshi Murayama\altaffilmark{2,16,17},
Kimihiko Nakajima\altaffilmark{7},
Fumiaki Nakata\altaffilmark{7},
Atsushi J. Nishizawa\altaffilmark{18,19},
Masamune Oguri\altaffilmark{20,1,2},
Nobuhiro Okabe\altaffilmark{21},
Yuki Okura\altaffilmark{7},
Yoshiaki Ono\altaffilmark{8},
Ken Osato\altaffilmark{22,23},
Masami Ouchi\altaffilmark{7,8,2},
Yen-Chen Pan\altaffilmark{24},
Andrés A. Plazas Malagón\altaffilmark{3},
Paul A. Price\altaffilmark{3},
Sophie L Reed\altaffilmark{3},
Eli S. Rykoff\altaffilmark{25,26},
Takatoshi Shibuya\altaffilmark{27},
Mirko Simunovic\altaffilmark{28},
Michael A. Strauss\altaffilmark{3},
Kanako Sugimori\altaffilmark{12,7},
Yasushi Suto\altaffilmark{1,20},
Nao Suzuki\altaffilmark{2},
Masahiro Takada\altaffilmark{2},
Yuhei Takagi\altaffilmark{28},
Tadafumi Takata\altaffilmark{7,12},
Satoshi Takita\altaffilmark{29},
Masayuki Tanaka\altaffilmark{7,12,*},
Shenli Tang\altaffilmark{1,2,8},
Dan S. Taranu\altaffilmark{3},
Tsuyoshi Terai\altaffilmark{28},
Yoshiki Toba\altaffilmark{30,10},
Edwin L. Turner\altaffilmark{3,2},
Hisakazu Uchiyama\altaffilmark{31},
Bovornpratch Vijarnwannaluk\altaffilmark{13},
Christopher Z Waters\altaffilmark{3},
Yoshihiko Yamada\altaffilmark{32,7},
Naoaki Yamamoto\altaffilmark{13},
Takuji Yamashita\altaffilmark{7,31}
}
\altaffiltext{1}{Department of Physics,  7-3-1 Hongo, Bunkyo-ku, The University of Tokyo, Tokyo 113-0033, Japan }
\altaffiltext{2}{Kavli Institute for the Physics and Mathematics of the Universe (Kavli IPMU, WPI), The University of Tokyo, 5-1-5 Kashiwanoha, Kashiwa, Chiba 277-8583, Japan}
\altaffiltext{3}{Department of Astrophysical Sciences, Princeton University, 4 Ivy Lane, Princeton, NJ 08544, USA}
\altaffiltext{4}{Department of Astronomy, Graduate School of Science, The University of Tokyo, 7-3-1 Hongo, Bunkyo, Tokyo, 113-0033, Japan}
\altaffiltext{5}{Lawrence Livermore National Laboratory, Livermore, CA 94551, USA}
\altaffiltext{6}{Steward Observatory, The University of Arizona, 933 North Cherry Avenue, Tucson, AZ 85721-0065, USA}
\altaffiltext{7}{National Astronomical Observatory of Japan, 2-21-1 Osawa, Mitaka, Tokyo 181-8588, Japan}
\altaffiltext{8}{Institute for Cosmic Ray Research, The University of Tokyo, 5-1-5 Kashiwanoha, Kashiwa, Chiba 277-8582, Japan}
\altaffiltext{9}{Department of Physics and Astronomy  University College London  Gower Street  London WC1E 6BT  UK}
\altaffiltext{10}{Institute of Astronomy and Astrophysics, Academia Sinica, 11F of Astronomy-Mathematics Building, AS/NTU No.1, Sec. 4, Roosevelt Rd, Taipei 10617, Taiwan, R.O.C.}
\altaffiltext{11}{National Institute of Technology, Wakayama College, 77 Noshima, Nada-cho, Gobo, Wakayama 644-0023, Japan}
\altaffiltext{12}{Department of Astronomical Science, The Graduate University for Advanced Studies, SOKENDAI, 2-21-1 Osawa, Mitaka, Tokyo, 181-8588, Japan}
\altaffiltext{13}{Astronomical Institute, Tohoku University,  6-3, Aramaki, Aoba-ku, Sendai, Miyagi, 980-8578, Japan}
\altaffiltext{14}{Kobayashi-Maskawa Institute for the Origin of Particles and the Universe (KMI), Nagoya University, Nagoya, 464-8602, Japan}
\altaffiltext{15}{Inter University Centre for Astronomy and Astrophysics, Ganeshkhind, Pune 411007, India}
\altaffiltext{16}{Department of Physics and Center for Japanese Studies, University of California, Berkeley, CA 94720, USA}
\altaffiltext{17}{Theoretical Physics Group, Lawrence Berkeley National Laboratory, MS 50A-5104, Berkeley, CA 94720}
\altaffiltext{18}{Institute for Advanced Research, Nagoya University, Furocho, Chikusa-ku, Nagoya, 464-8602 Japan}
\altaffiltext{19}{Division of Particle and Astrophysical Science, Graduate School of Science, Nagoya University, Furo-cho, Nagoya 464-8602, Japan}
\altaffiltext{20}{Research Center for the Early Universe, The University of Tokyo, 7-3-1 Hongo, Bunkyo-ku, Tokyo 113-0033, Japan}
\altaffiltext{21}{Physics Program, Graduate School of Advanced Science and Engineering, Hiroshima University  1-3-1 Kagamiyama  Higashi-Hiroshima  Hiroshima 739-8526  Japan}
\altaffiltext{22}{Center for Gravitational Physics, Yukawa Institute for Theoretical Physics, Kyoto University, Kyoto 606-8502, Japan}
\altaffiltext{23}{D\'{e}partment de Physique  \'{E}cole Normale Sup\'{e}rieure  Universit\'{e} PSL  CNRS  Sorbonne Universit\'{e}  Universit\'{e} de Paris  75005 Paris  France}
\altaffiltext{24}{Graduate Institute of Astronomy, National Central University, 300 Jhongda Road, 32001 Jhongli, Taiwan}
\altaffiltext{25}{Kavli Institute for Particle Astrophysics and Cosmology, P. O. Box 2450  Stanford University  Stanford  CA 94305  USA}
\altaffiltext{26}{Kavli Institute for Particle Astrophysics and Cosmology, SLAC National Accelerator Laboratory, Stanford University, 2575 Sand Hill Road, M/S 29, Menlo Park, CA 94025}
\altaffiltext{27}{Kitami Institute of Technology, 165 Koen-cho, Kitami, Hokkaido 090-8507, Japan}
\altaffiltext{28}{Subaru Telescope, National Astronomical Observatory of Japan, 650 North A`ohoku Pl, Hilo, HI 96720, USA}
\altaffiltext{29}{Institute of Astronomy, The University of Tokyo, 2-21-1 Osawa, Mitaka, Tokyo 181-0015,Japan}
\altaffiltext{30}{Department of Astronomy, Kyoto University, Kitashirakawa-Oiwake-cho, Sakyo-ku, Kyoto 606-8502, Japan}
\altaffiltext{31}{Research Center for Space and Cosmic Evolution, Ehime University, 2-5 Bunkyo-cho, Matsuyama, Ehime 790-8577, Japan}
\altaffiltext{32}{Japan Space Forum, Shin-Ochanomizu Urban Trinity Bldg. 3F, 3-2-1, Kandasurugadai, Chiyoda, Tokyo 101-0062, Japan}

%% file: hsc_dr3_lowres.bbl
\begin{thebibliography}{}
\expandafter\ifx\csname natexlab\endcsname\relax\def\natexlab#1{#1}\fi

\bibitem[{{Ahumada} {et~al.}(2020){Ahumada}, {Prieto}, {Almeida}, {Anders},
  {Anderson}, {Andrews}, {Anguiano}, {Arcodia}, {Armengaud}, {Aubert}, \&
  et~al.}]{ahumada20}
{Ahumada}, R., {Prieto}, C.~A., {Almeida}, A., {et~al.} 2020, \apjs, 249, 3

\bibitem[{{Aihara} {et~al.}(2018{\natexlab{a}}){Aihara}, {Armstrong},
  {Bickerton}, {Bosch}, {Coupon}, {Furusawa}, {Hayashi}, {Ikeda}, {Kamata},
  {Karoji}, {Kawanomoto}, {Koike}, {Komiyama}, {Lang}, {Lupton}, {Mineo},
  {Miyatake}, {Miyazaki}, {Morokuma}, {Obuchi}, {Oishi}, {Okura}, {Price},
  {Takata}, {Tanaka}, {Tanaka}, {Tanaka}, {Uchida}, {Uraguchi}, {Utsumi},
  {Wang}, {Yamada}, {Yamanoi}, {Yasuda}, {Arimoto}, {Chiba}, {Finet},
  {Fujimori}, {Fujimoto}, {Furusawa}, {Goto}, {Goulding}, {Gunn}, {Harikane},
  {Hattori}, {Hayashi}, {He{\l}miniak}, {Higuchi}, {Hikage}, {Ho}, {Hsieh},
  {Huang}, {Huang}, {Imanishi}, {Iwata}, {Jaelani}, {Jian}, {Kashikawa},
  {Katayama}, {Kojima}, {Konno}, {Koshida}, {Kusakabe}, {Leauthaud}, {Lee},
  {Lin}, {Lin}, {Mandelbaum}, {Matsuoka}, {Medezinski}, {Miyama}, {Momose},
  {More}, {More}, {Mukae}, {Murata}, {Murayama}, {Nagao}, {Nakata}, {Niida},
  {Niikura}, {Nishizawa}, {Oguri}, {Okabe}, {Ono}, {Onodera}, {Onoue}, {Ouchi},
  {Pyo}, {Shibuya}, {Shimasaku}, {Simet}, {Speagle}, {Spergel}, {Strauss},
  {Sugahara}, {Sugiyama}, {Suto}, {Suzuki}, {Tait}, {Takada}, {Terai}, {Toba},
  {Turner}, {Uchiyama}, {Umetsu}, {Urata}, {Usuda}, {Yeh}, \&
  {Yuma}}]{aihara18a}
{Aihara}, H., {Armstrong}, R., {Bickerton}, S., {et~al.} 2018{\natexlab{a}},
  \pasj, 70, S8

\bibitem[{{Aihara} {et~al.}(2018{\natexlab{b}}){Aihara}, {Arimoto},
  {Armstrong}, {Arnouts}, {Bahcall}, {Bickerton}, {Bosch}, {Bundy}, {Capak},
  {Chan}, {Chiba}, {Coupon}, {Egami}, {Enoki}, {Finet}, {Fujimori}, {Fujimoto},
  {Furusawa}, {Furusawa}, {Goto}, {Goulding}, {Greco}, {Greene}, {Gunn},
  {Hamana}, {Harikane}, {Hashimoto}, {Hattori}, {Hayashi}, {Hayashi},
  {He{\l}miniak}, {Higuchi}, {Hikage}, {Ho}, {Hsieh}, {Huang}, {Huang},
  {Ikeda}, {Imanishi}, {Inoue}, {Iwasawa}, {Iwata}, {Jaelani}, {Jian},
  {Kamata}, {Karoji}, {Kashikawa}, {Katayama}, {Kawanomoto}, {Kayo}, {Koda},
  {Koike}, {Kojima}, {Komiyama}, {Konno}, {Koshida}, {Koyama}, {Kusakabe},
  {Leauthaud}, {Lee}, {Lin}, {Lin}, {Lupton}, {Mandelbaum}, {Matsuoka},
  {Medezinski}, {Mineo}, {Miyama}, {Miyatake}, {Miyazaki}, {Momose}, {More},
  {More}, {Moritani}, {Moriya}, {Morokuma}, {Mukae}, {Murata}, {Murayama},
  {Nagao}, {Nakata}, {Niida}, {Niikura}, {Nishizawa}, {Obuchi}, {Oguri},
  {Oishi}, {Okabe}, {Okamoto}, {Okura}, {Ono}, {Onodera}, {Onoue}, {Osato},
  {Ouchi}, {Price}, {Pyo}, {Sako}, {Sawicki}, {Shibuya}, {Shimasaku},
  {Shimono}, {Shirasaki}, {Silverman}, {Simet}, {Speagle}, {Spergel},
  {Strauss}, {Sugahara}, {Sugiyama}, {Suto}, {Suyu}, {Suzuki}, {Tait},
  {Takada}, {Takata}, {Tamura}, {Tanaka}, {Tanaka}, {Tanaka}, {Tanaka},
  {Terai}, {Terashima}, {Toba}, {Tominaga}, {Toshikawa}, {Turner}, {Uchida},
  {Uchiyama}, {Umetsu}, {Uraguchi}, {Urata}, {Usuda}, {Utsumi}, {Wang}, {Wang},
  {Wong}, {Yabe}, {Yamada}, {Yamanoi}, {Yasuda}, {Yeh}, {Yonehara}, \&
  {Yuma}}]{aihara18b}
{Aihara}, H., {Arimoto}, N., {Armstrong}, R., {et~al.} 2018{\natexlab{b}},
  \pasj, 70, S4

\bibitem[{{Aihara} {et~al.}(2019){Aihara}, {AlSayyad}, {Ando}, {Armstrong},
  {Bosch}, {Egami}, {Furusawa}, {Furusawa}, {Goulding}, {Harikane}, {Hikage},
  {Ho}, {Hsieh}, {Huang}, {Ikeda}, {Imanishi}, {Ito}, {Iwata}, {Jaelani},
  {Kakuma}, {Kawana}, {Kikuta}, {Kobayashi}, {Koike}, {Komiyama}, {Li},
  {Liang}, {Lin}, {Luo}, {Lupton}, {Lust}, {MacArthur}, {Matsuoka}, {Mineo},
  {Miyatake}, {Miyazaki}, {More}, {Murata}, {Namiki}, {Nishizawa}, {Oguri},
  {Okabe}, {Okamoto}, {Okura}, {Ono}, {Onodera}, {Onoue}, {Osato}, {Ouchi},
  {Shibuya}, {Strauss}, {Sugiyama}, {Suto}, {Takada}, {Takagi}, {Takata},
  {Takita}, {Tanaka}, {Terai}, {Toba}, {Uchiyama}, {Utsumi}, {Wang}, {Wang}, \&
  {Yamada}}]{aihara19}
{Aihara}, H., {AlSayyad}, Y., {Ando}, M., {et~al.} 2019, \pasj, 71, 114

\bibitem[{{AlSayyad}(2018)}]{Alsayyad18}
{AlSayyad}, Y. 2018, {Coaddition Artifact Rejection and CompareWarp}, {LSST
  Data Management Technical Note} {DMTN-080}, {LSST Data Management},
  doi:10.5281/zenodo.2605418

\bibitem[{{Berk} {et~al.}(1999){Berk}, {Anderson}, {Bernstein}, {Acharya},
  {Dothe}, {Matthew}, {Adler-Golden}, {Chetwynd}, {Richtsmeier}, {Pukall},
  {Allred}, {Jeong}, \& {Hoke}}]{berk99}
{Berk}, A., {Anderson}, G.~P., {Bernstein}, L.~S., {et~al.} 1999, in Society of
  Photo-Optical Instrumentation Engineers (SPIE) Conference Series, Vol. 3756,
  Optical Spectroscopic Techniques and Instrumentation for Atmospheric and
  Space Research III, ed. A.~M. {Larar}, 348--353

\bibitem[{{Bertin} \& {Arnouts}(1996)}]{bertin96}
{Bertin}, E., \& {Arnouts}, S. 1996, \aaps, 117, 393

\bibitem[{{Bosch} {et~al.}(2018{\natexlab{a}}){Bosch}, {Armstrong},
  {Bickerton}, {Furusawa}, {Ikeda}, {Koike}, {Lupton}, {Mineo}, {Price},
  {Takata}, {Tanaka}, {Yasuda}, {AlSayyad}, {Becker}, {Coulton}, {Coupon},
  {Garmilla}, {Huang}, {Krughoff}, {Lang}, {Leauthaud}, {Lim}, {Lust},
  {MacArthur}, {Mandelbaum}, {Miyatake}, {Miyazaki}, {Murata}, {More}, {Okura},
  {Owen}, {Swinbank}, {Strauss}, {Yamada}, \& {Yamanoi}}]{bosch18}
{Bosch}, J., {Armstrong}, R., {Bickerton}, S., {et~al.} 2018{\natexlab{a}},
  \pasj, 70, S5

\bibitem[{{Bosch} {et~al.}(2018{\natexlab{b}}){Bosch}, {Armstrong},
  {Bickerton}, {Furusawa}, {Ikeda}, {Koike}, {Lupton}, {Mineo}, {Price},
  {Takata}, {Tanaka}, {Yasuda}, {AlSayyad}, {Becker}, {Coulton}, {Coupon},
  {Garmilla}, {Huang}, {Krughoff}, {Lang}, {Leauthaud}, {Lim}, {Lust},
  {MacArthur}, {Mand elbaum}, {Miyatake}, {Miyazaki}, {Murata}, {More},
  {Okura}, {Owen}, {Swinbank}, {Strauss}, {Yamada}, \&
  {Yamanoi}}]{2018PASJ...70S...5B}
---. 2018{\natexlab{b}}, \pasj, 70, S5

\bibitem[{{Bosch} {et~al.}(2019){Bosch}, {AlSayyad}, {Armstrong}, {Bellm},
  {Chiang}, {Eggl}, {Findeisen}, {Fisher-Levine}, {Guy}, {Guyonnet},
  {Ivezi{\'c}}, {Jenness}, {Kov{\'a}cs}, {Krughoff}, {Lupton}, {Lust},
  {MacArthur}, {Meyers}, {Moolekamp}, {Morrison}, {Morton}, {O'Mullane},
  {Parejko}, {Plazas}, {Price}, {Rawls}, {Reed}, {Schellart}, {Slater},
  {Sullivan}, {Swinbank}, {Taranu}, {Waters}, \& {Wood-Vasey}}]{bosch19}
{Bosch}, J., {AlSayyad}, Y., {Armstrong}, R., {et~al.} 2019, in Astronomical
  Society of the Pacific Conference Series, Vol. 523, Astronomical Data
  Analysis Software and Systems XXVII, ed. P.~J. {Teuben}, M.~W. {Pound}, B.~A.
  {Thomas}, \& E.~M. {Warner}, 521

\bibitem[{{Bradshaw} {et~al.}(2013){Bradshaw}, {Almaini}, {Hartley}, {Smith},
  {Conselice}, {Dunlop}, {Simpson}, {Chuter}, {Cirasuolo}, {Foucaud}, {McLure},
  {Mortlock}, \& {Pearce}}]{bradshaw13}
{Bradshaw}, E.~J., {Almaini}, O., {Hartley}, W.~G., {et~al.} 2013, \mnras, 433,
  194

\bibitem[{{Burke} {et~al.}(2018){Burke}, {Rykoff}, {Allam}, {Annis}, {Bechtol},
  {Bernstein}, {Drlica-Wagner}, {Finley}, {Gruendl}, {James}, {Kent},
  {Kessler}, {Kuhlmann}, {Lasker}, {Li}, {Scolnic}, {Smith}, {Tucker},
  {Wester}, {Yanny}, {Abbott}, {Abdalla}, {Benoit-L{\'e}vy}, {Bertin}, {Carnero
  Rosell}, {Carrasco Kind}, {Carretero}, {Cunha}, {D'Andrea}, {da Costa},
  {Desai}, {Diehl}, {Doel}, {Estrada}, {Garc{\'\i}a-Bellido}, {Gruen},
  {Gutierrez}, {Honscheid}, {Kuehn}, {Kuropatkin}, {Maia}, {March}, {Marshall},
  {Melchior}, {Menanteau}, {Miquel}, {Plazas}, {Sako}, {Sanchez}, {Scarpine},
  {Schindler}, {Sevilla-Noarbe}, {Smith}, {Smith}, {Soares-Santos}, {Sobreira},
  {Suchyta}, {Tarle}, {Walker}, \& {DES Collaboration}}]{burke18}
{Burke}, D.~L., {Rykoff}, E.~S., {Allam}, S., {et~al.} 2018, \aj, 155, 41

\bibitem[{{Chambers} {et~al.}(2016){Chambers}, {Magnier}, {Metcalfe},
  {Flewelling}, {Huber}, {Waters}, {Denneau}, {Draper}, {Farrow}, {Finkbeiner},
  {Holmberg}, {Koppenhoefer}, {Price}, {Rest}, {Saglia}, {Schlafly}, {Smartt},
  {Sweeney}, {Wainscoat}, {Burgett}, {Chastel}, {Grav}, {Heasley}, {Hodapp},
  {Jedicke}, {Kaiser}, {Kudritzki}, {Luppino}, {Lupton}, {Monet}, {Morgan},
  {Onaka}, {Shiao}, {Stubbs}, {Tonry}, {White}, {Ba{\~n}ados}, {Bell},
  {Bender}, {Bernard}, {Boegner}, {Boffi}, {Botticella}, {Calamida},
  {Casertano}, {Chen}, {Chen}, {Cole}, {Deacon}, {Frenk}, {Fitzsimmons},
  {Gezari}, {Gibbs}, {Goessl}, {Goggia}, {Gourgue}, {Goldman}, {Grant},
  {Grebel}, {Hambly}, {Hasinger}, {Heavens}, {Heckman}, {Henderson}, {Henning},
  {Holman}, {Hopp}, {Ip}, {Isani}, {Jackson}, {Keyes}, {Koekemoer}, {Kotak},
  {Le}, {Liska}, {Long}, {Lucey}, {Liu}, {Martin}, {Masci}, {McLean}, {Mindel},
  {Misra}, {Morganson}, {Murphy}, {Obaika}, {Narayan}, {Nieto-Santisteban},
  {Norberg}, {Peacock}, {Pier}, {Postman}, {Primak}, {Rae}, {Rai}, {Riess},
  {Riffeser}, {Rix}, {R{\"o}ser}, {Russel}, {Rutz}, {Schilbach}, {Schultz},
  {Scolnic}, {Strolger}, {Szalay}, {Seitz}, {Small}, {Smith}, {Soderblom},
  {Taylor}, {Thomson}, {Taylor}, {Thakar}, {Thiel}, {Thilker}, {Unger},
  {Urata}, {Valenti}, {Wagner}, {Walder}, {Walter}, {Watters}, {Werner},
  {Wood-Vasey}, \& {Wyse}}]{chambers16}
{Chambers}, K.~C., {Magnier}, E.~A., {Metcalfe}, N., {et~al.} 2016, arXiv
  e-prints, arXiv:1612.05560

\bibitem[{{Coil} {et~al.}(2011){Coil}, {Blanton}, {Burles}, {Cool},
  {Eisenstein}, {Moustakas}, {Wong}, {Zhu}, {Aird}, {Bernstein}, {Bolton}, \&
  {Hogg}}]{coil11}
{Coil}, A.~L., {Blanton}, M.~R., {Burles}, S.~M., {et~al.} 2011, \apj, 741, 8

\bibitem[{{Colless} {et~al.}(2003){Colless}, {Peterson}, {Jackson}, {Peacock},
  {Cole}, {Norberg}, {Baldry}, {Baugh}, {Bland-Hawthorn}, {Bridges}, {Cannon},
  {Collins}, {Couch}, {Cross}, {Dalton}, {De Propris}, {Driver}, {Efstathiou},
  {Ellis}, {Frenk}, {Glazebrook}, {Lahav}, {Lewis}, {Lumsden}, {Maddox},
  {Madgwick}, {Sutherland}, \& {Taylor}}]{colless03}
{Colless}, M., {Peterson}, B.~A., {Jackson}, C., {et~al.} 2003, arXiv e-prints,
  astro

\bibitem[{{Cool} {et~al.}(2013){Cool}, {Moustakas}, {Blanton}, {Burles},
  {Coil}, {Eisenstein}, {Wong}, {Zhu}, {Aird}, {Bernstein}, {Bolton}, {Hogg},
  \& {Mendez}}]{cool13}
{Cool}, R.~J., {Moustakas}, J., {Blanton}, M.~R., {et~al.} 2013, \apj, 767, 118

\bibitem[{{Cooper} {et~al.}(2011){Cooper}, {Aird}, {Coil}, {Davis}, {Faber},
  {Juneau}, {Lotz}, {Nandra}, {Newman}, {Willmer}, \& {Yan}}]{cooper11}
{Cooper}, M.~C., {Aird}, J.~A., {Coil}, A.~L., {et~al.} 2011, \apjs, 193, 14

\bibitem[{{Cooper} {et~al.}(2012){Cooper}, {Griffith}, {Newman}, {Coil},
  {Davis}, {Dutton}, {Faber}, {Guhathakurta}, {Koo}, {Lotz}, {Weiner},
  {Willmer}, \& {Yan}}]{cooper12}
{Cooper}, M.~C., {Griffith}, R.~L., {Newman}, J.~A., {et~al.} 2012, \mnras,
  419, 3018

\bibitem[{{Davis} {et~al.}(2003){Davis}, {Faber}, {Newman}, {Phillips},
  {Ellis}, {Steidel}, {Conselice}, {Coil}, {Finkbeiner}, {Koo}, {Guhathakurta},
  {Weiner}, {Schiavon}, {Willmer}, {Kaiser}, {Luppino}, {Wirth}, {Connolly},
  {Eisenhardt}, {Cooper}, \& {Gerke}}]{davis03}
{Davis}, M., {Faber}, S.~M., {Newman}, J., {et~al.} 2003, in \procspie, Vol.
  4834, Discoveries and Research Prospects from 6- to 10-Meter-Class Telescopes
  II, ed. P.~{Guhathakurta}, 161--172

\bibitem[{{Drinkwater} {et~al.}(2010){Drinkwater}, {Jurek}, {Blake}, {Woods},
  {Pimbblet}, {Glazebrook}, {Sharp}, {Pracy}, {Brough}, {Colless}, {Couch},
  {Croom}, {Davis}, {Forbes}, {Forster}, {Gilbank}, {Gladders}, {Jelliffe},
  {Jones}, {Li}, {Madore}, {Martin}, {Poole}, {Small}, {Wisnioski}, {Wyder}, \&
  {Yee}}]{drinkwater10}
{Drinkwater}, M.~J., {Jurek}, R.~J., {Blake}, C., {et~al.} 2010, \mnras, 401,
  1429

\bibitem[{{Gaia Collaboration} {et~al.}(2016{\natexlab{a}}){Gaia
  Collaboration}, {Brown}, {Vallenari}, {Prusti}, {de Bruijne}, {Mignard},
  {Drimmel}, {Babusiaux}, {Bailer-Jones}, {Bastian}, \& et~al.}]{gaia16a}
{Gaia Collaboration}, {Brown}, A.~G.~A., {Vallenari}, A., {et~al.}
  2016{\natexlab{a}}, \aap, 595, A2

\bibitem[{{Gaia Collaboration} {et~al.}(2016{\natexlab{b}}){Gaia
  Collaboration}, {Prusti}, {de Bruijne}, {Brown}, {Vallenari}, {Babusiaux},
  {Bailer-Jones}, {Bastian}, {Biermann}, {Evans}, \& et~al.}]{gaia16b}
{Gaia Collaboration}, {Prusti}, T., {de Bruijne}, J.~H.~J., {et~al.}
  2016{\natexlab{b}}, \aap, 595, A1

\bibitem[{{Gaia Collaboration} {et~al.}(2018){Gaia Collaboration}, {Brown},
  {Vallenari}, {Prusti}, {de Bruijne}, {Babusiaux}, {Bailer-Jones}, {Biermann},
  {Evans}, {Eyer}, \& et~al.}]{gaia18}
{Gaia Collaboration}, {Brown}, A.~G.~A., {Vallenari}, A., {et~al.} 2018, \aap,
  616, A1

\bibitem[{{Garilli} {et~al.}(2014){Garilli}, {Guzzo}, {Scodeggio},
  {Bolzonella}, {Abbas}, {Adami}, {Arnouts}, {Bel}, {Bottini}, {Branchini},
  {Cappi}, {Coupon}, {Cucciati}, {Davidzon}, {De Lucia}, {de la Torre},
  {Franzetti}, {Fritz}, {Fumana}, {Granett}, {Ilbert}, {Iovino}, {Krywult}, {Le
  Brun}, {Le F{\`e}vre}, {Maccagni}, {Ma{\l}ek}, {Marulli}, {McCracken},
  {Paioro}, {Polletta}, {Pollo}, {Schlagenhaufer}, {Tasca}, {Tojeiro},
  {Vergani}, {Zamorani}, {Zanichelli}, {Burden}, {Di Porto}, {Marchetti},
  {Marinoni}, {Mellier}, {Moscardini}, {Nichol}, {Peacock}, {Percival},
  {Phleps}, \& {Wolk}}]{garilli14}
{Garilli}, B., {Guzzo}, L., {Scodeggio}, M., {et~al.} 2014, \aap, 562, A23

\bibitem[{{Golob} {et~al.}(2021){Golob}, {Sawicki}, {Goulding}, \&
  {Coupon}}]{golob21}
{Golob}, A., {Sawicki}, M., {Goulding}, A.~D., \& {Coupon}, J. 2021, \mnras,
  503, 4136

\bibitem[{{Halevi} {et~al.}(2019){Halevi}, {Goulding}, {Greene}, {Coupon},
  {Golob}, {Gwyn}, {Johnson}, {Moutard}, {Sawicki}, {Suh}, \&
  {Toba}}]{halevi19}
{Halevi}, G., {Goulding}, A., {Greene}, J., {et~al.} 2019, \apjl, 885, L3

\bibitem[{{Harikane} {et~al.}(2021){Harikane}, {Ono}, {Ouchi}, {Liu},
  {Sawicki}, {Shibuya}, {Behroozi}, {He}, {Shimasaku}, {Arnouts}, {Coupon},
  {Fujimoto}, {Gwyn}, {Huang}, {Inoue}, {Kashikawa}, {Komiyama}, {Matsuoka}, \&
  {Willott}}]{harikane21}
{Harikane}, Y., {Ono}, Y., {Ouchi}, M., {et~al.} 2021, arXiv e-prints,
  arXiv:2108.01090

\bibitem[{{Hasinger} {et~al.}(2018){Hasinger}, {Capak}, {Salvato}, {Barger},
  {Cowie}, {Faisst}, {Hemmati}, {Kakazu}, {Kartaltepe}, {Masters}, {Mobasher},
  {Nayyeri}, {Sanders}, {Scoville}, {Suh}, {Steinhardt}, \&
  {Yang}}]{hasinger18}
{Hasinger}, G., {Capak}, P., {Salvato}, M., {et~al.} 2018, \apj, 858, 77

\bibitem[{{Hayashi} {et~al.}(2020){Hayashi}, {Shimakawa}, {Tanaka}, {Onodera},
  {Koyama}, {Inoue}, {Komiyama}, {Lee}, {Lin}, \& {Yabe}}]{hayashi20}
{Hayashi}, M., {Shimakawa}, R., {Tanaka}, M., {et~al.} 2020, \pasj, 72, 86

\bibitem[{{Huang} {et~al.}(2020){Huang}, {Urata}, {Huang}, {Lee}, {Tsai},
  {Shirasaki}, {Sawicki}, {Arnouts}, {Moutard}, {Gwyn}, {Wang}, {Foucaud},
  {Asada}, {Huber}, {Wainscoat}, \& {Chambers}}]{huang20}
{Huang}, Y.-J., {Urata}, Y., {Huang}, K., {et~al.} 2020, \apj, 897, 69

\bibitem[{{Inoue} {et~al.}(2020){Inoue}, {Yamanaka}, {Ouchi}, {Iwata},
  {Shimasaku}, {Taniguchi}, {Nagao}, {Kashikawa}, {Ono}, {Mawatari}, {Shibuya},
  {Hayashi}, {Ikeda}, {Zhang}, {Liang}, {Lee}, {Hilmi}, {Kikuta}, {Kusakabe},
  {Furusawa}, {Hayashino}, {Kajisawa}, {Matsuda}, {Nakajima}, {Momose},
  {Harikane}, {Saito}, {Kodama}, {Kikuchihara}, {Iye}, \& {Goto}}]{inoue20}
{Inoue}, A.~K., {Yamanaka}, S., {Ouchi}, M., {et~al.} 2020, \pasj, 72, 101

\bibitem[{{Ivezi{\'c}} {et~al.}(2019){Ivezi{\'c}}, {Kahn}, {Tyson}, {Abel},
  {Acosta}, {Allsman}, {Alonso}, {AlSayyad}, {Anderson}, {Andrew}, \&
  et~al.}]{ivezic19}
{Ivezi{\'c}}, {\v Z}., {Kahn}, S.~M., {Tyson}, J.~A., {et~al.} 2019, \apj, 873,
  111

\bibitem[{{Jarvis} {et~al.}(2013){Jarvis}, {Bonfield}, {Bruce}, {Geach},
  {McAlpine}, {McLure}, {Gonz{\'a}lez-Solares}, {Irwin}, {Lewis}, {Yoldas},
  {Andreon}, {Cross}, {Emerson}, {Dalton}, {Dunlop}, {Hodgkin}, {Le},
  {Karouzos}, {Meisenheimer}, {Oliver}, {Rawlings}, {Simpson}, {Smail},
  {Smith}, {Sullivan}, {Sutherland}, {White}, \& {Zwart}}]{jarvis13}
{Jarvis}, M.~J., {Bonfield}, D.~G., {Bruce}, V.~A., {et~al.} 2013, \mnras, 428,
  1281

\bibitem[{{Jones} {et~al.}(2004){Jones}, {Saunders}, {Colless}, {Read},
  {Parker}, {Watson}, {Campbell}, {Burkey}, {Mauch}, {Moore}, {Hartley},
  {Cass}, {James}, {Russell}, {Fiegert}, {Dawe}, {Huchra}, {Jarrett}, {Lahav},
  {Lucey}, {Mamon}, {Proust}, {Sadler}, \& {Wakamatsu}}]{jones04}
{Jones}, D.~H., {Saunders}, W., {Colless}, M., {et~al.} 2004, \mnras, 355, 747

\bibitem[{{Jones} {et~al.}(2009){Jones}, {Read}, {Saunders}, {Colless},
  {Jarrett}, {Parker}, {Fairall}, {Mauch}, {Sadler}, {Watson}, {Burton},
  {Campbell}, {Cass}, {Croom}, {Dawe}, {Fiegert}, {Frankcombe}, {Hartley},
  {Huchra}, {James}, {Kirby}, {Lahav}, {Lucey}, {Mamon}, {Moore}, {Peterson},
  {Prior}, {Proust}, {Russell}, {Safouris}, {Wakamatsu}, {Westra}, \&
  {Williams}}]{jones09}
{Jones}, D.~H., {Read}, M.~A., {Saunders}, W., {et~al.} 2009, \mnras, 399, 683

\bibitem[{{Juri{\'c}} {et~al.}(2017){Juri{\'c}}, {Kantor}, {Lim}, {Lupton},
  {Dubois-Felsmann}, {Jenness}, {Axelrod}, {Aleksi{\'c}}, {Allsman},
  {AlSayyad}, {Alt}, {Armstrong}, {Basney}, {Becker}, {Becla}, {Biswas},
  {Bosch}, {Boutigny}, {Kind}, {Ciardi}, {Connolly}, {Daniel}, {Daues},
  {Economou}, {Chiang}, {Fausti}, {Fisher-Levine}, {Freemon}, {Gris},
  {Hernandez}, {Hoblitt}, {Ivezi{\'c}}, {Jammes}, {Jevremovi{\'c}}, {Jones},
  {Kalmbach}, {Kasliwal}, {Krughoff}, {Lurie}, {Lust}, {MacArthur}, {Melchior},
  {Moeyens}, {Nidever}, {Owen}, {Parejko}, {Peterson}, {Petravick},
  {Pietrowicz}, {Price}, {Reiss}, {Shaw}, {Sick}, {Slater}, {Strauss},
  {Sullivan}, {Swinbank}, {Van Dyk}, {Vuj{\v c}i{\'c}}, {Withers}, \&
  {Yoachim}}]{juric17}
{Juri{\'c}}, M., {Kantor}, J., {Lim}, K.-T., {et~al.} 2017, in Astronomical
  Society of the Pacific Conference Series, Vol. 512, Astronomical Data
  Analysis Software and Systems XXV, ed. N.~P.~F. {Lorente}, K.~{Shortridge},
  \& R.~{Wayth}, 279

\bibitem[{{Kamata} {et~al.}(2014){Kamata}, {Nakaya}, {Kawanomoto}, \&
  {Miyazaki}}]{kamata14}
{Kamata}, Y., {Nakaya}, H., {Kawanomoto}, S., \& {Miyazaki}, S. 2014, in
  Society of Photo-Optical Instrumentation Engineers (SPIE) Conference Series,
  Vol. 9154, High Energy, Optical, and Infrared Detectors for Astronomy VI, ed.
  A.~D. {Holland} \& J.~{Beletic}, 91541Z

\bibitem[{{Kashino} {et~al.}(2019){Kashino}, {Silverman}, {Sanders},
  {Kartaltepe}, {Daddi}, {Renzini}, {Rodighiero}, {Puglisi}, {Valentino},
  {Juneau}, {Arimoto}, {Nagao}, {Ilbert}, {Le F{\`e}vre}, \&
  {Koekemoer}}]{kashino19}
{Kashino}, D., {Silverman}, J.~D., {Sanders}, D., {et~al.} 2019, \apjs, 241, 10

\bibitem[{{Kawanomoto} {et~al.}(2018){Kawanomoto}, {Uraguchi}, {Komiyama},
  {Miyazaki}, {Furusawa}, {Finet}, {Hattori}, {Wang}, {Yasuda}, \&
  {Suzuki}}]{kawanomoto18}
{Kawanomoto}, S., {Uraguchi}, F., {Komiyama}, Y., {et~al.} 2018, \pasj, 70, 66

\bibitem[{{Kim} {et~al.}(2011){Kim}, {Edge}, {Wake}, \& {Stott}}]{kim11}
{Kim}, J.~W., {Edge}, A.~C., {Wake}, D.~A., \& {Stott}, J.~P. 2011, \mnras,
  410, 241

\bibitem[{{Lacy} {et~al.}(2021){Lacy}, {Surace}, {Farrah}, {Nyland}, {Afonso},
  {Brandt}, {Clements}, {Lagos}, {Maraston}, {Pforr}, {Sajina}, {Sako},
  {Vaccari}, {Wilson}, {Ballantyne}, {Barkhouse}, {Brunner}, {Cane}, {Clarke},
  {Cooper}, {Cooray}, {Covone}, {D'Andrea}, {Evrard}, {Ferguson}, {Frieman},
  {Gonzalez-Perez}, {Gupta}, {Hatziminaoglou}, {Huang}, {Jagannathan},
  {Jarvis}, {Jones}, {Kimball}, {Lidman}, {Lubin}, {Marchetti}, {Martini},
  {McMahon}, {Mei}, {Messias}, {Murphy}, {Newman}, {Nichol}, {Norris},
  {Oliver}, {Perez-Fournon}, {Peters}, {Pierre}, {Polisensky}, {Richards},
  {Ridgway}, {R{\"o}ttgering}, {Seymour}, {Shirley}, {Somerville}, {Strauss},
  {Suntzeff}, {Thorman}, {van Kampen}, {Verma}, {Wechsler}, \&
  {Wood-Vasey}}]{lacy21}
{Lacy}, M., {Surace}, J.~A., {Farrah}, D., {et~al.} 2021, \mnras, 501, 892

\bibitem[{{Lawrence} {et~al.}(2007){Lawrence}, {Warren}, {Almaini}, {Edge},
  {Hambly}, {Jameson}, {Lucas}, {Casali}, {Adamson}, {Dye}, {Emerson},
  {Foucaud}, {Hewett}, {Hirst}, {Hodgkin}, {Irwin}, {Lodieu}, {McMahon},
  {Simpson}, {Smail}, {Mortlock}, \& {Folger}}]{lawrence07}
{Lawrence}, A., {Warren}, S.~J., {Almaini}, O., {et~al.} 2007, \mnras, 379,
  1599

\bibitem[{{Le F{\`e}vre} {et~al.}(2013){Le F{\`e}vre}, {Cassata}, {Cucciati},
  {Garilli}, {Ilbert}, {Le Brun}, {Maccagni}, {Moreau}, {Scodeggio}, {Tresse},
  {Zamorani}, {Adami}, {Arnouts}, {Bardelli}, {Bolzonella}, {Bondi},
  {Bongiorno}, {Bottini}, {Cappi}, {Charlot}, {Ciliegi}, {Contini}, {de la
  Torre}, {Foucaud}, {Franzetti}, {Gavignaud}, {Guzzo}, {Iovino}, {Lemaux},
  {L{\'o}pez-Sanjuan}, {McCracken}, {Marano}, {Marinoni}, {Mazure}, {Mellier},
  {Merighi}, {Merluzzi}, {Paltani}, {Pell{\`o}}, {Pollo}, {Pozzetti},
  {Scaramella}, {Tasca}, {Vergani}, {Vettolani}, {Zanichelli}, \&
  {Zucca}}]{lefevre13}
{Le F{\`e}vre}, O., {Cassata}, P., {Cucciati}, O., {et~al.} 2013, \aap, 559,
  A14

\bibitem[{{L{\'e}get} {et~al.}(2021){L{\'e}get}, {Astier}, {Regnault},
  {Jarvis}, {Antilogus}, {Roodman}, {Rubin}, \& {Saunders}}]{leget21}
{L{\'e}get}, P.~F., {Astier}, P., {Regnault}, N., {et~al.} 2021, \aap, 650, A81

\bibitem[{{Li} {et~al.}(2021){Li}, {Miyatake}, {Luo}, {More}, {Oguri},
  {Hamana}, {Mandelbaum}, {Shirasaki}, {Takada}, {Armstrong}, {Kannawadi},
  {Takita}, {Miyazaki}, {Nishizawa}, {Plazas Malag{\'o}n}, {Strauss}, {Tanaka},
  \& {Yoshida}}]{li21}
{Li}, X., {Miyatake}, H., {Luo}, W., {et~al.} 2021, arXiv e-prints,
  arXiv:2107.00136

\bibitem[{{Lilly} {et~al.}(2009){Lilly}, {Le Brun}, {Maier}, {Mainieri},
  {Mignoli}, {Scodeggio}, {Zamorani}, {Carollo}, {Contini}, {Kneib}, {Le
  F{\`e}vre}, {Renzini}, {Bardelli}, {Bolzonella}, {Bongiorno}, {Caputi},
  {Coppa}, {Cucciati}, {de la Torre}, {de Ravel}, {Franzetti}, {Garilli},
  {Iovino}, {Kampczyk}, {Kovac}, {Knobel}, {Lamareille}, {Le Borgne}, {Pello},
  {Peng}, {P{\'e}rez-Montero}, {Ricciardelli}, {Silverman}, {Tanaka}, {Tasca},
  {Tresse}, {Vergani}, {Zucca}, {Ilbert}, {Salvato}, {Oesch}, {Abbas},
  {Bottini}, {Capak}, {Cappi}, {Cassata}, {Cimatti}, {Elvis}, {Fumana},
  {Guzzo}, {Hasinger}, {Koekemoer}, {Leauthaud}, {Maccagni}, {Marinoni},
  {McCracken}, {Memeo}, {Meneux}, {Porciani}, {Pozzetti}, {Sanders},
  {Scaramella}, {Scarlata}, {Scoville}, {Shopbell}, \& {Taniguchi}}]{lilly09}
{Lilly}, S.~J., {Le Brun}, V., {Maier}, C., {et~al.} 2009, \apjs, 184, 218

\bibitem[{{Liske} {et~al.}(2015){Liske}, {Baldry}, {Driver}, {Tuffs},
  {Alpaslan}, {Andrae}, {Brough}, {Cluver}, {Grootes}, {Gunawardhana},
  {Kelvin}, {Loveday}, {Robotham}, {Taylor}, {Bamford}, {Bland-Hawthorn},
  {Brown}, {Drinkwater}, {Hopkins}, {Meyer}, {Norberg}, {Peacock}, {Agius},
  {Andrews}, {Bauer}, {Ching}, {Colless}, {Conselice}, {Croom}, {Davies}, {De
  Propris}, {Dunne}, {Eardley}, {Ellis}, {Foster}, {Frenk}, {H{\"a}u{\ss}ler},
  {Holwerda}, {Howlett}, {Ibarra}, {Jarvis}, {Jones}, {Kafle}, {Lacey},
  {Lange}, {Lara-L{\'o}pez}, {L{\'o}pez-S{\'a}nchez}, {Maddox}, {Madore},
  {McNaught-Roberts}, {Moffett}, {Nichol}, {Owers}, {Palamara}, {Penny},
  {Phillipps}, {Pimbblet}, {Popescu}, {Prescott}, {Proctor}, {Sadler},
  {Sansom}, {Seibert}, {Sharp}, {Sutherland}, {V{\'a}zquez-Mata}, {van Kampen},
  {Wilkins}, {Williams}, \& {Wright}}]{liske15}
{Liske}, J., {Baldry}, I.~K., {Driver}, S.~P., {et~al.} 2015, \mnras, 452, 2087

\bibitem[{{Magnier} {et~al.}(2013){Magnier}, {Schlafly}, {Finkbeiner}, {Juric},
  {Tonry}, {Burgett}, {Chambers}, {Flewelling}, {Kaiser}, {Kudritzki},
  {Morgan}, {Price}, {Sweeney}, \& {Stubbs}}]{magnier13}
{Magnier}, E.~A., {Schlafly}, E., {Finkbeiner}, D., {et~al.} 2013, \apjs, 205,
  20

\bibitem[{{Mandelbaum} {et~al.}(2018){Mandelbaum}, {Miyatake}, {Hamana},
  {Oguri}, {Simet}, {Armstrong}, {Bosch}, {Murata}, {Lanusse}, {Leauthaud},
  {Coupon}, {More}, {Takada}, {Miyazaki}, {Speagle}, {Shirasaki}, {Sif{\'o}n},
  {Huang}, {Nishizawa}, {Medezinski}, {Okura}, {Okabe}, {Czakon}, {Takahashi},
  {Coulton}, {Hikage}, {Komiyama}, {Lupton}, {Strauss}, {Tanaka}, \&
  {Utsumi}}]{mandelbaum18}
{Mandelbaum}, R., {Miyatake}, H., {Hamana}, T., {et~al.} 2018, \pasj, 70, S25

\bibitem[{{Masters} {et~al.}(2017){Masters}, {Stern}, {Cohen}, {Capak},
  {Rhodes}, {Castander}, \& {Paltani}}]{masters17}
{Masters}, D.~C., {Stern}, D.~K., {Cohen}, J.~G., {et~al.} 2017, \apj, 841, 111

\bibitem[{{Masters} {et~al.}(2019){Masters}, {Stern}, {Cohen}, {Capak},
  {Stanford}, {Hernitschek}, {Galametz}, {Davidzon}, {Rhodes}, {Sanders},
  {Mobasher}, {Castander}, {Pruett}, \& {Fotopoulou}}]{masters19}
---. 2019, \apj, 877, 81

\bibitem[{{Mauduit} {et~al.}(2012){Mauduit}, {Lacy}, {Farrah}, {Surace},
  {Jarvis}, {Oliver}, {Maraston}, {Vaccari}, {Marchetti}, {Zeimann},
  {Gonz{\'a}les-Solares}, {Pforr}, {Petric}, {Henriques}, {Thomas}, {Afonso},
  {Rettura}, {Wilson}, {Falder}, {Geach}, {Huynh}, {Norris}, {Seymour},
  {Richards}, {Stanford}, {Alexander}, {Becker}, {Best}, {Bizzocchi},
  {Bonfield}, {Castro}, {Cava}, {Chapman}, {Christopher}, {Clements}, {Covone},
  {Dubois}, {Dunlop}, {Dyke}, {Edge}, {Ferguson}, {Foucaud}, {Franceschini},
  {Gal}, {Grant}, {Grossi}, {Hatziminaoglou}, {Hickey}, {Hodge}, {Huang},
  {Ivison}, {Kim}, {LeFevre}, {Lehnert}, {Lonsdale}, {Lubin}, {McLure},
  {Messias}, {Mart{\'\i}nez-Sansigre}, {Mortier}, {Nielsen}, {Ouchi}, {Parish},
  {Perez-Fournon}, {Pierre}, {Rawlings}, {Readhead}, {Ridgway}, {Rigopoulou},
  {Romer}, {Rosebloom}, {Rottgering}, {Rowan-Robinson}, {Sajina}, {Simpson},
  {Smail}, {Squires}, {Stevens}, {Taylor}, {Trichas}, {Urrutia}, {van Kampen},
  {Verma}, \& {Xu}}]{mauduit12}
{Mauduit}, J.~C., {Lacy}, M., {Farrah}, D., {et~al.} 2012, \pasp, 124, 714

\bibitem[{{McCracken} {et~al.}(2012){McCracken}, {Milvang-Jensen}, {Dunlop},
  {Franx}, {Fynbo}, {Le F{\`e}vre}, {Holt}, {Caputi}, {Goranova}, {Buitrago},
  {Emerson}, {Freudling}, {Hudelot}, {L{\'o}pez-Sanjuan}, {Magnard}, {Mellier},
  {M{\o}ller}, {Nilsson}, {Sutherland}, {Tasca}, \& {Zabl}}]{mccracken12}
{McCracken}, H.~J., {Milvang-Jensen}, B., {Dunlop}, J., {et~al.} 2012, \aap,
  544, A156

\bibitem[{{McLure} {et~al.}(2013){McLure}, {Pearce}, {Dunlop}, {Cirasuolo},
  {Curtis-Lake}, {Bruce}, {Caputi}, {Almaini}, {Bonfield}, {Bradshaw},
  {Buitrago}, {Chuter}, {Foucaud}, {Hartley}, \& {Jarvis}}]{mclure13}
{McLure}, R.~J., {Pearce}, H.~J., {Dunlop}, J.~S., {et~al.} 2013, \mnras, 428,
  1088

\bibitem[{{Mehta} {et~al.}(2018){Mehta}, {Scarlata}, {Capak}, {Davidzon},
  {Faisst}, {Hsieh}, {Ilbert}, {Jarvis}, {Laigle}, {Phillips}, {Silverman},
  {Strauss}, {Tanaka}, {Bowler}, {Coupon}, {Foucaud}, {Hemmati}, {Masters},
  {McCracken}, {Mobasher}, {Ouchi}, {Shibuya}, \& {Wang}}]{mehta18}
{Mehta}, V., {Scarlata}, C., {Capak}, P., {et~al.} 2018, \apjs, 235, 36

\bibitem[{{Melchior} {et~al.}(2018){Melchior}, {Moolekamp}, {Jerdee},
  {Armstrong}, {Sun}, {Bosch}, \& {Lupton}}]{melchior18}
{Melchior}, P., {Moolekamp}, F., {Jerdee}, M., {et~al.} 2018, Astronomy and
  Computing, 24, 129

\bibitem[{{Miyazaki} {et~al.}(2018){Miyazaki}, {Komiyama}, {Kawanomoto}, {Doi},
  {Furusawa}, {Hamana}, {Hayashi}, {Ikeda}, {Kamata}, {Karoji}, {Koike},
  {Kurakami}, {Miyama}, {Morokuma}, {Nakata}, {Namikawa}, {Nakaya}, {Nariai},
  {Obuchi}, {Oishi}, {Okada}, {Okura}, {Tait}, {Takata}, {Tanaka}, {Tanaka},
  {Terai}, {Tomono}, {Uraguchi}, {Usuda}, {Utsumi}, {Yamada}, {Yamanoi},
  {Aihara}, {Fujimori}, {Mineo}, {Miyatake}, {Oguri}, {Uchida}, {Tanaka},
  {Yasuda}, {Takada}, {Murayama}, {Nishizawa}, {Sugiyama}, {Chiba}, {Futamase},
  {Wang}, {Chen}, {Ho}, {Liaw}, {Chiu}, {Ho}, {Lai}, {Lee}, {Jeng}, {Iwamura},
  {Armstrong}, {Bickerton}, {Bosch}, {Gunn}, {Lupton}, {Loomis}, {Price},
  {Smith}, {Strauss}, {Turner}, {Suzuki}, {Miyazaki}, {Muramatsu}, {Yamamoto},
  {Endo}, {Ezaki}, {Ito}, {Kawaguchi}, {Sofuku}, {Taniike}, {Akutsu}, {Dojo},
  {Kasumi}, {Matsuda}, {Imoto}, {Miwa}, {Suzuki}, {Takeshi}, \&
  {Yokota}}]{miyazaki18}
{Miyazaki}, S., {Komiyama}, Y., {Kawanomoto}, S., {et~al.} 2018, \pasj, 70, S1

\bibitem[{{Momcheva} {et~al.}(2016){Momcheva}, {Brammer}, {van Dokkum},
  {Skelton}, {Whitaker}, {Nelson}, {Fumagalli}, {Maseda}, {Leja}, {Franx},
  {Rix}, {Bezanson}, {Da Cunha}, {Dickey}, {F{\"o}rster Schreiber},
  {Illingworth}, {Kriek}, {Labb{\'e}}, {Ulf Lange}, {Lundgren}, {Magee},
  {Marchesini}, {Oesch}, {Pacifici}, {Patel}, {Price}, {Tal}, {Wake}, {van der
  Wel}, \& {Wuyts}}]{momcheva16}
{Momcheva}, I.~G., {Brammer}, G.~B., {van Dokkum}, P.~G., {et~al.} 2016, \apjs,
  225, 27

\bibitem[{{Moutard} {et~al.}(2020){Moutard}, {Sawicki}, {Arnouts}, {Golob},
  {Coupon}, {Ilbert}, {Yang}, \& {Gwyn}}]{moutard20}
{Moutard}, T., {Sawicki}, M., {Arnouts}, S., {et~al.} 2020, \mnras, 494, 1894

\bibitem[{{Newman} {et~al.}(2013){Newman}, {Cooper}, {Davis}, {Faber}, {Coil},
  {Guhathakurta}, {Koo}, {Phillips}, {Conroy}, {Dutton}, {Finkbeiner}, {Gerke},
  {Rosario}, {Weiner}, {Willmer}, {Yan}, {Harker}, {Kassin}, {Konidaris},
  {Lai}, {Madgwick}, {Noeske}, {Wirth}, {Connolly}, {Kaiser}, {Kirby},
  {Lemaux}, {Lin}, {Lotz}, {Luppino}, {Marinoni}, {Matthews}, {Metevier}, \&
  {Schiavon}}]{newman13}
{Newman}, J.~A., {Cooper}, M.~C., {Davis}, M., {et~al.} 2013, \apjs, 208, 5

\bibitem[{{Nishizawa} {et~al.}(2020){Nishizawa}, {Hsieh}, {Tanaka}, \&
  {Takata}}]{nishizawa20}
{Nishizawa}, A.~J., {Hsieh}, B.-C., {Tanaka}, M., \& {Takata}, T. 2020, arXiv
  e-prints, arXiv:2003.01511

\bibitem[{{Oke} \& {Gunn}(1983)}]{oke83}
{Oke}, J.~B., \& {Gunn}, J.~E. 1983, \apj, 266, 713

\bibitem[{{P{\^a}ris} {et~al.}(2018){P{\^a}ris}, {Petitjean}, {Aubourg},
  {Myers}, {Streblyanska}, {Lyke}, {Anderson}, {Armengaud}, {Bautista},
  {Blanton}, {Blomqvist}, {Brinkmann}, {Brownstein}, {Brandt}, {Burtin},
  {Dawson}, {de la Torre}, {Georgakakis}, {Gil-Mar{\'{\i}}n}, {Green}, {Hall},
  {Kneib}, {LaMassa}, {Le Goff}, {MacLeod}, {Mariappan}, {McGreer}, {Merloni},
  {Noterdaeme}, {Palanque-Delabrouille}, {Percival}, {Ross}, {Rossi},
  {Schneider}, {Seo}, {Tojeiro}, {Weaver}, {Weijmans}, {Y{\`e}che}, {Zarrouk},
  \& {Zhao}}]{paris18}
{P{\^a}ris}, I., {Petitjean}, P., {Aubourg}, {\'E}., {et~al.} 2018, \aap, 613,
  A51

\bibitem[{{Pentericci} {et~al.}(2018){Pentericci}, {McLure}, {Garilli},
  {Cucciati}, {Franzetti}, {Iovino}, {Amorin}, {Bolzonella}, {Bongiorno},
  {Carnall}, {Castellano}, {Cimatti}, {Cirasuolo}, {Cullen}, {De Barros},
  {Dunlop}, {Elbaz}, {Finkelstein}, {Fontana}, {Fontanot}, {Fumana},
  {Gargiulo}, {Guaita}, {Hartley}, {Jarvis}, {Juneau}, {Karman}, {Maccagni},
  {Marchi}, {Marmol-Queralto}, {Nandra}, {Pompei}, {Pozzetti}, {Scodeggio},
  {Sommariva}, {Talia}, {Almaini}, {Balestra}, {Bardelli}, {Bell}, {Bourne},
  {Bowler}, {Brusa}, {Buitrago}, {Caputi}, {Cassata}, {Charlot}, {Citro},
  {Cresci}, {Cristiani}, {Curtis-Lake}, {Dickinson}, {Fazio}, {Ferguson},
  {Fiore}, {Franco}, {Fynbo}, {Galametz}, {Georgakakis}, {Giavalisco},
  {Grazian}, {Hathi}, {Jung}, {Kim}, {Koekemoer}, {Khusanova}, {Le F{\`e}vre},
  {Lotz}, {Mannucci}, {Maltby}, {Matsuoka}, {McLeod}, {Mendez-Hernandez},
  {Mendez-Abreu}, {Mignoli}, {Moresco}, {Mortlock}, {Nonino}, {Pannella},
  {Papovich}, {Popesso}, {Rosario}, {Salvato}, {Santini}, {Schaerer},
  {Schreiber}, {Stark}, {Tasca}, {Thomas}, {Treu}, {Vanzella}, {Wild},
  {Williams}, {Zamorani}, \& {Zucca}}]{pentericci18}
{Pentericci}, L., {McLure}, R.~J., {Garilli}, B., {et~al.} 2018, \aap, 616,
  A174

\bibitem[{{Pickles}(1998)}]{pickles98}
{Pickles}, A.~J. 1998, \pasp, 110, 863

\bibitem[{{Sanders} {et~al.}(2007){Sanders}, {Salvato}, {Aussel}, {Ilbert},
  {Scoville}, {Surace}, {Frayer}, {Sheth}, {Helou}, {Brooke}, {Bhattacharya},
  {Yan}, {Kartaltepe}, {Barnes}, {Blain}, {Calzetti}, {Capak}, {Carilli},
  {Carollo}, {Comastri}, {Daddi}, {Ellis}, {Elvis}, {Fall}, {Franceschini},
  {Giavalisco}, {Hasinger}, {Impey}, {Koekemoer}, {Le F{\`e}vre}, {Lilly},
  {Liu}, {McCracken}, {Mobasher}, {Renzini}, {Rich}, {Schinnerer}, {Shopbell},
  {Taniguchi}, {Thompson}, {Urry}, \& {Williams}}]{sanders07}
{Sanders}, D.~B., {Salvato}, M., {Aussel}, H., {et~al.} 2007, \apjs, 172, 86

\bibitem[{{Sawicki} {et~al.}(2019){Sawicki}, {Arnouts}, {Huang}, {Coupon},
  {Golob}, {Gwyn}, {Foucaud}, {Moutard}, {Iwata}, {Liu}, {Chen}, {Desprez},
  {Harikane}, {Ono}, {Strauss}, {Tanaka}, {Thibert}, {Balogh}, {Bundy},
  {Chapman}, {Gunn}, {Hsieh}, {Ilbert}, {Jing}, {LeF{\`e}vre}, {Li}, {Matsuda},
  {Miyazaki}, {Nagao}, {Nishizawa}, {Ouchi}, {Shimasaku}, {Silverman}, {de la
  Torre}, {Tresse}, {Wang}, {Willott}, {Yamada}, {Yang}, \& {Yee}}]{sawicki19}
{Sawicki}, M., {Arnouts}, S., {Huang}, J., {et~al.} 2019, \mnras, 489, 5202

\bibitem[{{Schlafly} {et~al.}(2012){Schlafly}, {Finkbeiner}, {Juri{\'c}},
  {Magnier}, {Burgett}, {Chambers}, {Grav}, {Hodapp}, {Kaiser}, {Kudritzki},
  {Martin}, {Morgan}, {Price}, {Rix}, {Stubbs}, {Tonry}, \&
  {Wainscoat}}]{schlafly12}
{Schlafly}, E.~F., {Finkbeiner}, D.~P., {Juri{\'c}}, M., {et~al.} 2012, \apj,
  756, 158

\bibitem[{{Schlegel} {et~al.}(1998){Schlegel}, {Finkbeiner}, \&
  {Davis}}]{schlegel98}
{Schlegel}, D.~J., {Finkbeiner}, D.~P., \& {Davis}, M. 1998, \apj, 500, 525

\bibitem[{{Sevilla-Noarbe} {et~al.}(2021){Sevilla-Noarbe}, {Bechtol}, {Carrasco
  Kind}, {Carnero Rosell}, {Becker}, {Drlica-Wagner}, {Gruendl}, {Rykoff},
  {Sheldon}, {Yanny}, {Alarcon}, {Allam}, {Amon}, {Benoit-L{\'e}vy},
  {Bernstein}, {Bertin}, {Burke}, {Carretero}, {Choi}, {Diehl}, {Everett},
  {Flaugher}, {Gaztanaga}, {Gschwend}, {Harrison}, {Hartley}, {Hoyle},
  {Jarvis}, {Johnson}, {Kessler}, {Kron}, {Kuropatkin}, {Leistedt}, {Li},
  {Menanteau}, {Morganson}, {Ogando}, {Palmese}, {Paz-Chinch{\'o}n}, {Pieres},
  {Pond}, {Rodriguez-Monroy}, {Smith}, {Stringer}, {Troxel}, {Tucker}, {de
  Vicente}, {Wester}, {Zhang}, {Abbott}, {Aguena}, {Annis}, {Avila},
  {Bhargava}, {Bridle}, {Brooks}, {Brout}, {Castander}, {Cawthon}, {Chang},
  {Conselice}, {Costanzi}, {Crocce}, {da Costa}, {Pereira}, {Davis}, {Desai},
  {Dietrich}, {Doel}, {Eckert}, {Evrard}, {Ferrero}, {Fosalba},
  {Garc{\'\i}a-Bellido}, {Gerdes}, {Giannantonio}, {Gruen}, {Gutierrez},
  {Hinton}, {Hollowood}, {Honscheid}, {Huff}, {Huterer}, {James}, {Jeltema},
  {Kuehn}, {Lahav}, {Lidman}, {Lima}, {Lin}, {Maia}, {Marshall}, {Martini},
  {Melchior}, {Miquel}, {Mohr}, {Morgan}, {Neilsen}, {Plazas}, {Romer},
  {Roodman}, {Sanchez}, {Scarpine}, {Schubnell}, {Serrano}, {Smith}, {Suchyta},
  {Tarle}, {Thomas}, {To}, {Varga}, {Wechsler}, {Weller}, {Wilkinson}, \& {DES
  Collaboration}}]{sevilla21}
{Sevilla-Noarbe}, I., {Bechtol}, K., {Carrasco Kind}, M., {et~al.} 2021, \apjs,
  254, 24

\bibitem[{{Shimakawa} {et~al.}(2021){Shimakawa}, {Higuchi}, {Shirasaki},
  {Tanaka}, {Lin}, {Hayashi}, {Momose}, {Lee}, {Kusakabe}, {Kodama}, \&
  {Yamamoto}}]{shimakawa21}
{Shimakawa}, R., {Higuchi}, Y., {Shirasaki}, M., {et~al.} 2021, \mnras, 503,
  3896

\bibitem[{{Silverman} {et~al.}(2015){Silverman}, {Kashino}, {Sanders},
  {Kartaltepe}, {Arimoto}, {Renzini}, {Rodighiero}, {Daddi}, {Zahid}, {Nagao},
  {Kewley}, {Lilly}, {Sugiyama}, {Baronchelli}, {Capak}, {Carollo}, {Chu},
  {Hasinger}, {Ilbert}, {Juneau}, {Kajisawa}, {Koekemoer}, {Kovac}, {Le
  F{\`e}vre}, {Masters}, {McCracken}, {Onodera}, {Schulze}, {Scoville},
  {Strazzullo}, \& {Taniguchi}}]{silverman15}
{Silverman}, J.~D., {Kashino}, D., {Sanders}, D., {et~al.} 2015, \apjs, 220, 12

\bibitem[{{Skelton} {et~al.}(2014){Skelton}, {Whitaker}, {Momcheva}, {Brammer},
  {van Dokkum}, {Labb{\'e}}, {Franx}, {van der Wel}, {Bezanson}, {Da Cunha},
  {Fumagalli}, {F{\"o}rster Schreiber}, {Kriek}, {Leja}, {Lundgren}, {Magee},
  {Marchesini}, {Maseda}, {Nelson}, {Oesch}, {Pacifici}, {Patel}, {Price},
  {Rix}, {Tal}, {Wake}, \& {Wuyts}}]{skelton14}
{Skelton}, R.~E., {Whitaker}, K.~E., {Momcheva}, I.~G., {et~al.} 2014, \apjs,
  214, 24

\bibitem[{{Straatman} {et~al.}(2018){Straatman}, {van der Wel}, {Bezanson},
  {Pacifici}, {Gallazzi}, {Wu}, {Noeske}, {Bari{\v{s}}i{\'c}}, {Bell},
  {Brammer}, {Calhau}, {Chauke}, {Franx}, {van Houdt}, {Labb{\'e}}, {Maseda},
  {Mu{\~n}oz-Mateos}, {Muzzin}, {van de Sande}, {Sobral}, \&
  {Spilker}}]{straatman18}
{Straatman}, C. M.~S., {van der Wel}, A., {Bezanson}, R., {et~al.} 2018, \apjs,
  239, 27

\bibitem[{{Tamura} {et~al.}(2018){Tamura}, {Takato}, {Shimono}, {Moritani},
  {Yabe}, {Ishizuka}, {Kamata}, {Ueda}, {Aghazarian}, {Arnouts}, {Barkhouser},
  {Balard}, {Barette}, {Belhadi}, {Burnham}, {Caplar}, {Carr}, {Chabaud},
  {Chang}, {Chen}, {Chou}, {Chu}, {Cohen}, {de Almeida}, {de Oliveira}, {de
  Oliveira}, {Dekany}, {Dohlen}, {dos Santos}, {dos Santos}, {Ellis},
  {Fabricius}, {Ferreira}, {Furusawa}, {Garcia-Carpio}, {Golebiowski}, {Gross},
  {Gunn}, {Hammond}, {Harding}, {Hart}, {Heckman}, {Ho}, {Hope}, {Hover},
  {Hsu}, {Hu}, {Huang}, {Jamal}, {Jaquet}, {Jeschke}, {Jing}, {Kado-Fong},
  {Karr}, {Kimura}, {King}, {Koike}, {Komatsu}, {Le Brun}, {Le F{\`e}vre}, {Le
  Fur}, {Le Mignant}, {Ling}, {Loomis}, {Lupton}, {Madec}, {Mao}, {Marchesini},
  {Marrara}, {Medvedev}, {Mineo}, {Minowa}, {Murayama}, {Murray}, {Ohyama},
  {Onodera}, {Orndorff}, {Pascal}, {Peebles}, {Pernot}, {Pourcelot}, {Reiley},
  {Reinecke}, {Roberts}, {Rosa}, {Rousselle}, {Schmitt}, {Schwochert},
  {Seiffert}, {Siddiqui}, {Smee}, {Sodr{\'e}}, {Steinkraus}, {Strauss},
  {Surace}, {Tait}, {Takada}, {Tamura}, {Tanaka}, {Tanaka}, {Thakar},
  {Verducci}, {Vibert}, {Wang}, {Wang}, {Wen}, {Werner}, {Yamada}, {Yan},
  {Yasuda}, {Yoshida}, \& {Yoshida}}]{tamura18}
{Tamura}, N., {Takato}, N., {Shimono}, A., {et~al.} 2018, in Society of
  Photo-Optical Instrumentation Engineers (SPIE) Conference Series, Vol. 10702,
  Ground-based and Airborne Instrumentation for Astronomy VII, ed. C.~J.
  {Evans}, L.~{Simard}, \& H.~{Takami}, 107021C

\bibitem[{{Tanaka}(2015)}]{tanaka15}
{Tanaka}, M. 2015, \apj, 801, 20

\bibitem[{{Tanaka} {et~al.}(2018){Tanaka}, {Coupon}, {Hsieh}, {Mineo},
  {Nishizawa}, {Speagle}, {Furusawa}, {Miyazaki}, \& {Murayama}}]{tanaka18}
{Tanaka}, M., {Coupon}, J., {Hsieh}, B.-C., {et~al.} 2018, \pasj, 70, S9

\bibitem[{{Thibert} {et~al.}(2021){Thibert}, {Sawicki}, {Goulding}, {Arnouts},
  {Coupon}, \& {Gwyn}}]{thibert21}
{Thibert}, N., {Sawicki}, M., {Goulding}, A., {et~al.} 2021, Research Notes of
  the American Astronomical Society, 5, 144

\bibitem[{{Timlin} {et~al.}(2016){Timlin}, {Ross}, {Richards}, {Lacy}, {Ryan},
  {Stone}, {Bauer}, {Brandt}, {Fan}, {Glikman}, {Haggard}, {Jiang}, {LaMassa},
  {Lin}, {Makler}, {McGehee}, {Myers}, {Schneider}, {Urry}, {Wollack}, \&
  {Zakamska}}]{timlin16}
{Timlin}, J.~D., {Ross}, N.~P., {Richards}, G.~T., {et~al.} 2016, \apjs, 225, 1

\bibitem[{{Tonry} {et~al.}(2012){Tonry}, {Stubbs}, {Lykke}, {Doherty},
  {Shivvers}, {Burgett}, {Chambers}, {Hodapp}, {Kaiser}, {Kudritzki},
  {Magnier}, {Morgan}, {Price}, \& {Wainscoat}}]{tonry12}
{Tonry}, J.~L., {Stubbs}, C.~W., {Lykke}, K.~R., {et~al.} 2012, \apj, 750, 99

\bibitem[{{Yasuda} {et~al.}(2019){Yasuda}, {Tanaka}, {Tominaga}, {Jiang},
  {Moriya}, {Morokuma}, {Suzuki}, {Takahashi}, {Yamaguchi}, {Maeda}, {Sako},
  {Ikeda}, {Kimura}, {Morii}, {Ueda}, {Yoshida}, {Lee}, {Suyu}, {Komiyama},
  {Regnault}, \& {Rubin}}]{yasuda19}
{Yasuda}, N., {Tanaka}, M., {Tominaga}, N., {et~al.} 2019, \pasj, 71, 74

\end{thebibliography}
